\documentclass[structabstract]{aa}
\usepackage{txfonts,epsfig,graphicx,natbib,url,twoopt}
\usepackage{aalongtable}
\usepackage[breaklinks=true]{hyperref} 
\usepackage[labelfont=footnotesize,textfont=footnotesize]{caption}  

\bibpunct{(}{)}{;}{a}{}{,}    
\newcommandtwoopt{\citeads}[3][][]{\href{http://adsabs.harvard.edu/abs/#3}%
                                        {\citealp[#1][#2]{#3}}}
\newcommandtwoopt{\citepads}[3][][]{\href{http://adsabs.harvard.edu/abs/#3}%
                                         {\citep[#1][#2]{#3}}}
\newcommandtwoopt{\citetads}[3][][]{\href{http://adsabs.harvard.edu/abs/#3}%
                                         {\citet[#1][#2]{#3}}}
\newcommandtwoopt{\citeyearrads}%
  [3][][]{\href{http://adsabs.harvard.edu/abs/#3}%
                                         {\citeyear[#1][#2]{#3}}}

\begin{document}   

\title{A sample of relatively unstudied star clusters in the Large Magellanic Cloud: fundamental parameters determined from Washington photometry\thanks{Table 3 is only available in electronic form at the CDS via anonymous ftp to cdsarc.u-strasbg.fr (130.79.128.5) or via http://cdsweb.u-strasbg.fr/cgi-bin/qcat?J/A+A/}} 
\author{T. Palma\inst{1,2} \and J.J. Clari\'a\inst{1,2} \and D. Geisler\inst{3} \and A.E. Piatti\inst{2,4} \and A.V. Ahumada\inst{1,2} } 

\institute{Observatorio Astron\'mico, Universidad Nacional de C\'ordoba, Laprida 854, C\'ordoba, CP 5000\\ \email{tali@mail.oac.uncor.edu} 
\and 
Consejo Nacional de Investigaciones Cient\'ificas y T\'ecnicas (CONICET)
\and
Grupo de Astronom\'ia, Departamento de Astronom\'ia, Universidad de Concepci\'on, Casilla 160-C, Concepci\'on, Chile 
\and
Instituto de Astronom\'ia y F\'isica del Espacio, CC67, Suc. 28, 1428, Ciudad de Buenos Aires, Argentina 
 }
\date{ Received November 26, 2012; accepted April 16, 2013}

\abstract
   {}
   {To enlarge our growing sample of well-studied star clusters in the Large Magellanic Cloud (LMC), we present CCD Washington 
   $CT_1$ photometry to $T_1 \sim$ 23 in the fields of twenty-three mostly unstudied clusters located in the inner disc and 
   outer regions of the LMC.}
   {We estimated cluster radii from star counts. Using the cluster Washington $(T_1,C-T_1)$ colour-magnitude diagrams, statistically cleaned from field star contamination, we derived cluster ages and metallicities from a comparison with theoretical isochrones of 
   the Padova group. Whenever possible, we also derived ages using $\delta T_1$ - the magnitude difference between the red giant clump and the main sequence turn off - and estimated metallicities from the standard giant branch procedure. We enlarged our sample by 
   adding clusters with published ages and metallicities determined on a similar scale by applying the same methods. We examined relationships between their positions in the LMC, ages and metallicities.}
   {We find that the two methods for age and metallicity determination agree well with each other. Fourteen clusters are found to be intermediate-age clusters (1-2 Gyr), with [Fe/H] values ranging from -0.4 to -0.7. The remaining nine clusters turn out to be younger than 1 Gyr, with metallicities between 0.0 and -0.4. }
   {Our 23 clusters represent an increase of $\sim$ 30\% in the current total amount number of well-studied LMC clusters using 
   Washington photometry. In agreement with previous studies, we find no evidence for a metallicity gradient. We also find that the 
   younger clusters were formed closer to the LMC centre than the older ones.}


   \titlerunning{fundamental parameters of star clusters in the LMC }
   \authorrunning{T. Palma et al.}
   \maketitle
\section{Introduction}     \label{sec:introduction}

The Large Magellanic Cloud (LMC) has become one of the most-studied objects in the past three decades. In particular, studying the LMC star cluster system advances our understanding of the chemical enrichment and star-formation history of this galaxy as a whole \citepads[e.g.,][]{baum12}, with the important caveat that a large portion of the LMC's history cannot be studied via clusters because of the infamous age gap \citepads[]{g97}. However, the number of well-studied clusters in the LMC still constitutes a very small fraction of those that have been catalogued, and thorough investigations of even a handful of clusters can significantly improve our knowledge of the chemical enrichment history of this critical galaxy.  \\

The current study represents a step forward in the systematic study of LMC clusters carried out as uniformly as possible using 
the Washington photometric system. Although initially developed for late-type stars and old stellar populations \citepads[]{canterna}, 
the Washington system has been widely applied to intermediate-age and old clusters in the Galaxy and in the Magellanic Clouds 
\citepads[e.g.,][]{g97,gs99,p03a,p12}. It is our purpose to derive ages and metallicities for a sample of 23 mostly unstudied LMC star 
clusters with the aim of adding them to our growing sample of well-studied clusters. The reasons why we have chosen to work in 
this photometric system and its advantages for this type of study have already been described in previous papers 
\citep[e.g.,][]{g97,petal11}.\\

The cluster sample is presented in Section 2. The observations and reductions are described in Section 3. The procedure followed to 
estimate cluster radii from the stellar density profiles is described in Section 4. We also include in this section the method 
applied to minimize the field-star contamination in the Washington $(T_1,C-T_1)$ colour-magnitude diagrams (CMDs) and the 
estimation of the cluster fundamental properties. A brief analysis and discussion of the results is presented in Section 5, 
while Sect. 6 summarizes our results.\\   

\section{Cluster sample}  \label{sec:sample}

\begin{figure}
 \resizebox{\hsize}{!}{\includegraphics{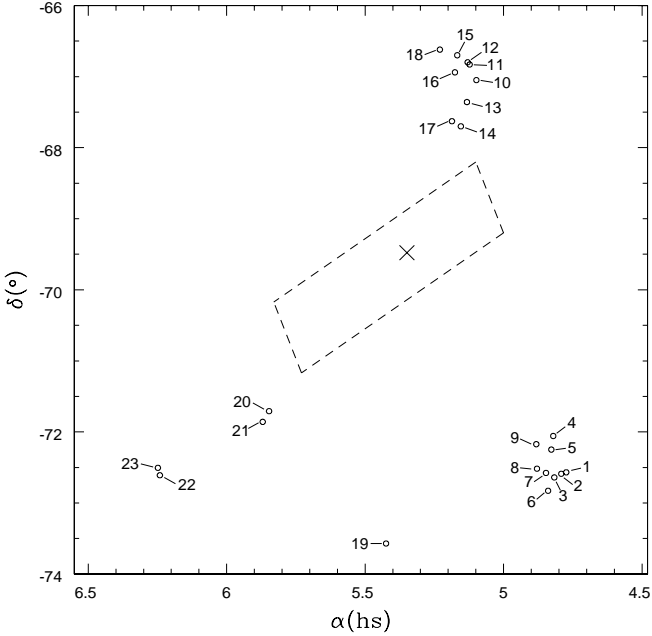}}
 \caption{Position of our target clusters (open circles) in relation to the LMC bar (dashed lines) and  geometrical centre (cross). (1) SL,33, (2) SL\,41, (3) KMHK\,123, (4) KMHK\,128, (5) LW\,69, (6) KMHK\,151, (7) SL\,54, (8) SL\,73, (9) SL\,72, (10) BSDL\,594, (11) BSDL\,654, (12) BSDL\,665, (13) BSDL\,675, (14) HS\,130, (15) BSDL\,761, (16) BSDL\,779, (17) HS\,156, (18) HS\,178, (19) LW\,211, (20) C\,11, (21) BSDL\,3158, (22) KMHK\,1702, (23) SL\,870.} 
 \label{f:fig1}
\end{figure}

After a careful revision of the Washington wide-field images of 21 LMC regions (see Section 3), we selected those star clusters located outside the bar for the present study that appeared to be unstudied or were only poorly studied. Our final sample includes a total of 23 mostly unstudied clusters, eleven of which lie in the inner disc of the LMC but outside the bar. The remaining twelve are located in the outer region (Fig. \ref{f:fig1}). Here we adopt the definition presented in \citetads[]{b98} in the sense that the inner disc is that region where the mean field turnoff becomes as bright as the clump. This takes place at a deprojected radius of $\sim$ 4$^{\circ}$. The cluster sample is presented in Table \ref{t:coord}, where we list the various star cluster designations from different catalogues (column 1), 2000.0 equatorial and galactic coordinates (columns 2-5), and the cluster radii given in \citetads[]{b08} (column 6). These radii constitute half of the mean apparent diameters obtained by computing the average between the major (a) and minor (b) axes. The last two columns of Table 1 (columns 7,8) list the cluster radii derived in the current study in pixels and parsecs, 
respectively (see Section 4.1). \\

\begin{table*}[!ht]
 \caption{Observed star clusters in the LMC.}
  \label{t:coord}
  \centering
  \begin{tabular}{lccccccc}
  \hline \hline
   Star cluster \tablefootmark{a}  &  $\alpha_{2000}$ & $\delta_{2000}$ & l & b & r \tablefootmark{b} & r$_{cls}$ & r$_{cls}$\\
        & (h m s) & ($\circ$ ' '') & ($\circ$) & ($\circ$) & (') & (px) & (pc)\\
 \hline
SL\,33, LW\,59, KMHK\,91 & 04 46 25 & -72 34 06 & 284.717 & -34.986 & 0.55 & 200 & 13.1 \\  
SL\,41, LW\,64, KMHK\,105 & 04 47 30 & -72 35 18 & 284.704 & -34.903 & 0.72 & 220 & 14.4 \\
KMHK\,123               & 04 49 00 & -72 38 24 & 284.713 & -34.780 & 0.30 & 110 & 7.2 \\
KMHK\,128               & 04 49 14 & -72 03 24 & 285.177 & -34.613 & 0.26 & 110 & 7.2 \\     
LW\,69, KMHK\,137       & 04 49 39 & -72 14 53 & 284.246 & -34.874 & 0.32 & 120 & 7.8 \\
KMHK\,151               & 04 50 21 & -72 49 39 & 284.881 & -34.619 & 0.27 & 170 & 11.1 \\
SL\,54, LW\,78, KMHK\,162 & 04 50 48 & -72 34 36 & 284.582 & -34.677 & 0.55 & 200 & 13.1 \\
SL\,73, LW\,86, KMHK\,214 & 04 52 45 & -72 31 04 & 284.454 & -34.561 & 0.34 & 190 & 12.4 \\
SL\,72, LW\,87, KMHK\,217 & 04 52 54 & -72 10 21 & 284.054 & -34.667 & 0.42 & 160 & 10.5\\
BSDL\,594, LOGLE\,87    & 05 05 53 & -67 02 58 & 277.678 & -35.039 & 0.42 & 140 & 9.2 \\
BSDL\,654,LOGLE\,123    & 05 07 21 & -66 49 45 & 277.377 & -34.949 & 0.22 & 75 & 4.9 \\
BSDL\,665, LOGLE\,130   & 05 07 47 & -66 47 53 & 277.329 & -34.914 & 0.22 & 60 & 3.9 \\ 
BSDL\,675, LOGLE\,134   & 05 07 56 & -67 21 28 & 277.990 & -34.776 & 0.29 & 90 & 5.9 \\
HS\,130, KMHK\,588      & 05 09 15 & -67 41 59 & 278.362 & -34.577 & 0.27 & 90 & 5.9 \\
BSDL\,761               & 05 10 02 & -66 41 57 & 277.155 & -34.717 & 0.32 & 90 & 5.9 \\
BSDL\,779, LOGLE\,182   & 05 10 32 & -66 56 24 & 277.428 & -34.619 & 0.22 & 80 & 5.2 \\
HS\,156, H88-190, KMHK\,632, LOGLE\,199 & 05 11 11 & -67 37 36 & 278.227 & -34.414 & 0.25 & 120 & 7.8 \\
HS\,178, KMHK\,667      & 05 13 48 & -66 37 12 & 276.970 & -34.367 & 0.33 & 120 & 7.8 \\
LW\,211, KMHK\,901      & 05 25 27 & -73 34 13 & 284.858 & -31.979 & 0.33 & 160 & 10.5 \\ 
C11                     & 05 50 48 & -71 42 28 & 282.371 & -30.397 & 0.20 & 150 & 9.8 \\
BSDL\,3158              & 05 52 11 & -71 51 30 & 282.533 & -30.276 & 0.46 & 220 & 14.4 \\
KMHK\,1702              & 06 13 56 & -72 30 19 & 283.190 & -28.586 & 0.31 & 100 & 6.5 \\ 
SL\,870, LW\,440, KMHK\,1705 & 06 14 28 & -72 36 34 & 283.310 & -28.546 & 0.58 & 230 & 15.0 \\ 
  
\hline 
\end{tabular}
\tablefoot{
\tablefoottext{a}{Cluster identifications from (SL): \citetads{sl}; (LW): \citetads{lw}; (HS): \citetads{hs}; (C): \citetads{h75}; (H88): \citetads{h88}; (KMHK): \citetads{kmhk}; (LOGLE): \citetads{logle98,logle99}; (BSDL): \citetads{bsdl} }\\
\tablefoottext{b}{Taken from \citetads{b08}}
}
\end{table*}

\section{Data collection and reduction} 

The observations of the selected clusters were carried out with the ``V\'ictor Blanco'' 4-m telescope at Cerro Tololo Inter-American Observatory (CTIO, Chile), during the nights of 2000 December 29 and 30. Washington wide-field images of about 21 LMC regions were 
taken with the MOSAIC II camera, which consists of an 8K$\times$8K CCD detector array. One pixel of the MOSAIC wide-field camera subtends 0.27'' on the sky, resulting in a 36'$\times$36' field of view. We used the Washington C \citepads[]{canterna} and Kron-Cousins R filters to be consistent with our previous studies. As shown in \citetads[]{g96}, the R filter has a significantly higher throughput than the standard Washington $T_1$ filter, so that R magnitudes can be accurately transformed to yield $T_1$ magnitudes. In particular, this filter combination allowed us to derive accurate metallicities based on the standard giant branch method described in  \citetads[]{gs99}. The Washington C and Kron-Cousins R filters were used with typical exposure times of 450 and 150 seconds, respectively. We obtained a series of bias, dome, and sky flat-field exposures per filter to calibrate the CCD instrumental signature. Standard stars of selected areas SA\,98 and SA\,101 from the list in \citetads[]{g96} were also nightly observed to standardize our photometry. SA\,98 and SA\,101 contain 15 and 9 standard stars, respectively, with a wide range in colour. \\

The MOSAIC data were reduced using the MSCRED package within IRAF\footnote{IRAF is distributed by the National Optical Astronomy 
Observatories, which is operated by AURA, Inc., under contract with the National Science Foundation.} following the guide for mosaic reduction of \citetads[]{j03}. Stellar photometry was performed using the stand-alone DAOPHOT II provided by Peter Stetson. 
For star-finding and point spread function (PSF) fitting routines, and to combine all independent measurements, we used the DAOPHOT, ALLSTAR, DAOMATCH, and DAOMASTER programs \citepads[]{stet87}. The calibration between instrumental and standard magnitudes was obtained using the following equations:
\begin{equation}
c= a_1+(C-T_1)+T_1+a_2X_C+a_3(C-T_1) ,
\end{equation}
\begin{equation}
r= b_1+T_1+b_2X_{T_1}+b_3(C-T_1) ,
\end{equation}
where X is the effective airmass. Upper and lower-case letters represent standard and instrumental magnitudes, respectively. The 
coefficients $a_i$ and $b_i$ were derived nightly through the IRAF routine FITPARAM. The resulting mean calibration coefficients 
together with their errors are shown in Table \ref{t:std}. The nightly rms errors from the transformation to the standard system 
were 0.006 and 0.007 mag for $C$ and $T_1$, respectively, indicating that the nights had excellent photometric quality. We finally used about 15 standard stars. \\

\begin{table}[h]
 \caption{Standard system mean calibration coefficients.}
 \label{t:std}
 \centering
 \begin{tabular}{cc}
 \hline \hline
$C$ & $T_1$ \\
\hline
 $a_1=(0.039\pm0.013)$ & $b_1=(-0.667\pm0.011)$  \\
 $a_2=(0.249\pm0.007)$ & $b_2=(0.049\pm0.006)$  \\
 $a_3=(-0.098\pm0.003)$ & $b_3=(-0.020\pm0.003)$ \\
\hline
\end{tabular}
\end{table}   

\begin{figure}
 \resizebox{\hsize}{!}{\includegraphics{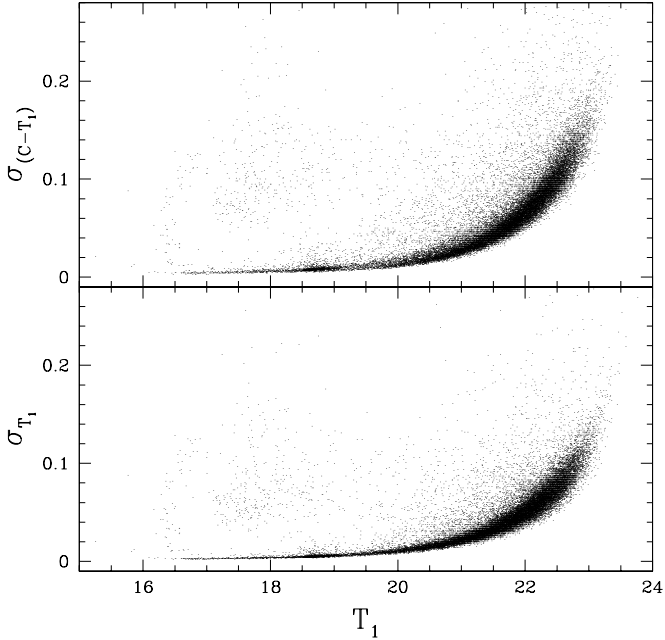}}
 \caption{$T_1$ magnitude and $C-T_1$ colour photometric errors as a function of $T_1$ for stars measured in the field of BSDL\,3158. } 
 \label{f:fig2}
\end{figure}

The full information gathered for each cluster consists of a running star number, the CCD x and y coordinates, the derived $T_1$ magnitude and $C-T_1$ colour, and the photometric errors $\sigma(T_1)$ and $\sigma(C-T_1)$. The behaviour of the $T_1$  and $C-T_1$ errors as a function of $T_1$ for the field of BSDL\,3158 is shown in Fig. \ref{f:fig2}. Only a portion of the Washington data obtained for the star cluster SL\,33 is shown here (see Table \ref{t:master}) for guidance regarding their form and content. 
The entire dataset for all  clusters can be obtained as supplementary material from the on-line version of the journal.   \\

\begin{table}[h]
 \caption{CCD $CT_1$ data of all stars in the field of SL\,33. }
 \label{t:master}
 \centering
 \begin{tabular}{ccccccc}
  \hline \hline
   ID  &  x & y &                                                                                                                                                                                             $T_1$ & $\sigma T_1$ & $C-T_1$ & $\sigma(C-T_1)$ \\
        & (px) & (px) & (mag) & (mag) & (mag) & (mag) \\
\hline
100 &  10.893 & 2847.002 & 22.363 & 0.060 & 0.689 & 0.071 \\ 
101 &  10.917 & 4041.597 & 21.874 & 0.038 & 1.165 & 0.064 \\
102 &  11.043 & 1675.679 & 20.334 & 0.012 & 1.679 & 0.022 \\
103 &  11.095 &  695.358 & 21.023 & 0.018 & 0.585 & 0.023 \\
104 &  11.106 & 3709.245 & 21.308 & 0.022 & 0.679 & 0.029 \\
105 &  11.135 & 3222.543 & 16.817 & 0.005 & 1.962 & 0.006 \\
106 &  11.155 & 1742.405 & 21.034 & 0.022 & 0.725 & 0.028 \\
107 &  11.211 &  723.802 & 21.527 & 0.025 & 0.670 & 0.035 \\
108 &  11.345 & 2543.743 & 21.005 & 0.020 & 0.526 & 0.025 \\
109 &  11.365 & 3961.926 & 21.809 & 0.035 & 0.943 & 0.050 \\
110 &  11.677 & 1627.802 & 18.183 & 0.004 & 2.153 & 0.007 \\
\hline 
\end{tabular}
\end{table}

\section{Data analysis}
\subsection{Cluster properties from star counts}

\begin{figure*}
\centering
 \includegraphics[width=30mm]{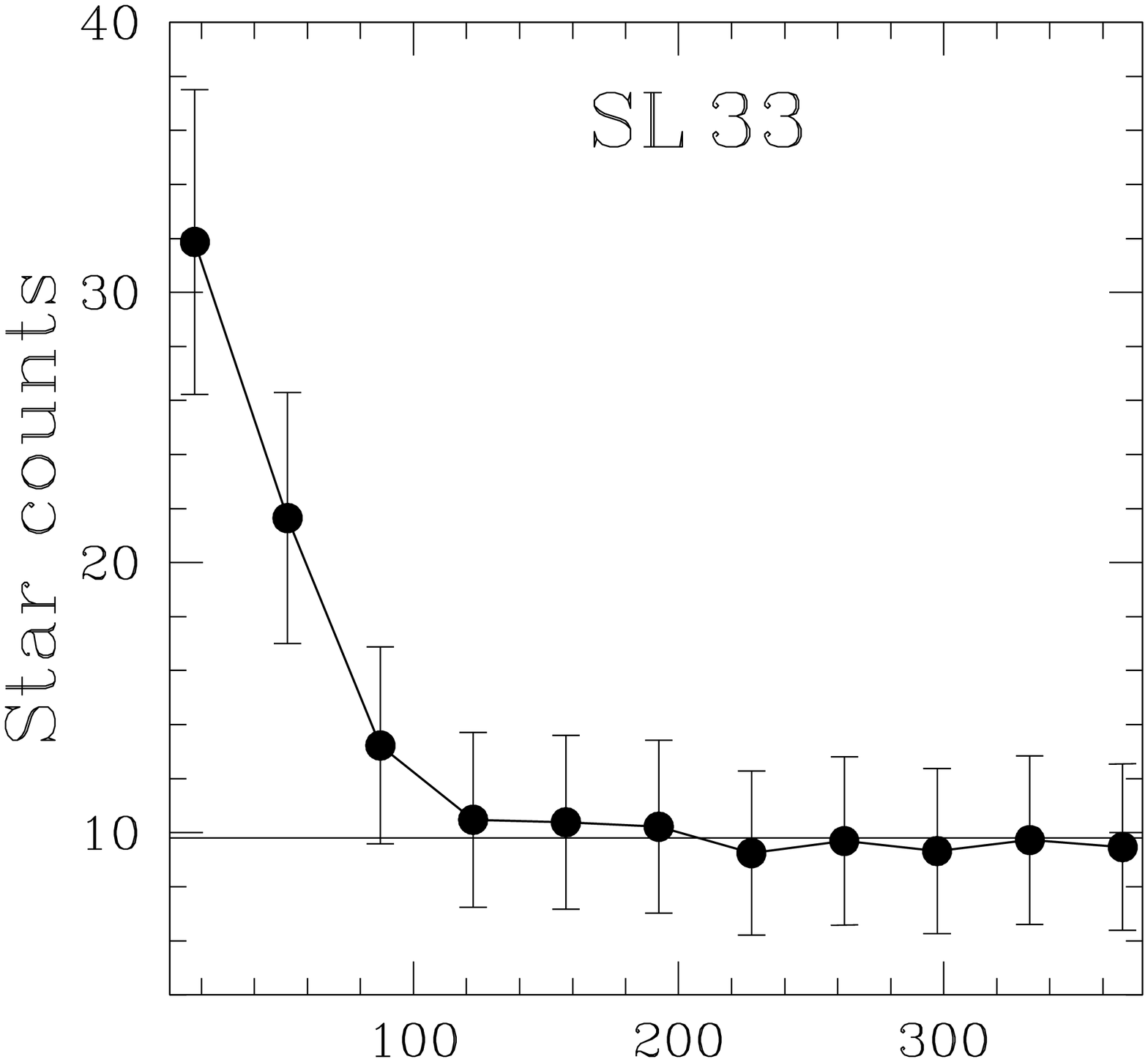}
 \includegraphics[width=30mm]{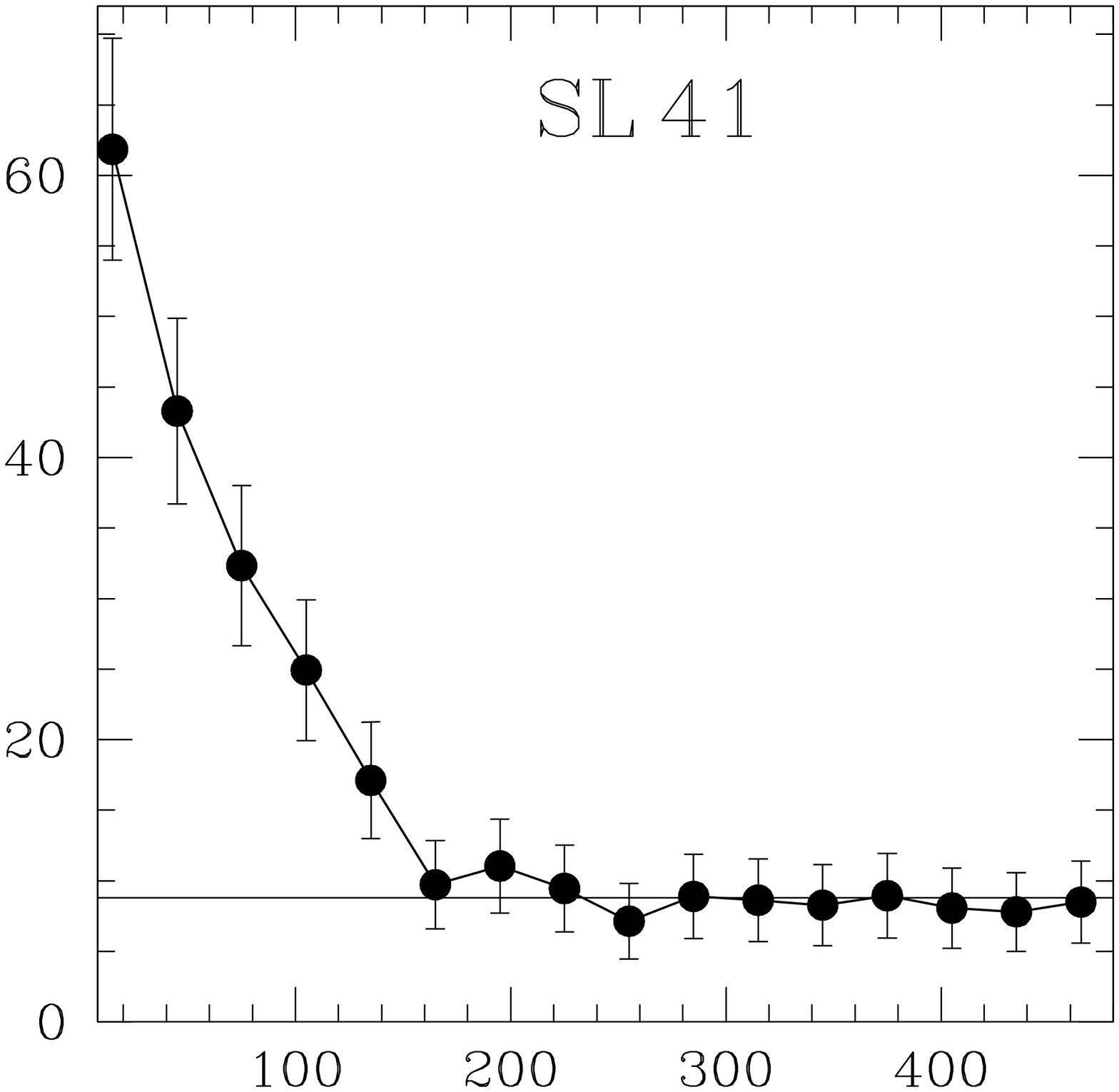}
 \includegraphics[width=30mm]{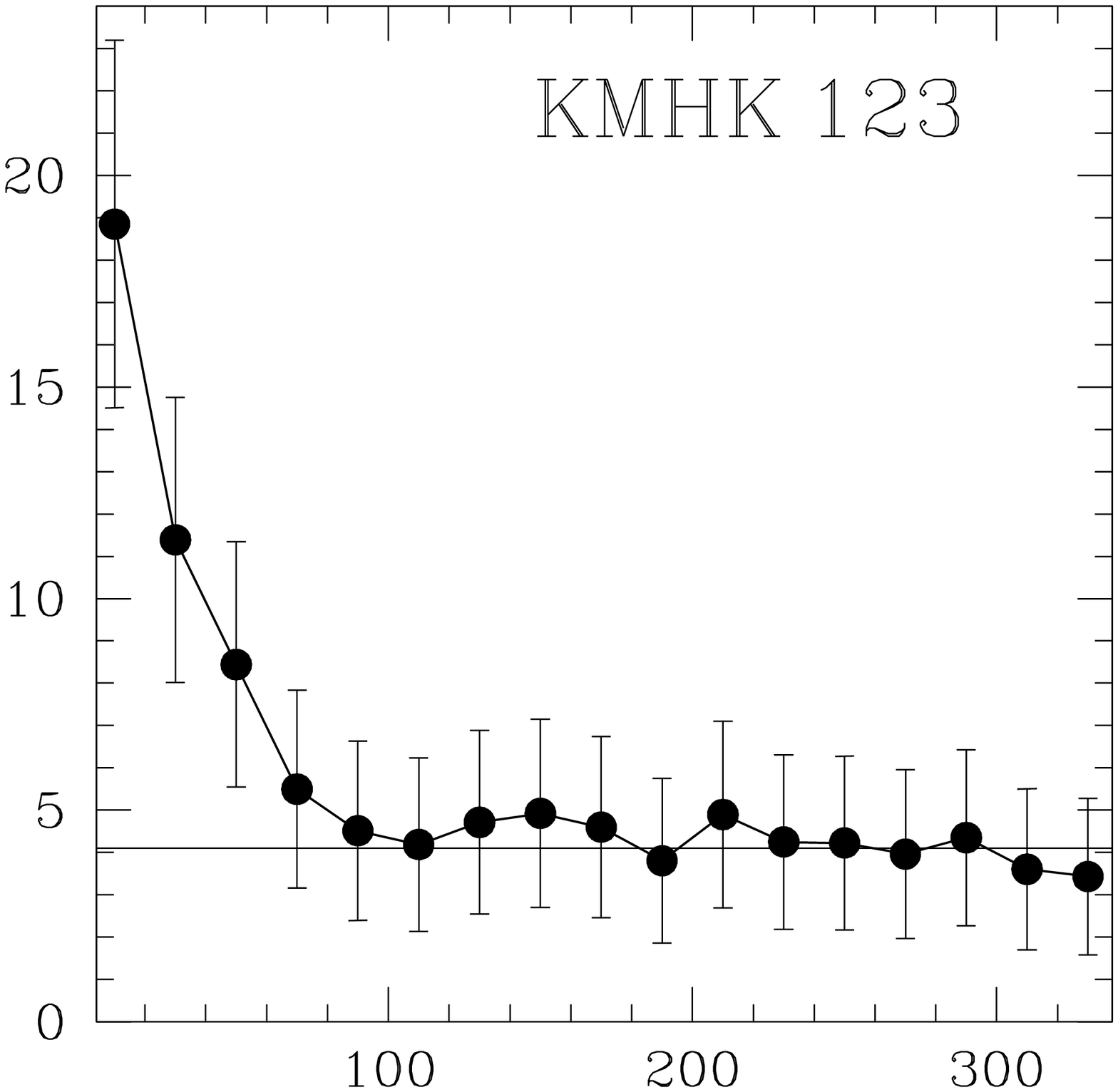}
 \includegraphics[width=30mm]{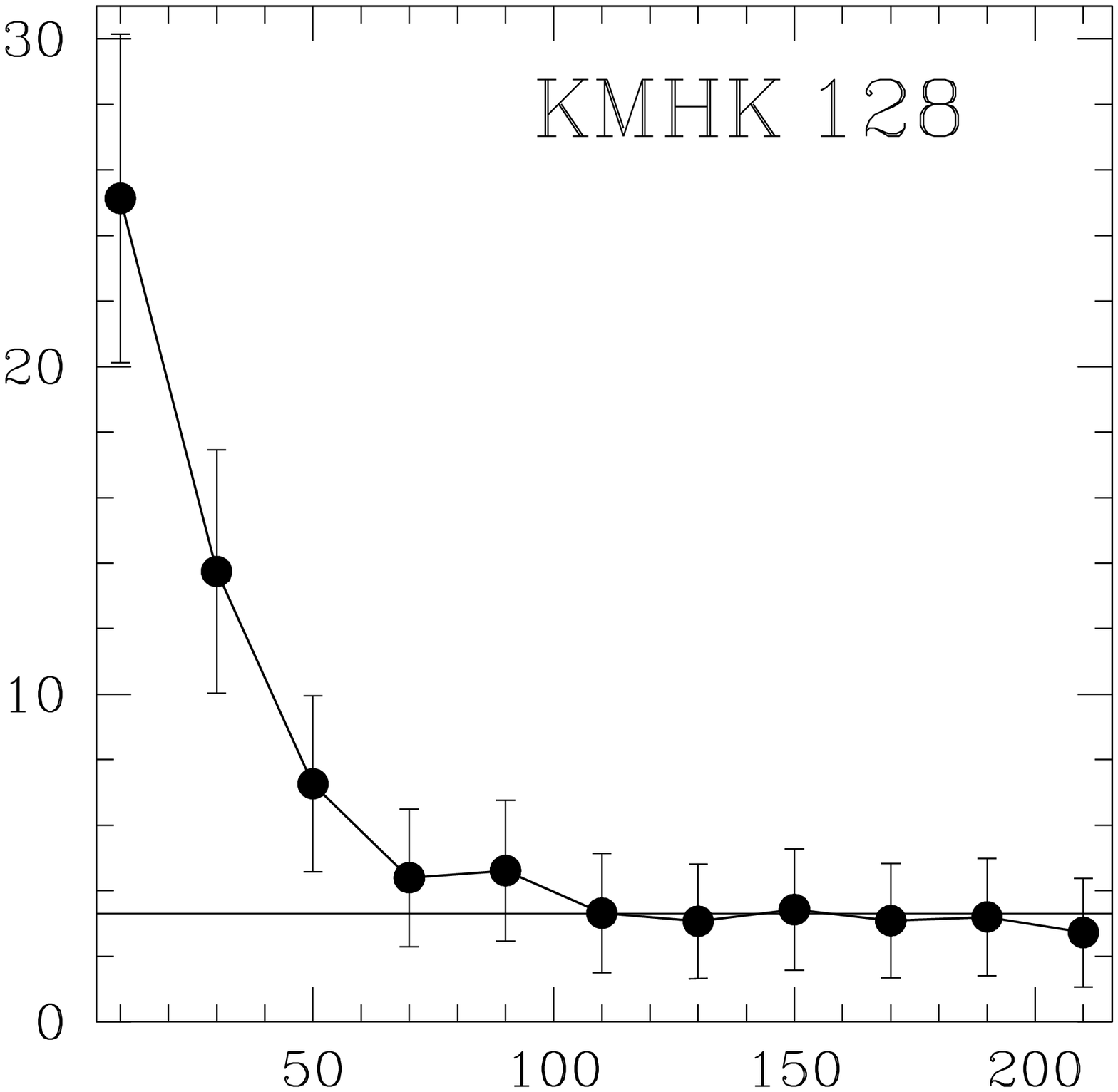}
 \includegraphics[width=30mm]{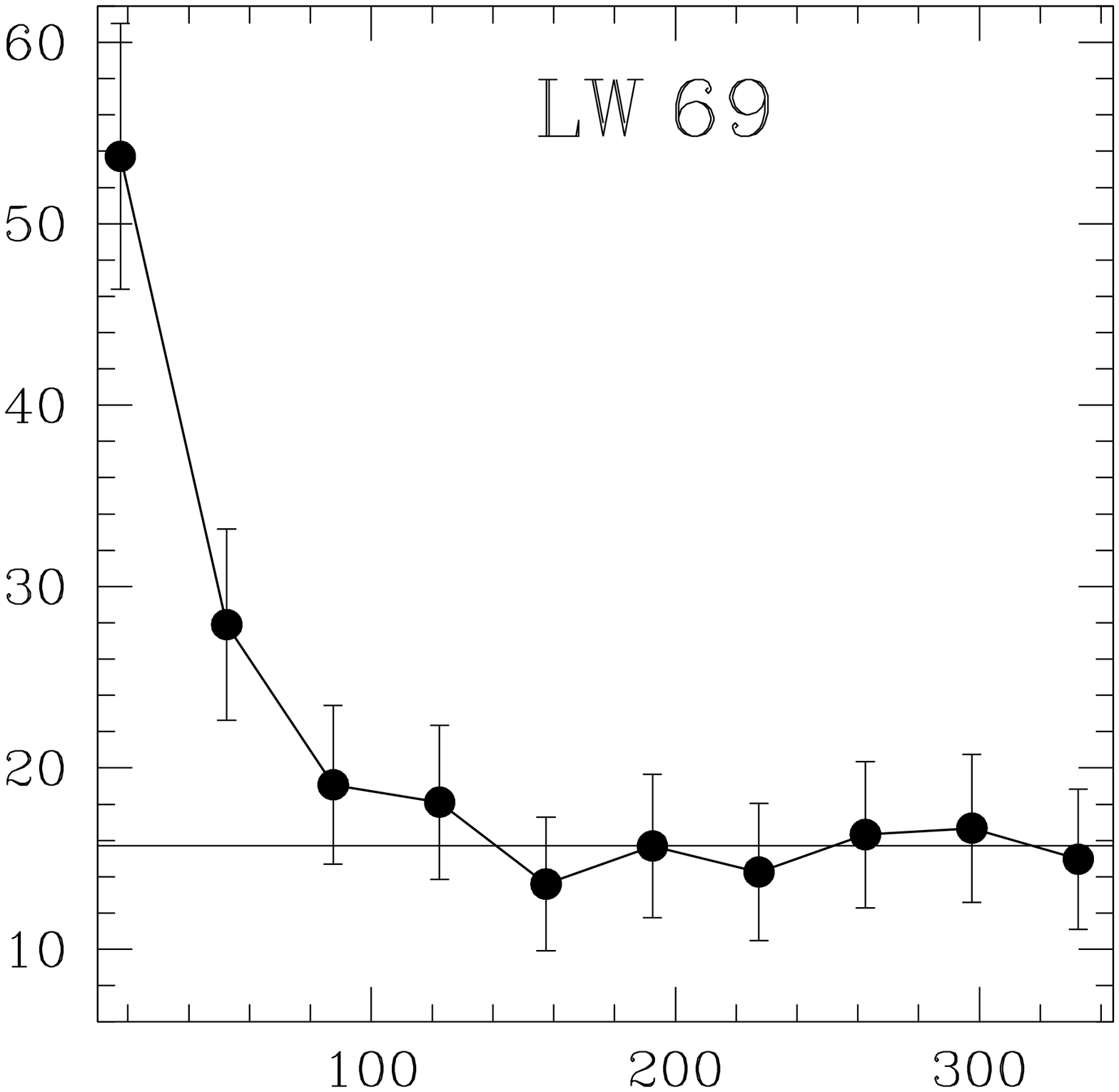}
 \includegraphics[width=30mm]{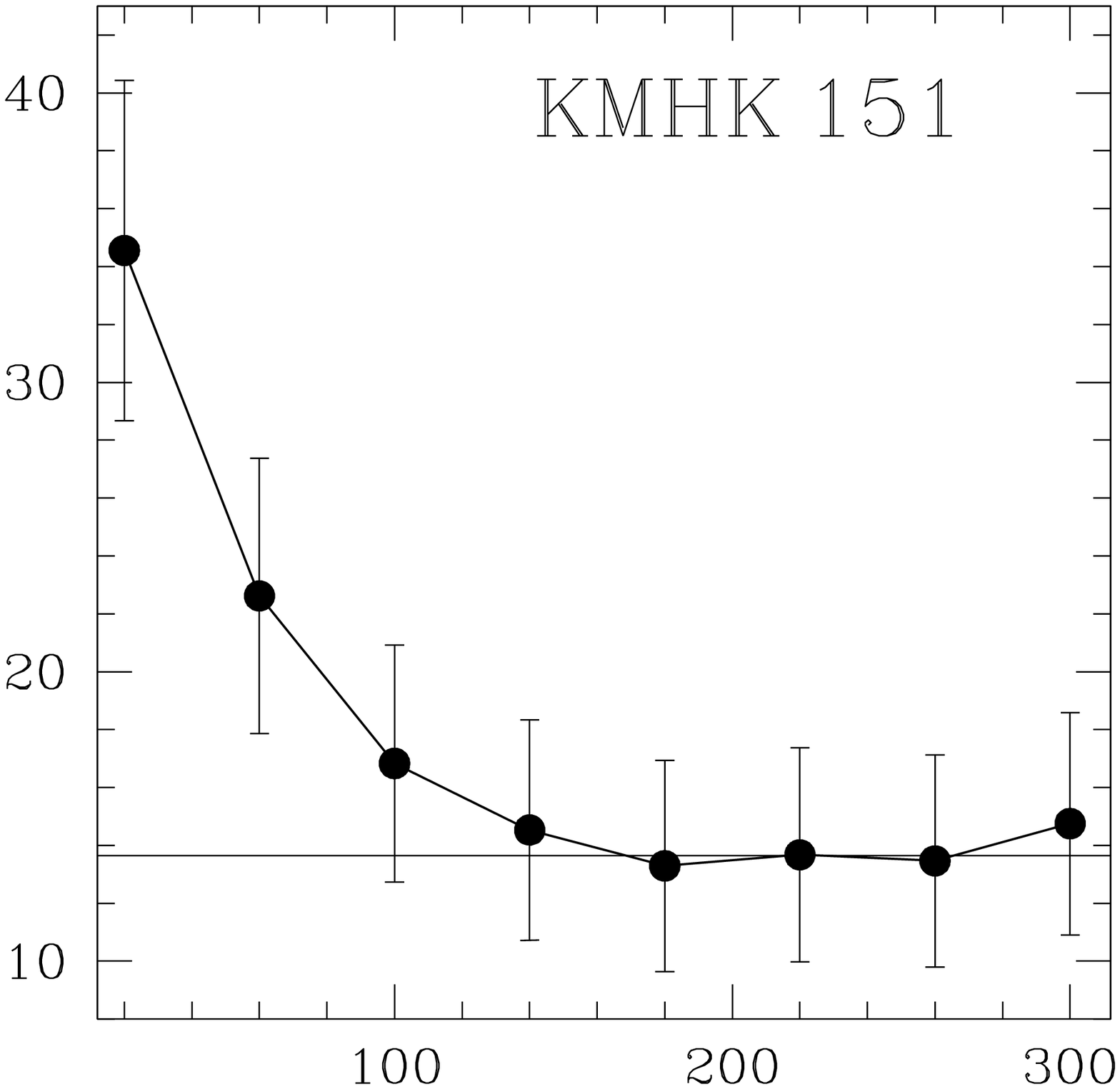}
 \includegraphics[width=30mm]{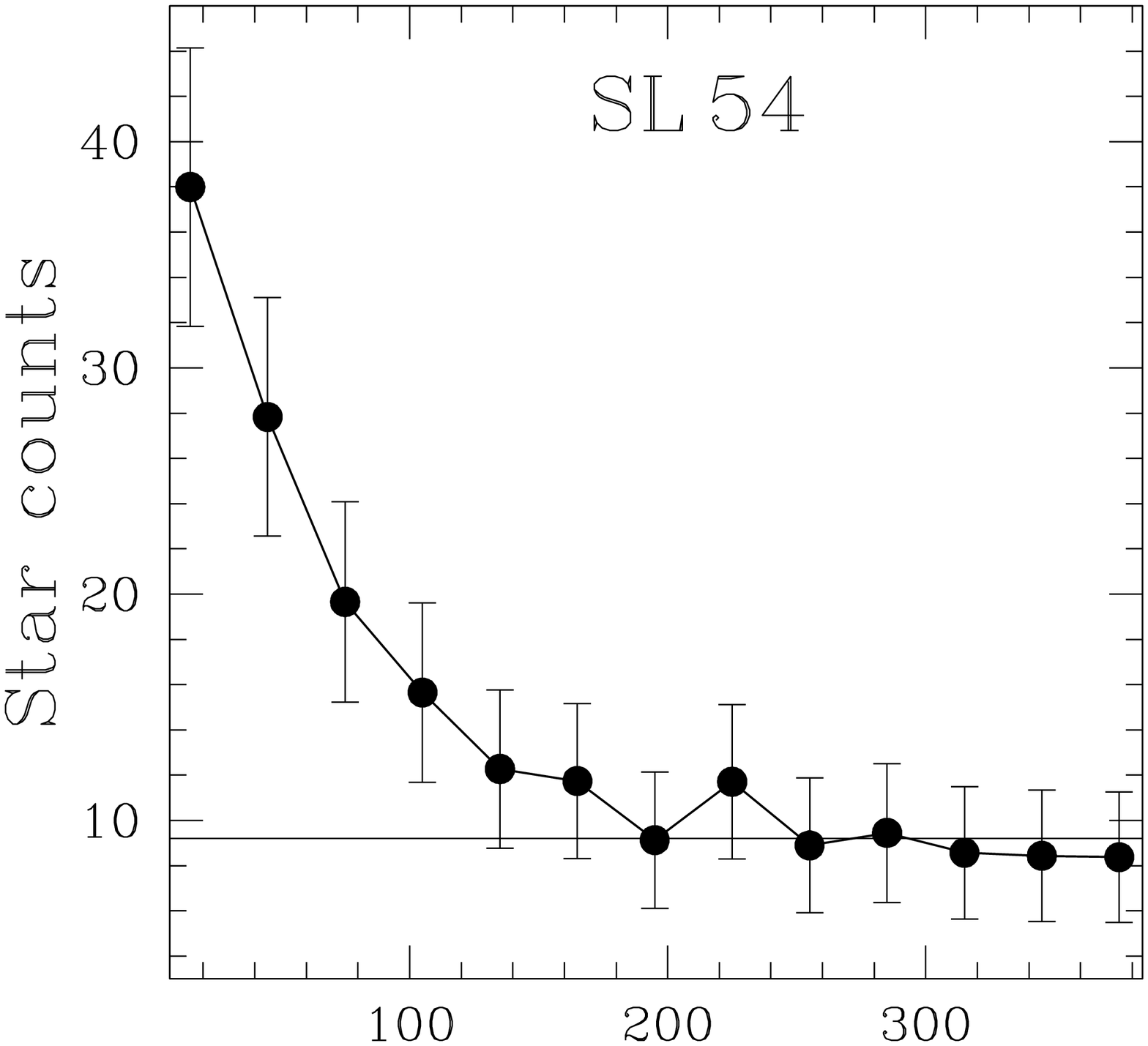}
 \includegraphics[width=30mm]{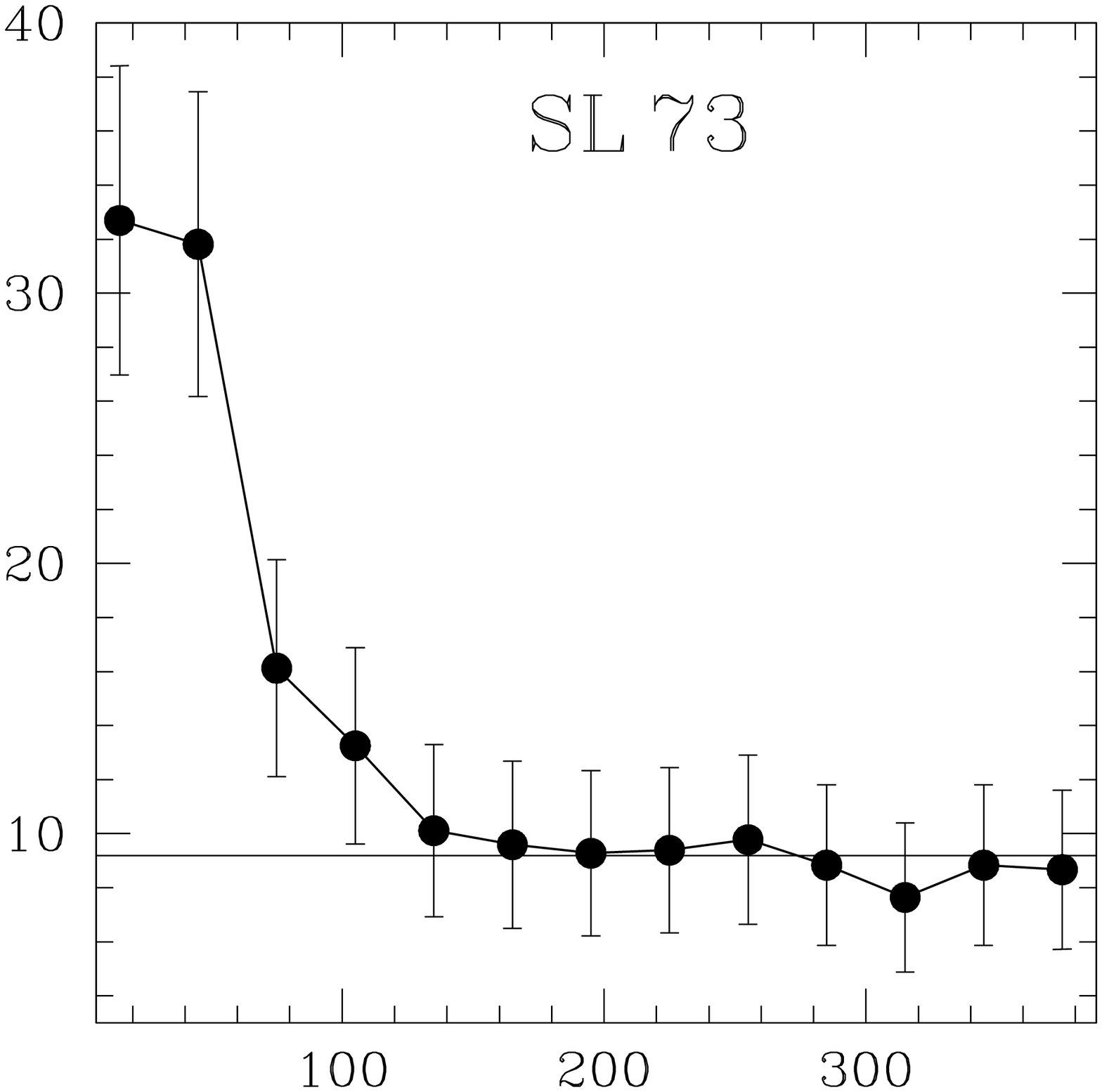}
 \includegraphics[width=30mm]{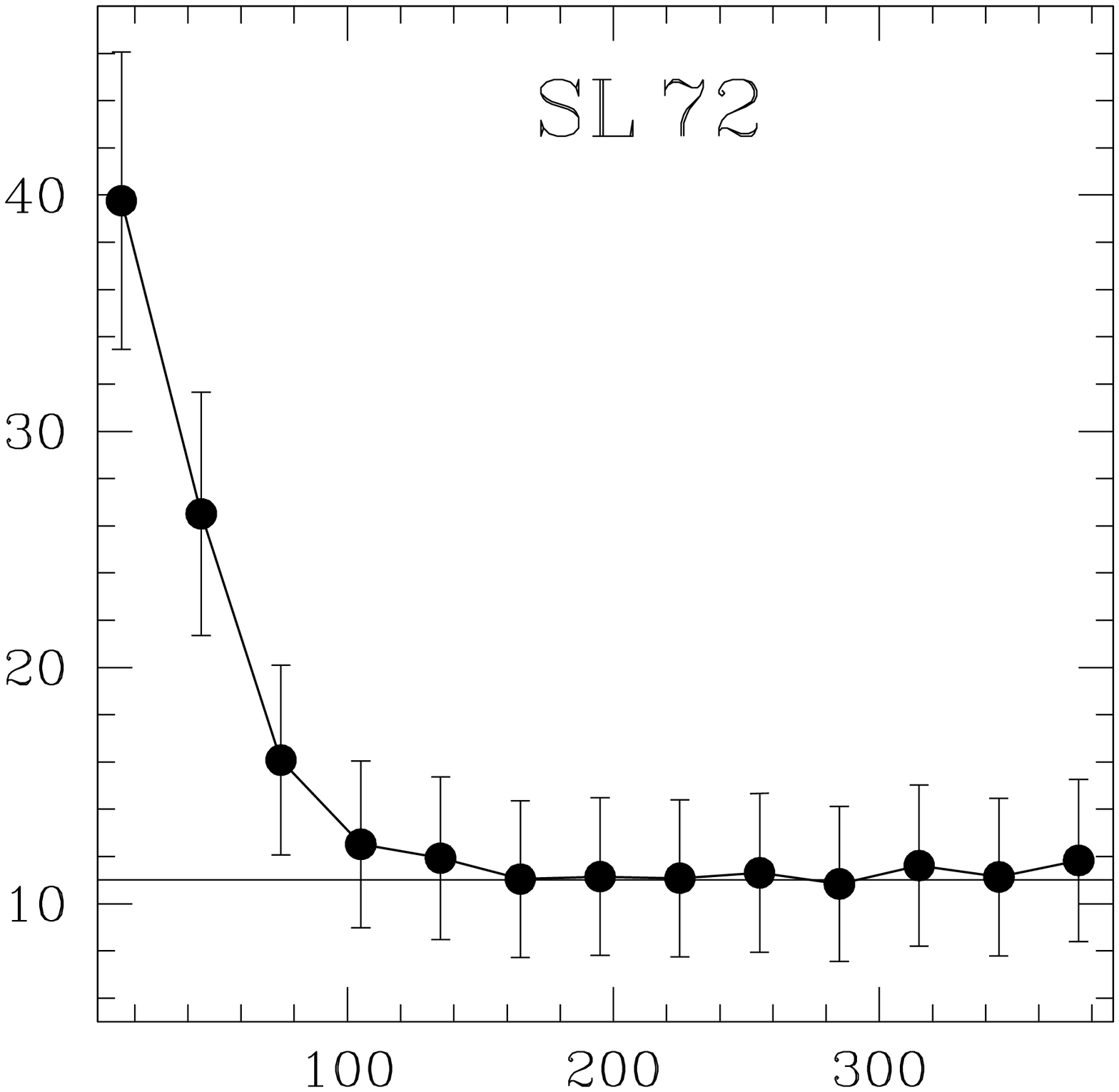}
 \includegraphics[width=30mm]{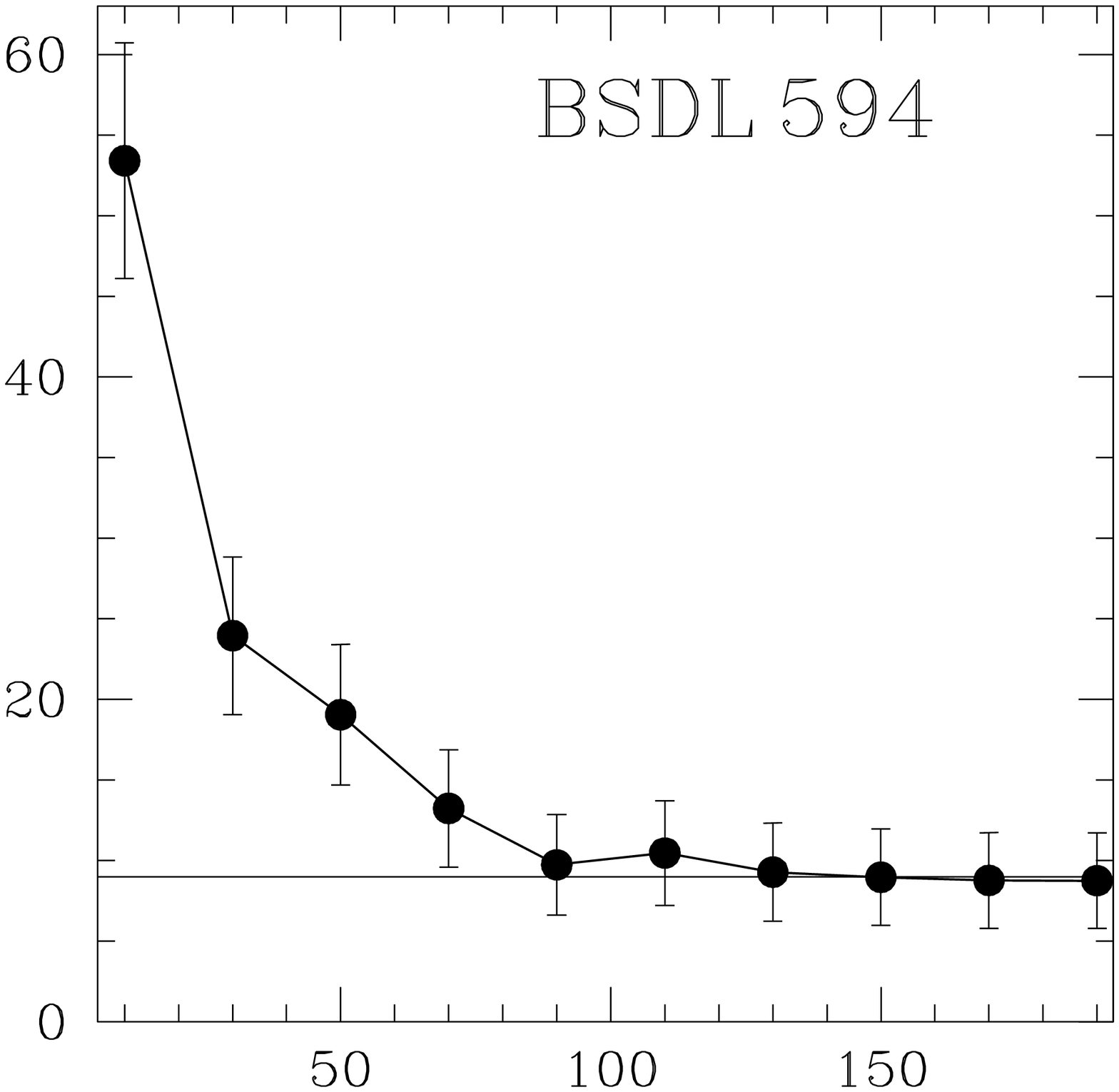}
 \includegraphics[width=30mm]{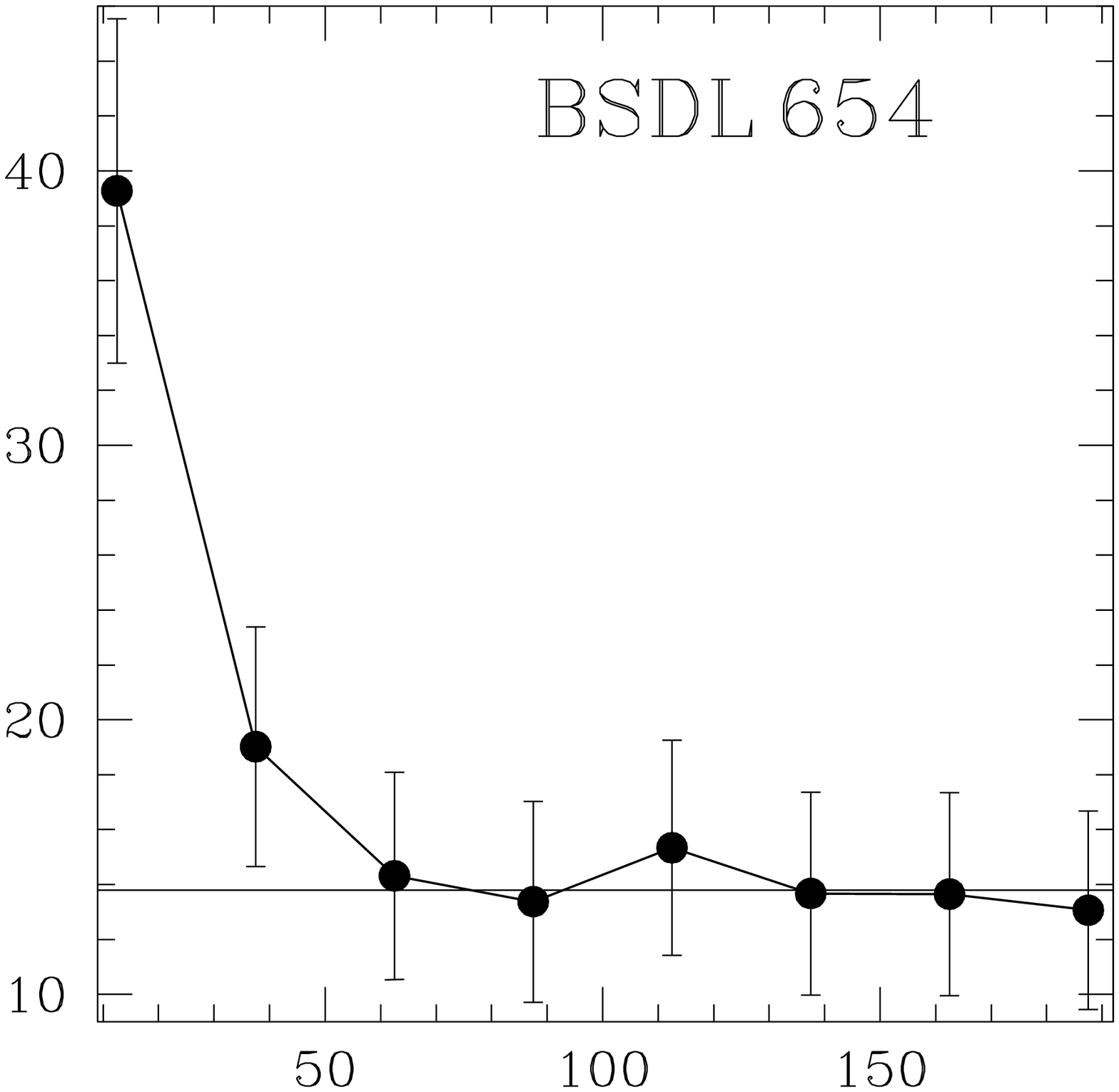}
 \includegraphics[width=30mm]{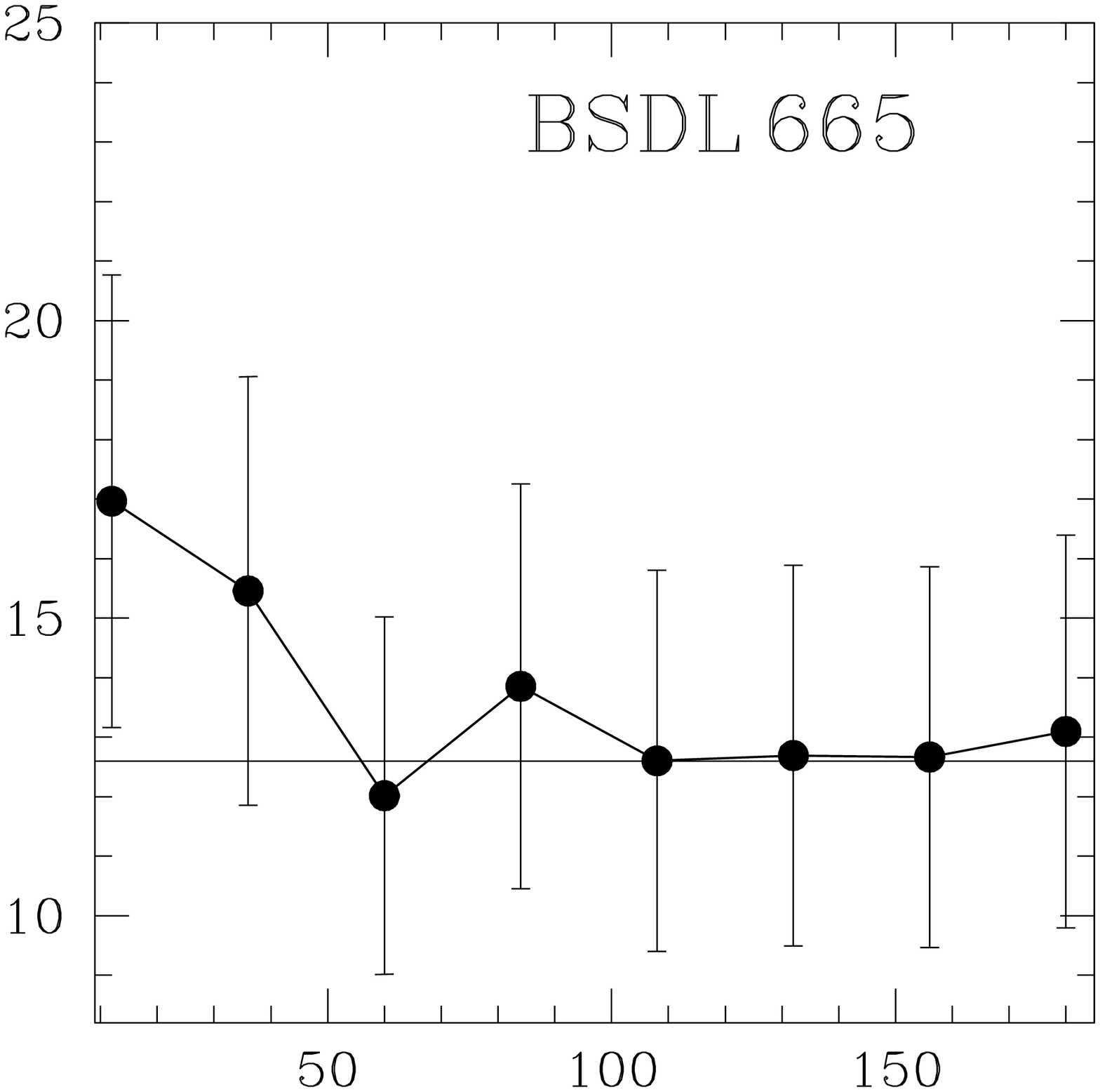}
 \includegraphics[width=30mm]{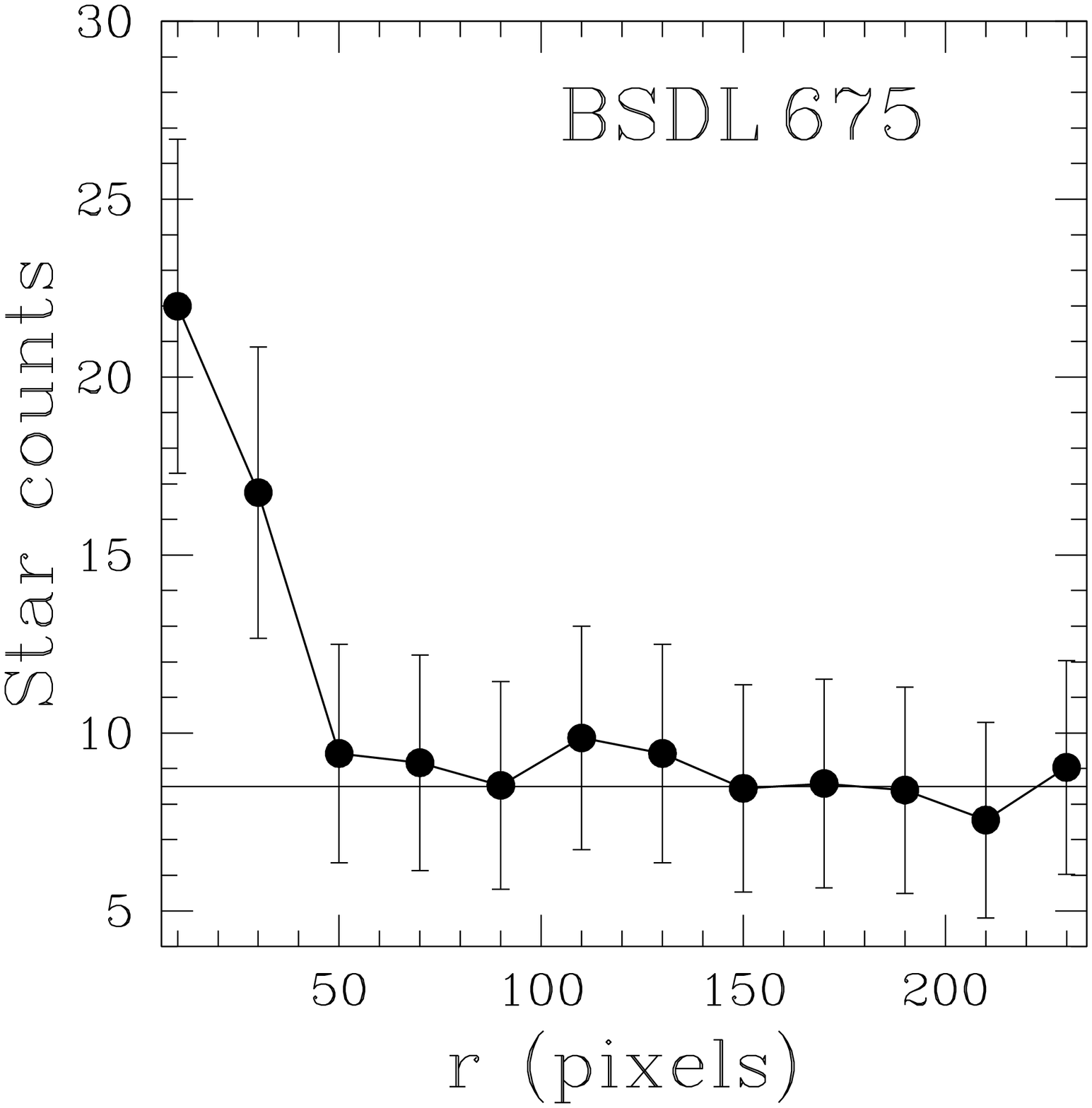}
 \includegraphics[width=30mm]{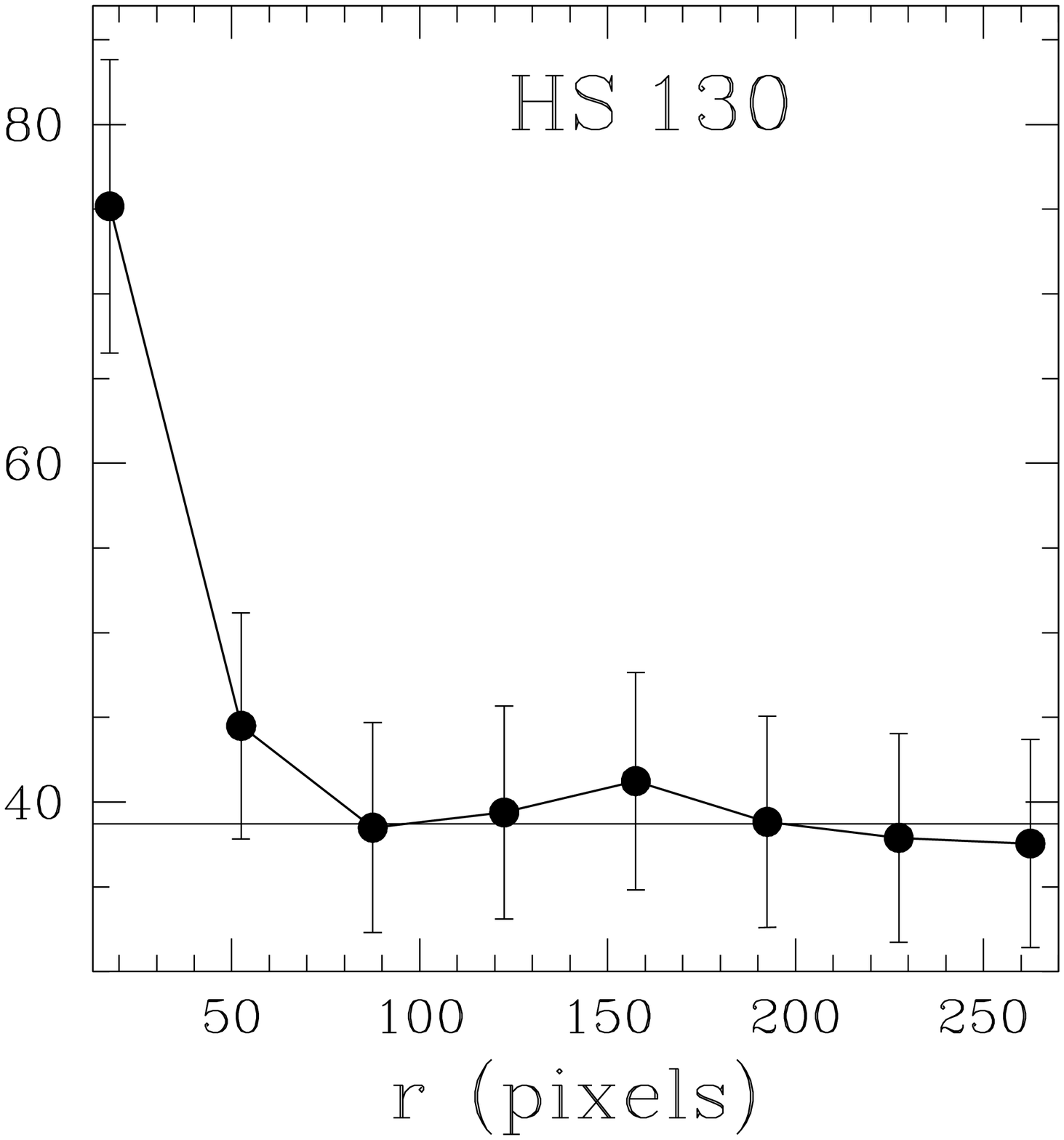}
 \includegraphics[width=30mm]{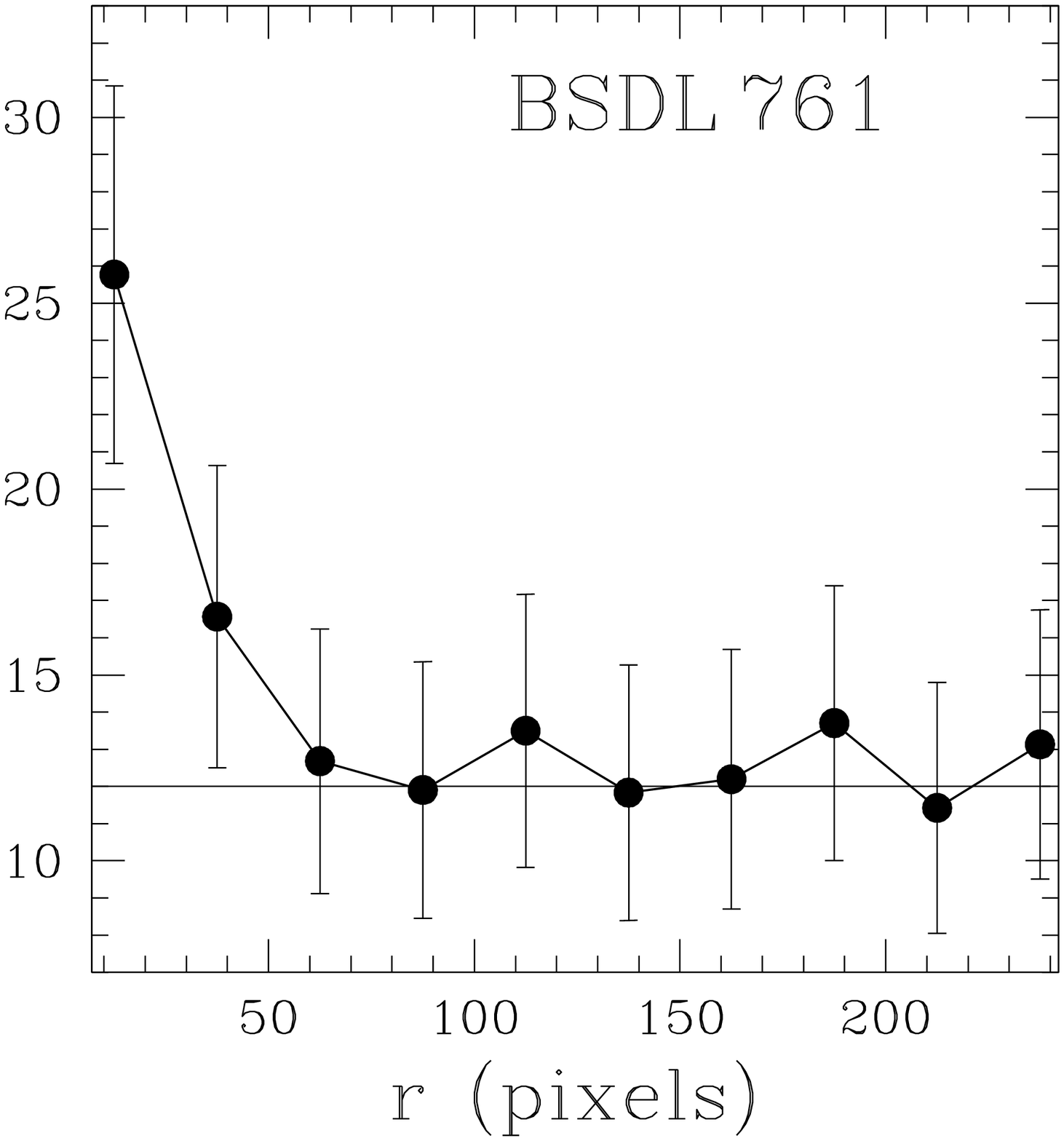}
 \includegraphics[width=30mm]{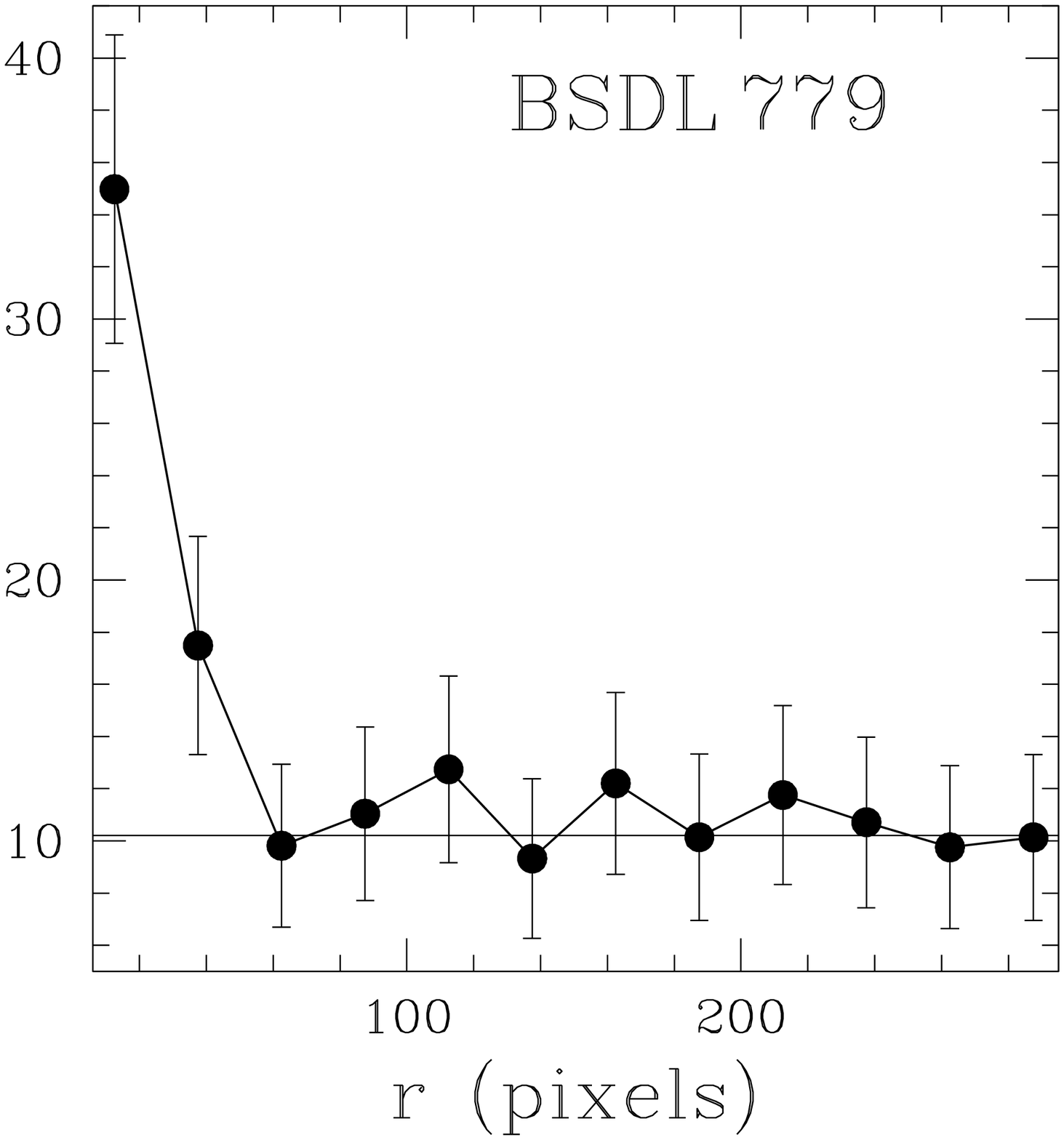}
 \includegraphics[width=30mm]{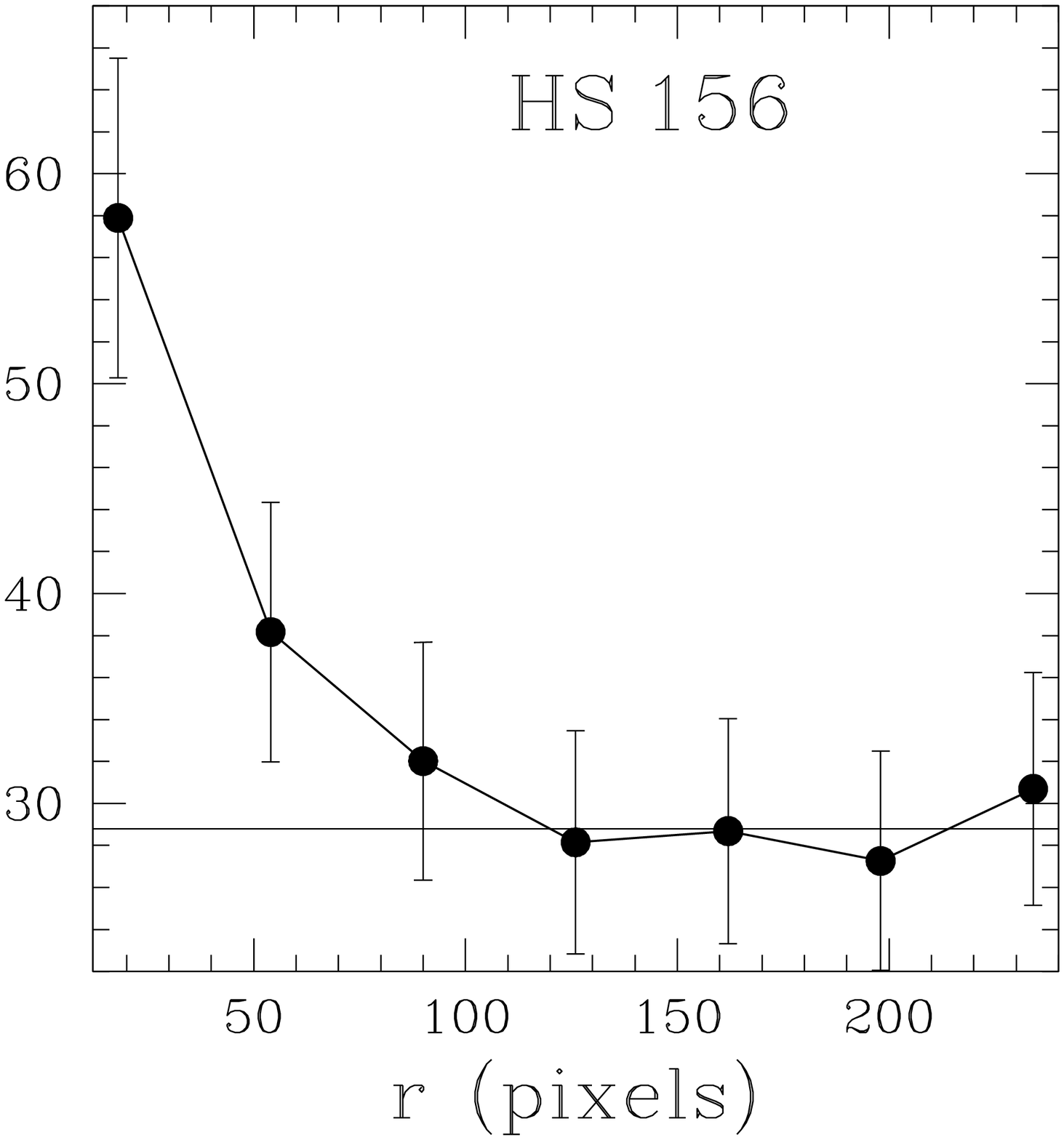}
 \includegraphics[width=30mm]{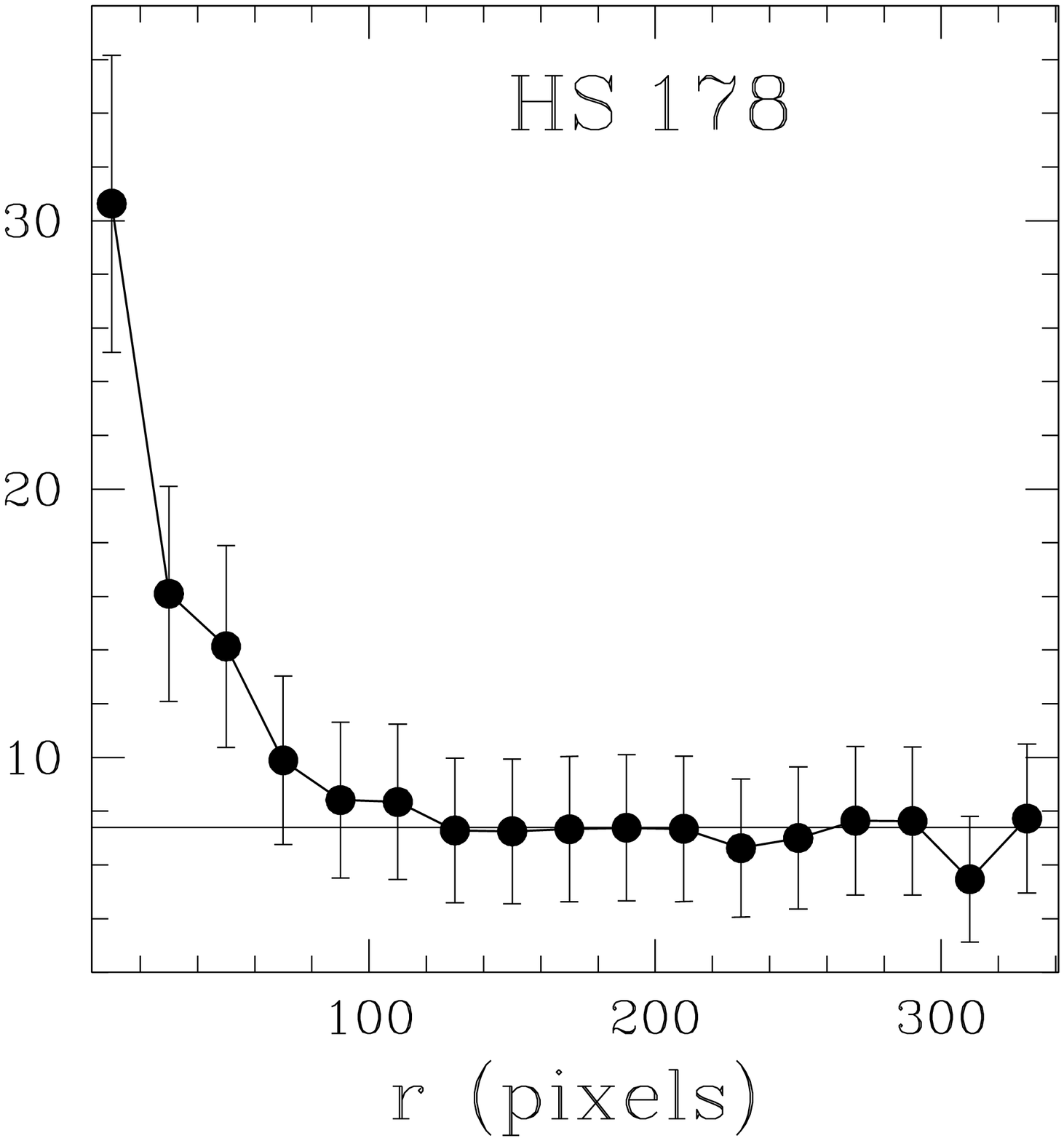}
 \includegraphics[width=30mm]{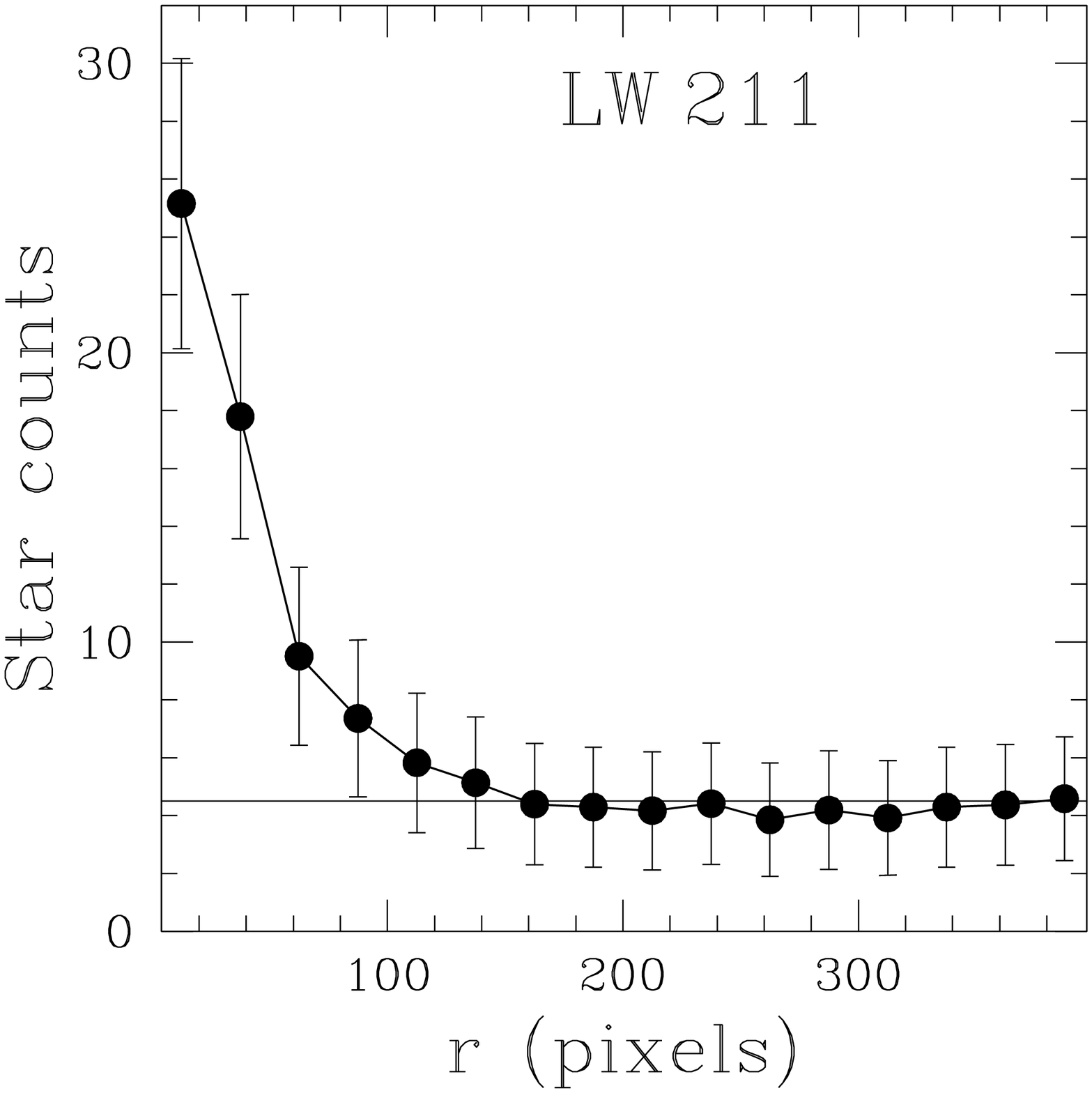}
 \includegraphics[width=30mm]{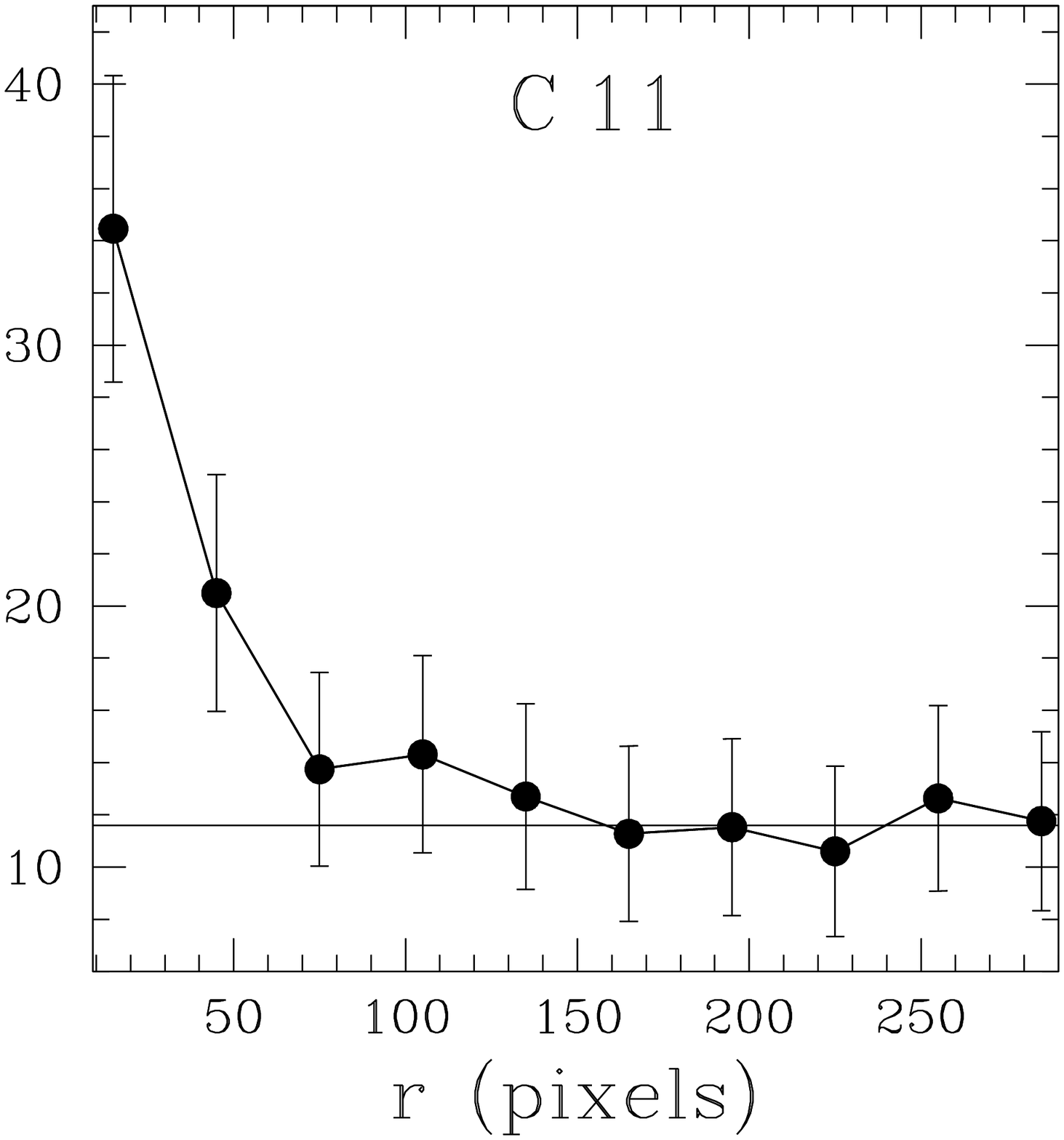}
 \includegraphics[width=30mm]{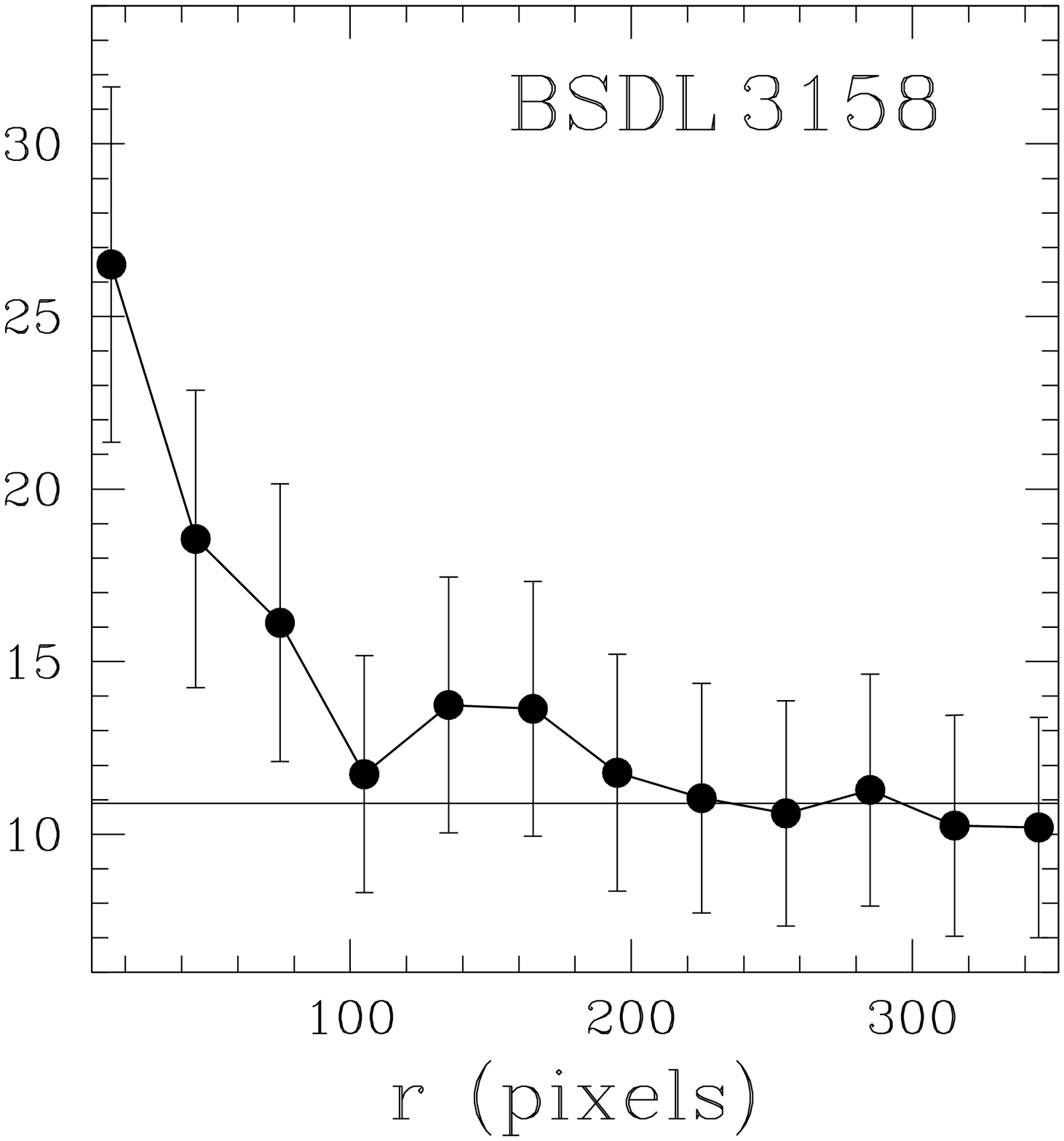} 
 \includegraphics[width=30mm]{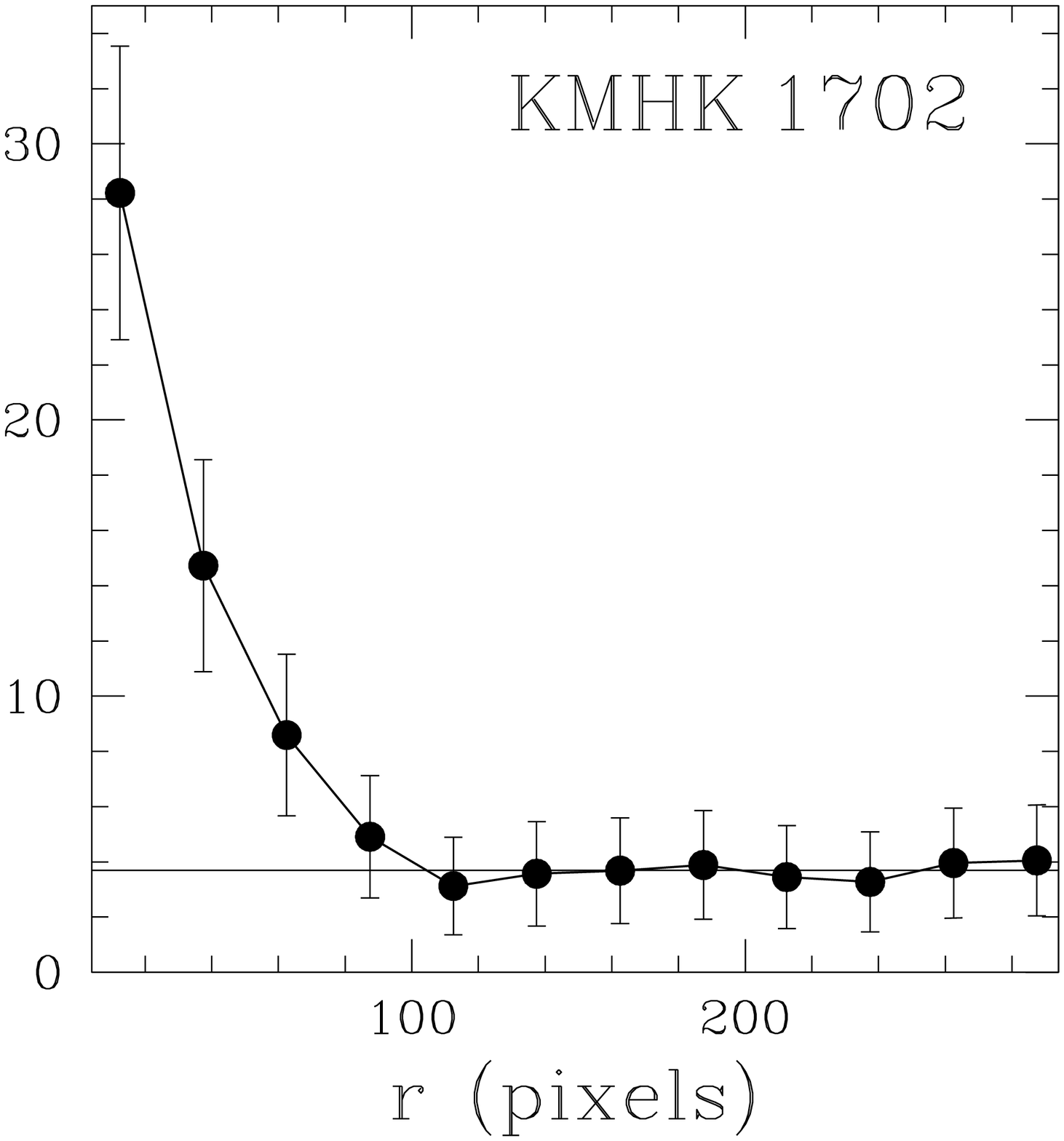}
 \includegraphics[width=30mm]{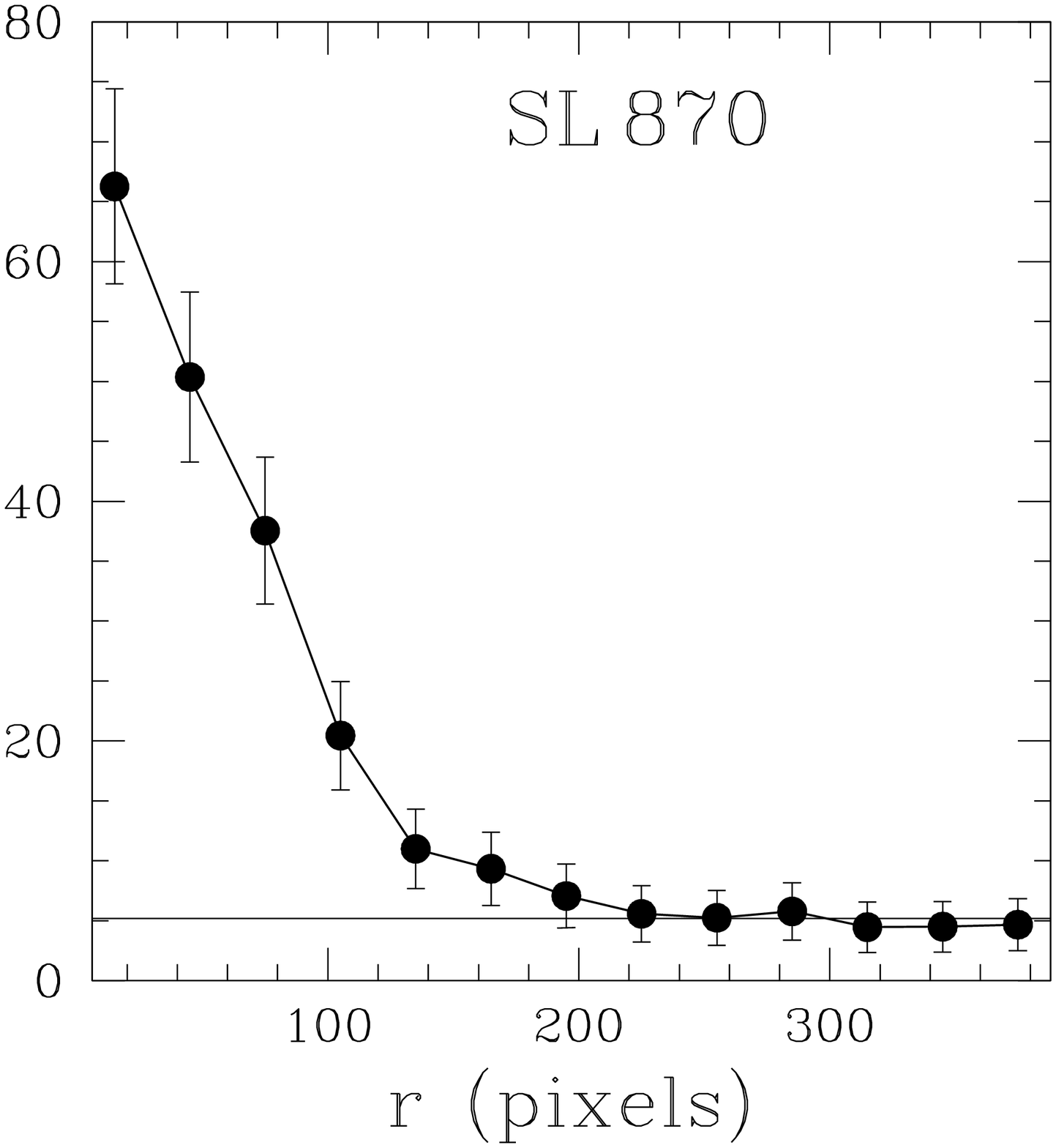}
 \caption{Stellar density profiles for the selected clusters, with the Poisson errors included. The horizontal lines correspond to the background levels far from the clusters. The background level was determined by estimating the mean stellar density at distances larger than $\sim$ 300 pixels from the centre of each cluster.}
  \label{f:rad}
 \end{figure*}

To construct density profiles of the observed clusters, we began by fitting Gaussian distributions to the star counts in the x and y directions to determine the coordinates of the cluster centres and their estimated uncertainties. The number of stars projected along these two directions were counted using five-pixel intervals, thus allowing us to statistically sample the spatial distributions. The fit of a single Gaussian for each projected density profile was performed using the NGAUSSFIT routine in the STSDAS/IRAF package. The cluster centres were determined with a typical standard deviation of $\pm$5 pixels ($\sim$ 1.35''). In the particular cases in which clusters showed peculiar morphologies, like an elongation in one direction or a marked sparsity, the standard deviation turned out to be slightly higher ($\pm$10 pixels or $\sim$ 2.7''). This is the case of clusters LW\,69, SL\,72, BSDL\,665, and BSDL\,675. Although the determination of cluster centres in these cases includes a higher degree of uncertainty, this uncertainty does not significantly change the final value obtained for the cluster radius because of the clusters' dimensions. We then built the cluster radial profiles by computing the number of stars per unit area at a given radius r, as shown in Fig. \ref{f:rad}. The cluster radius (r$_{cls}$), defined as the distance from the cluster's centre where the density of stars equals that of the background, is given in pixels in column 7 of Table \ref{t:coord}.  Column 8 of Table \ref{t:coord} lists the cluster linear radii in parsecs calculated assuming that LMC is located at a distance of 50 kpc \citepads[]{ss10}. \\

\begin{figure*}
\centering
 \includegraphics[width=30mm]{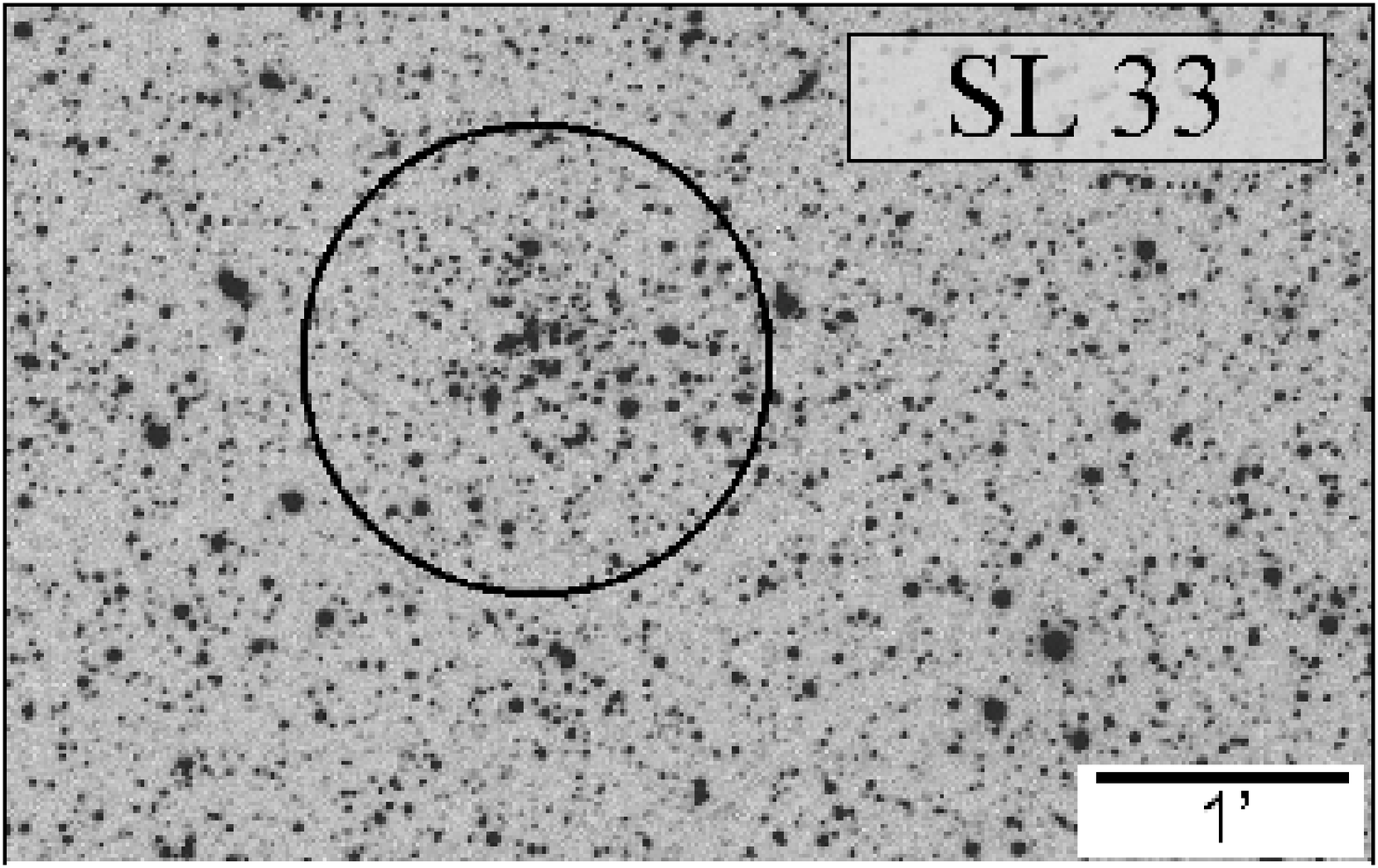}
 \includegraphics[width=30mm]{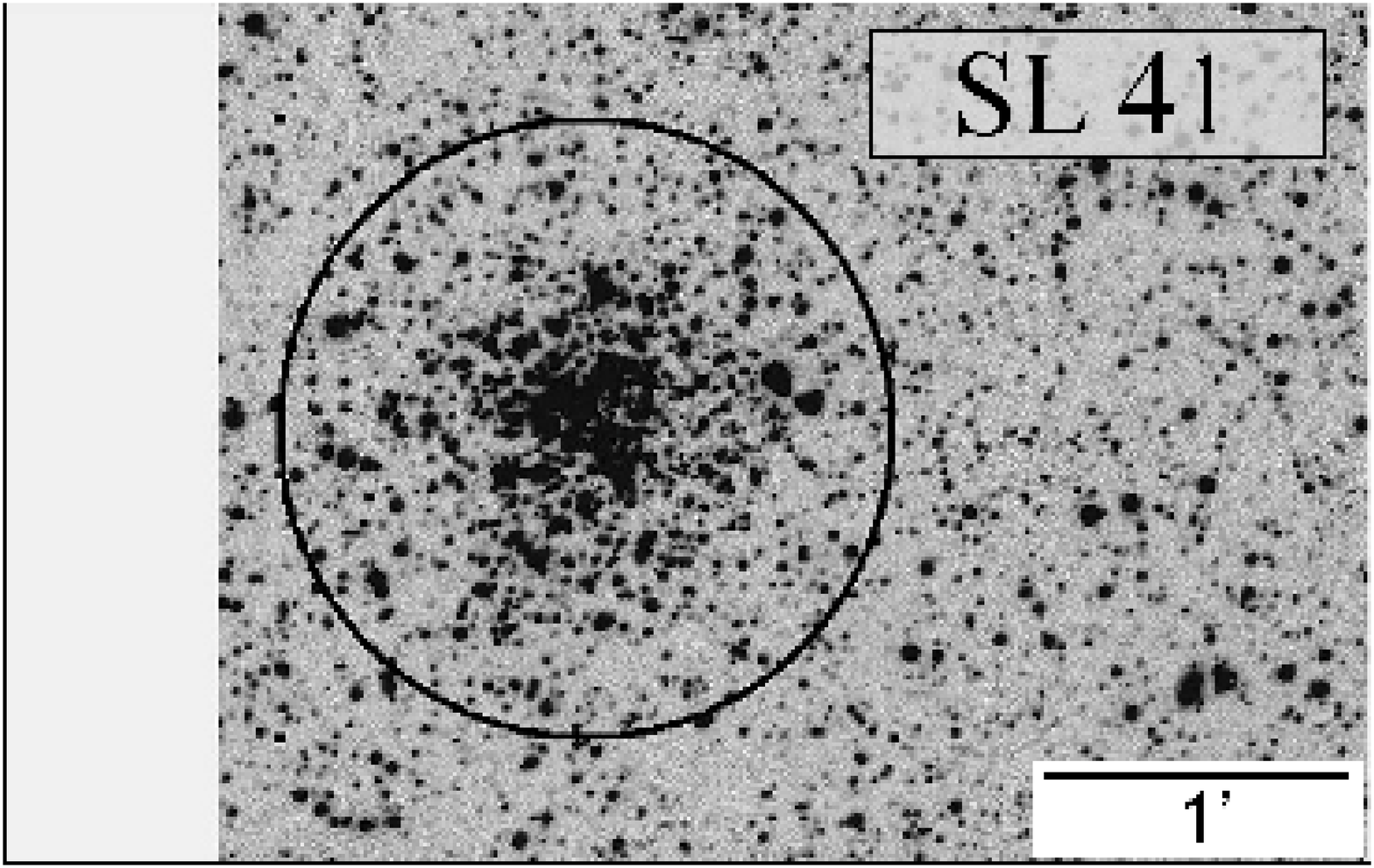}
 \includegraphics[width=30mm]{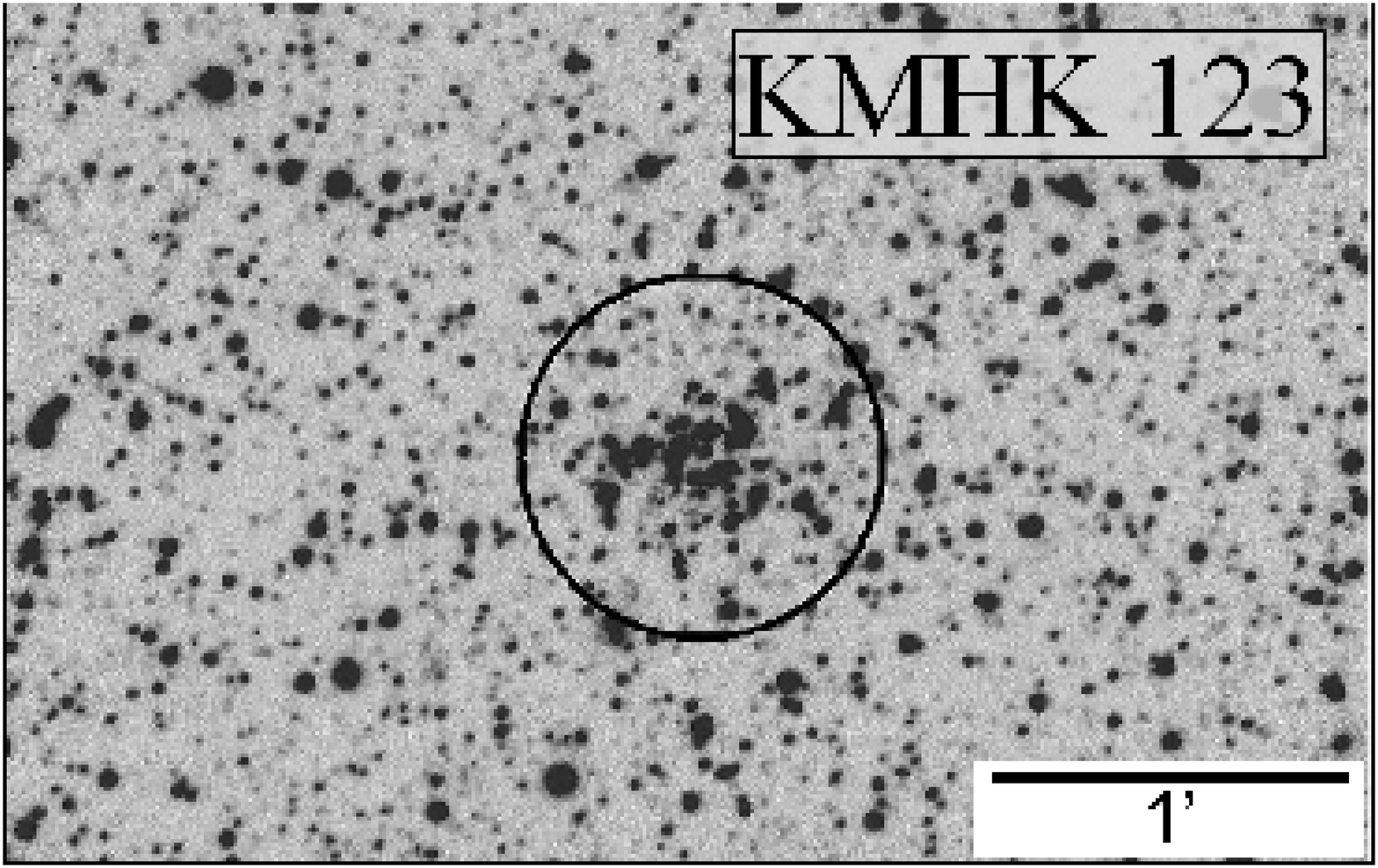}
 \includegraphics[width=30mm]{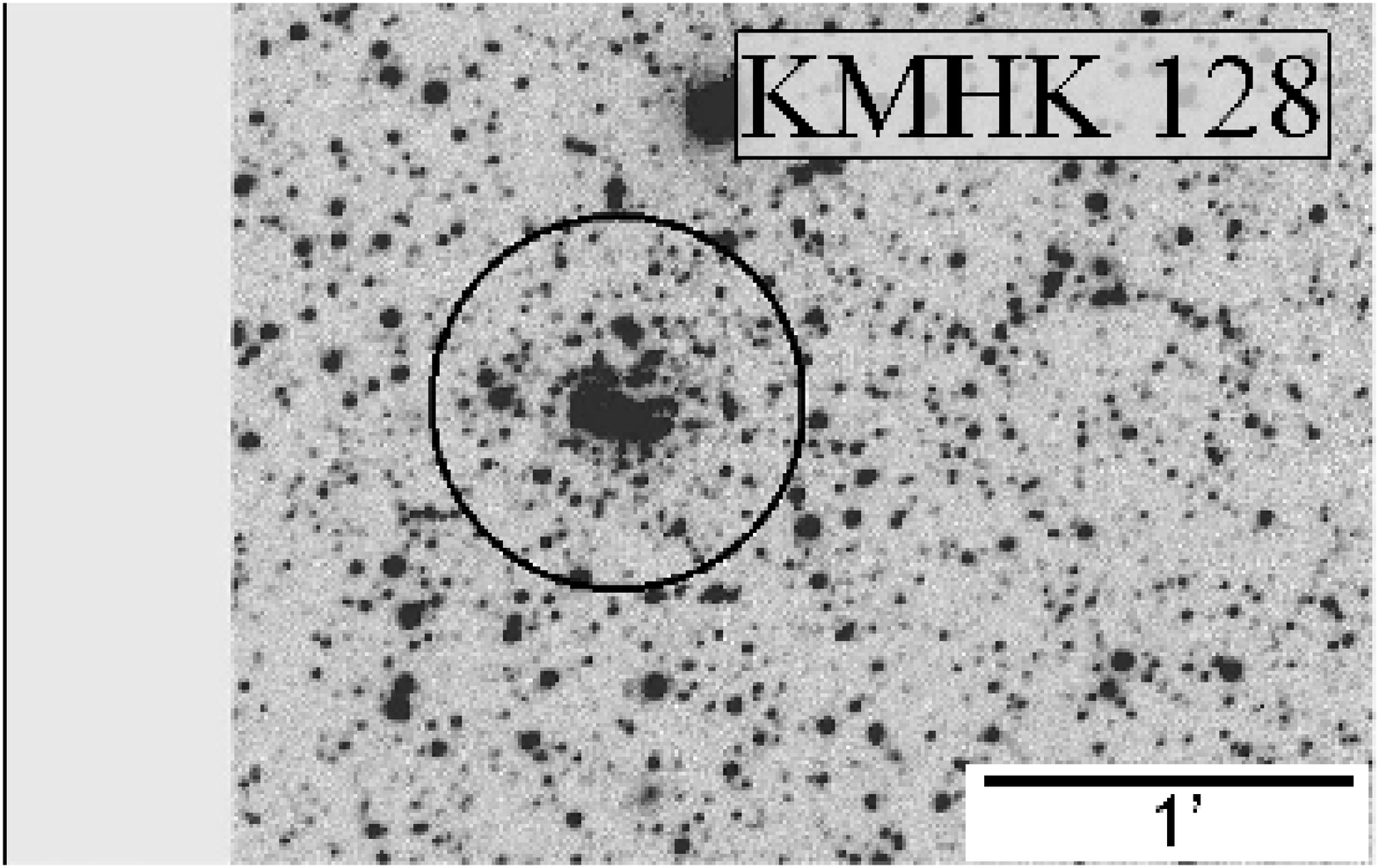}
 \includegraphics[width=30mm]{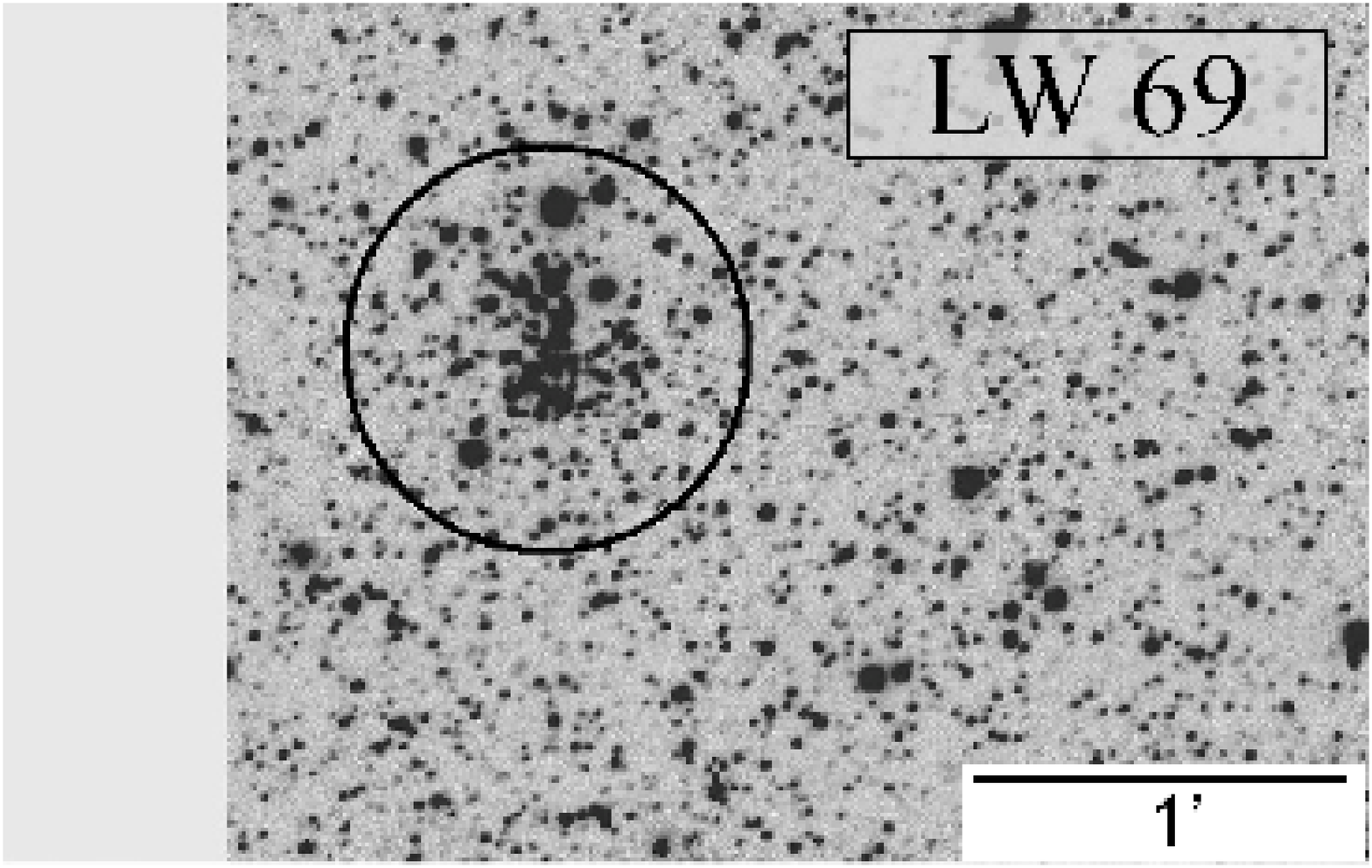}
 \includegraphics[width=30mm]{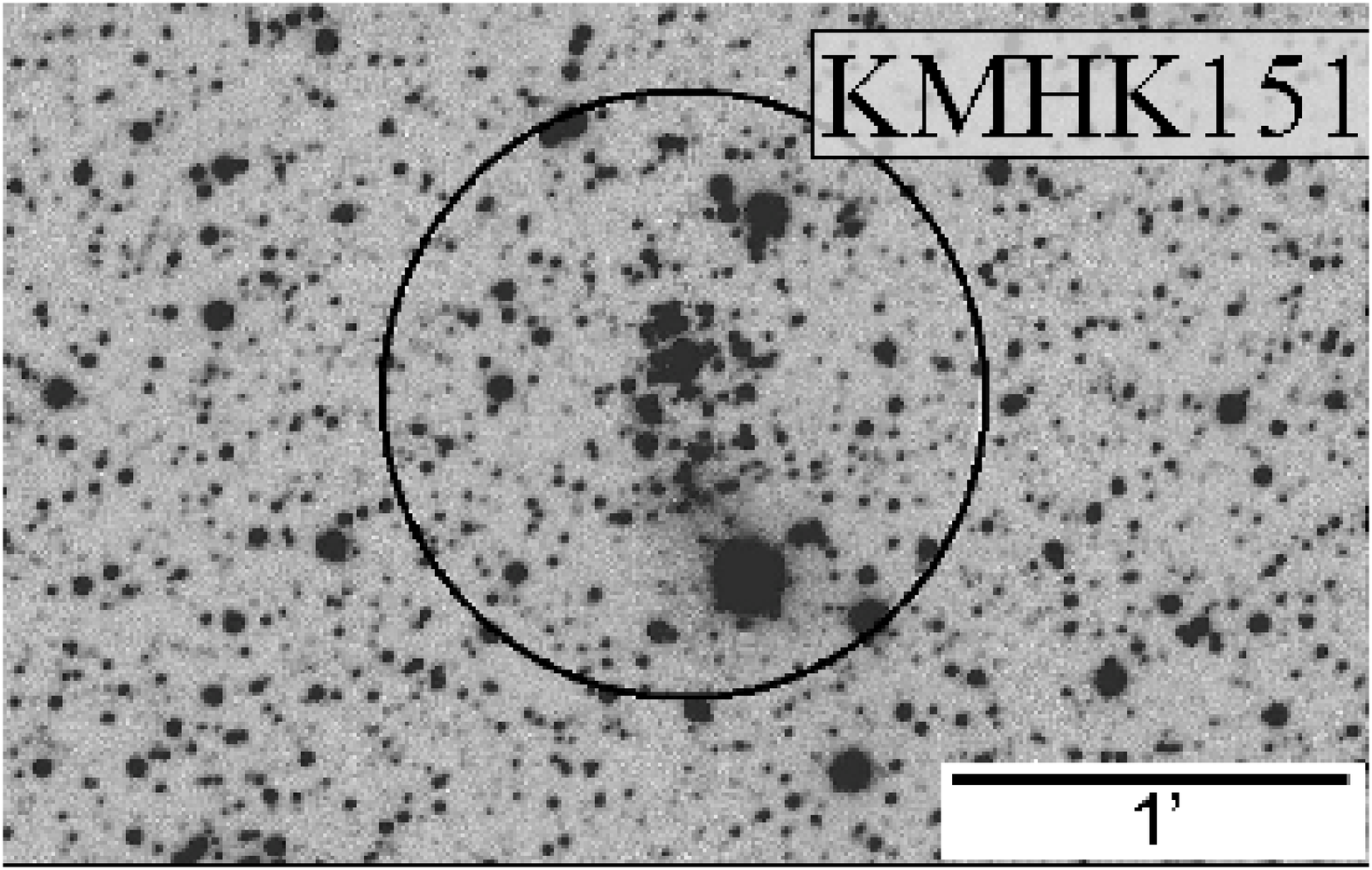}
 \includegraphics[width=30mm]{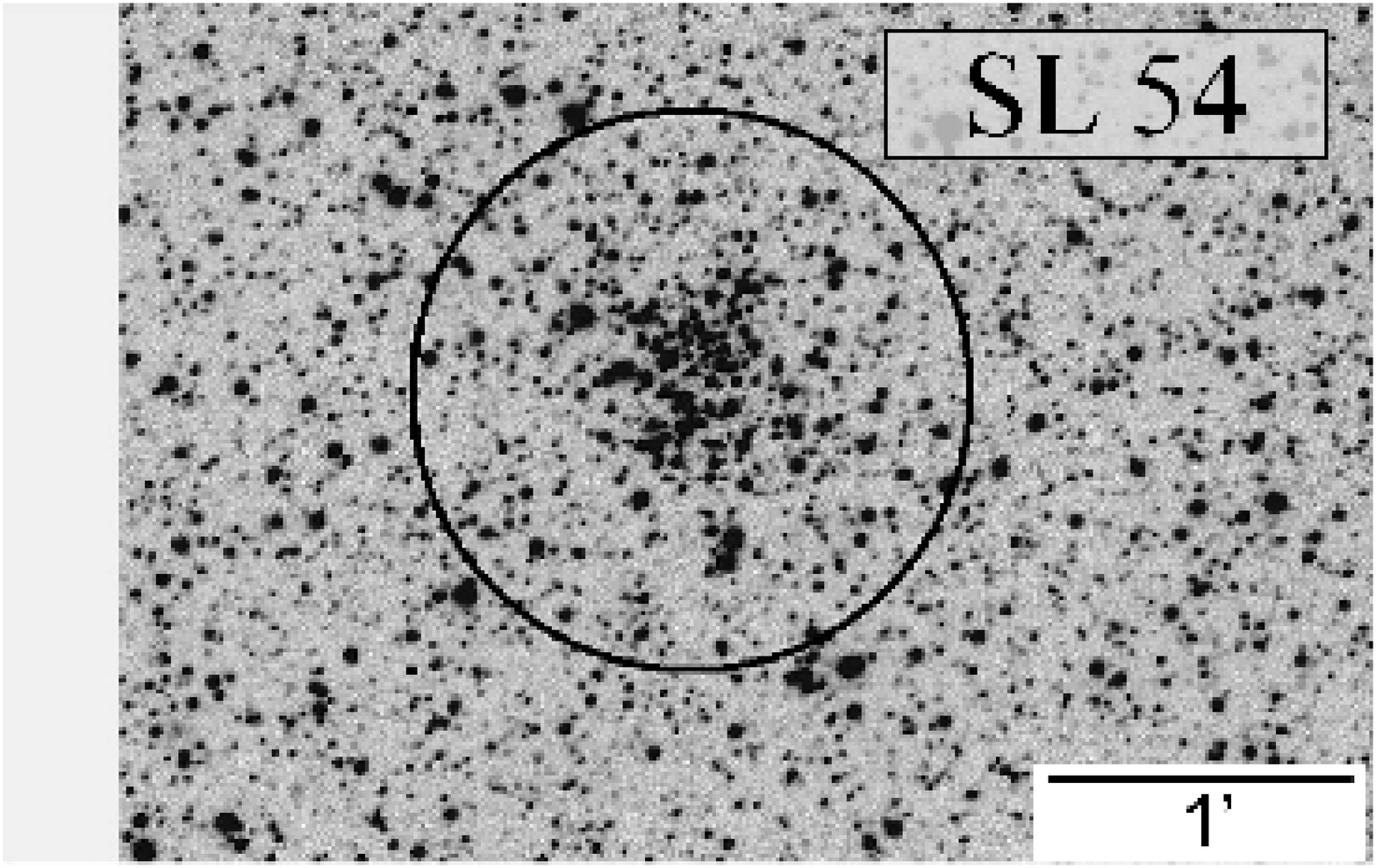}
 \includegraphics[width=30mm]{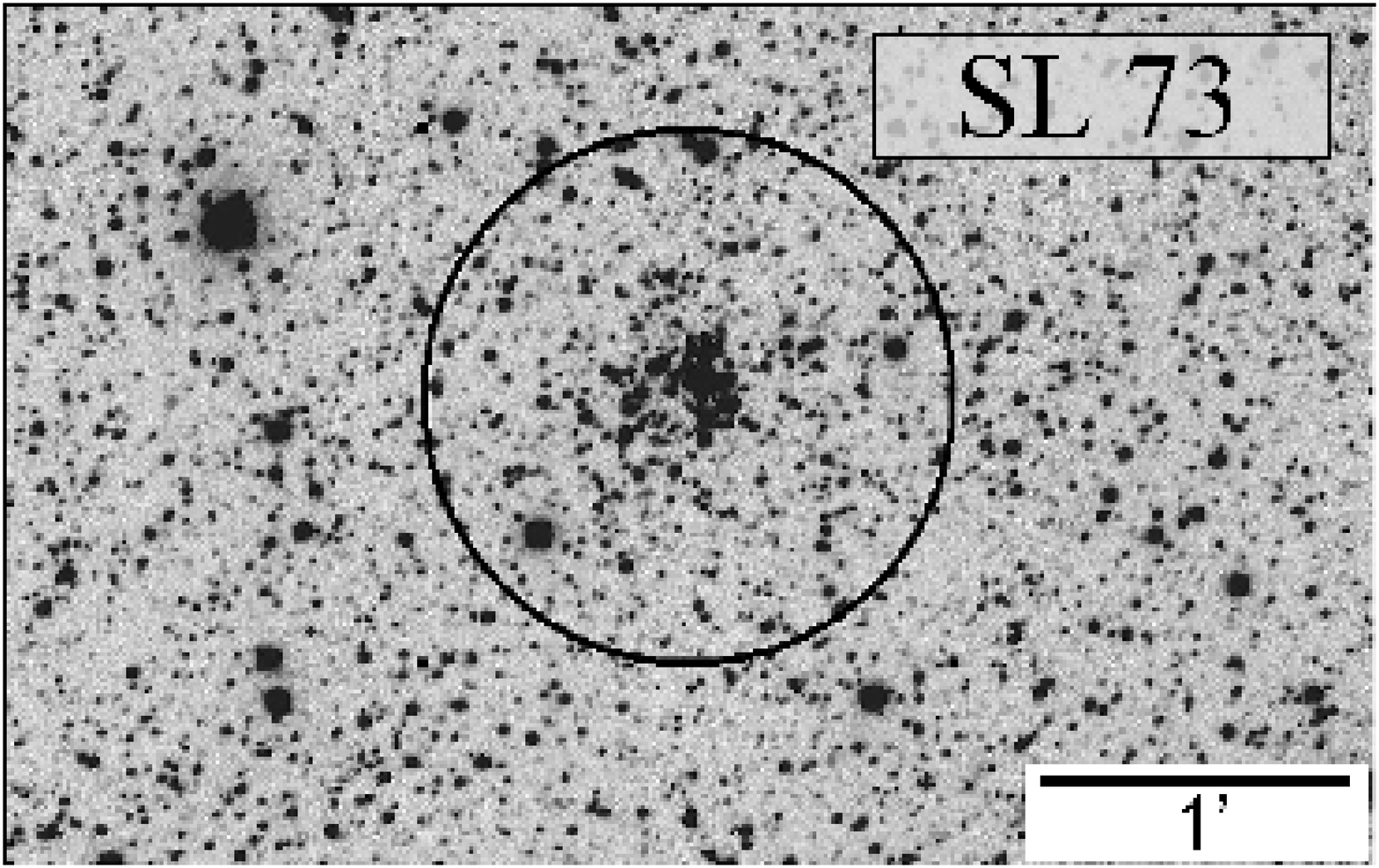}
 \includegraphics[width=30mm]{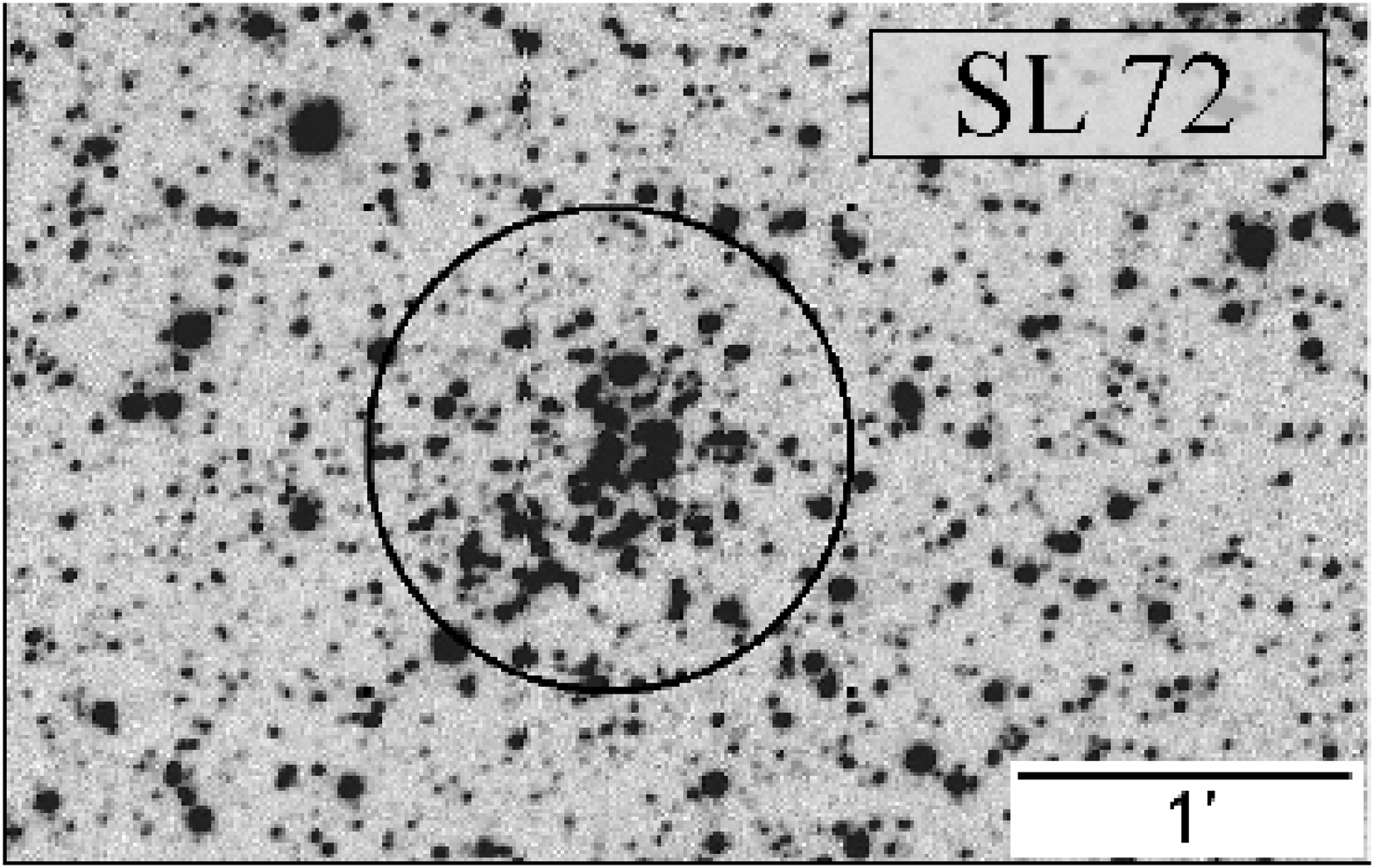}
 \includegraphics[width=30mm]{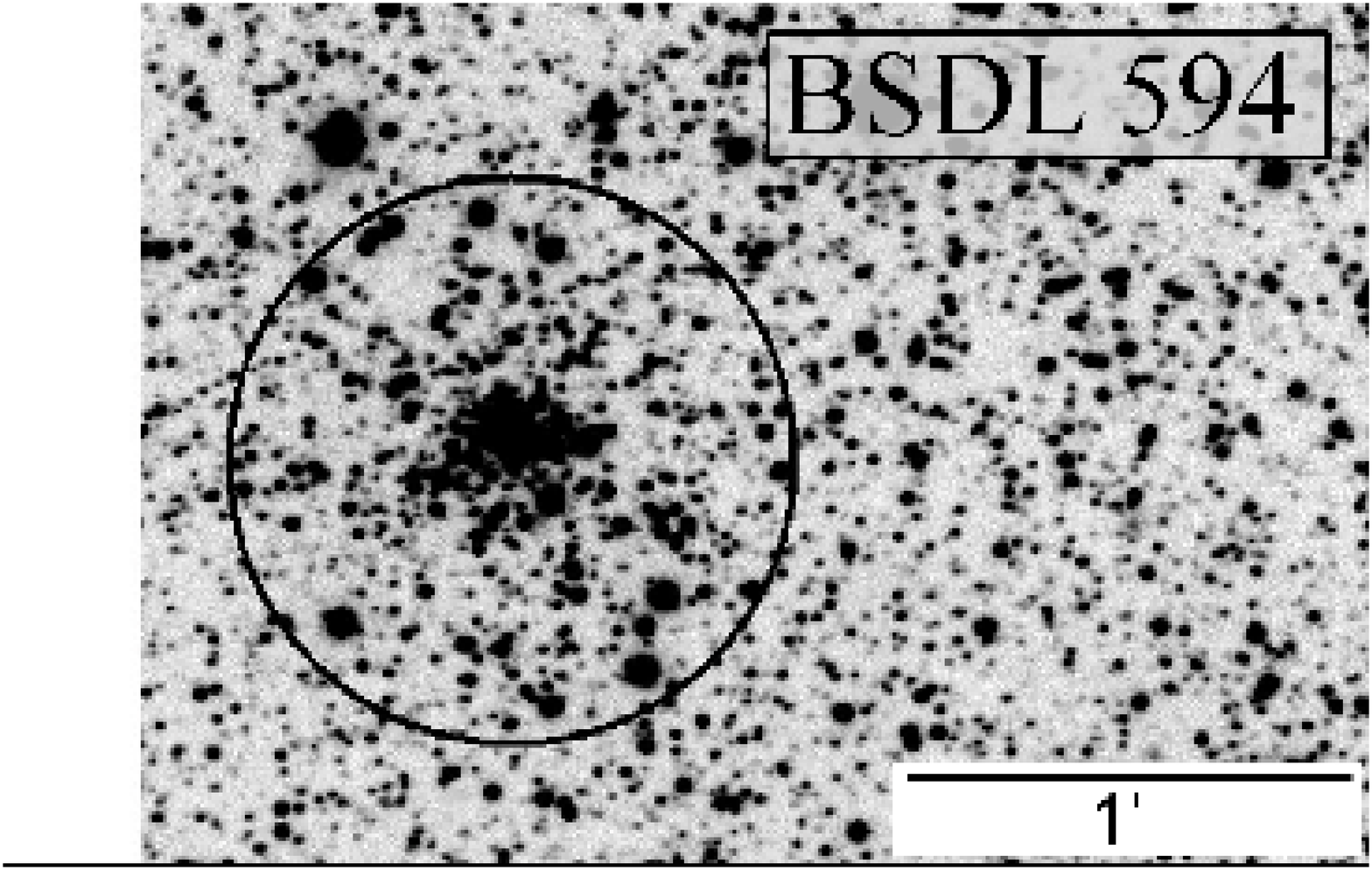}
 \includegraphics[width=30mm]{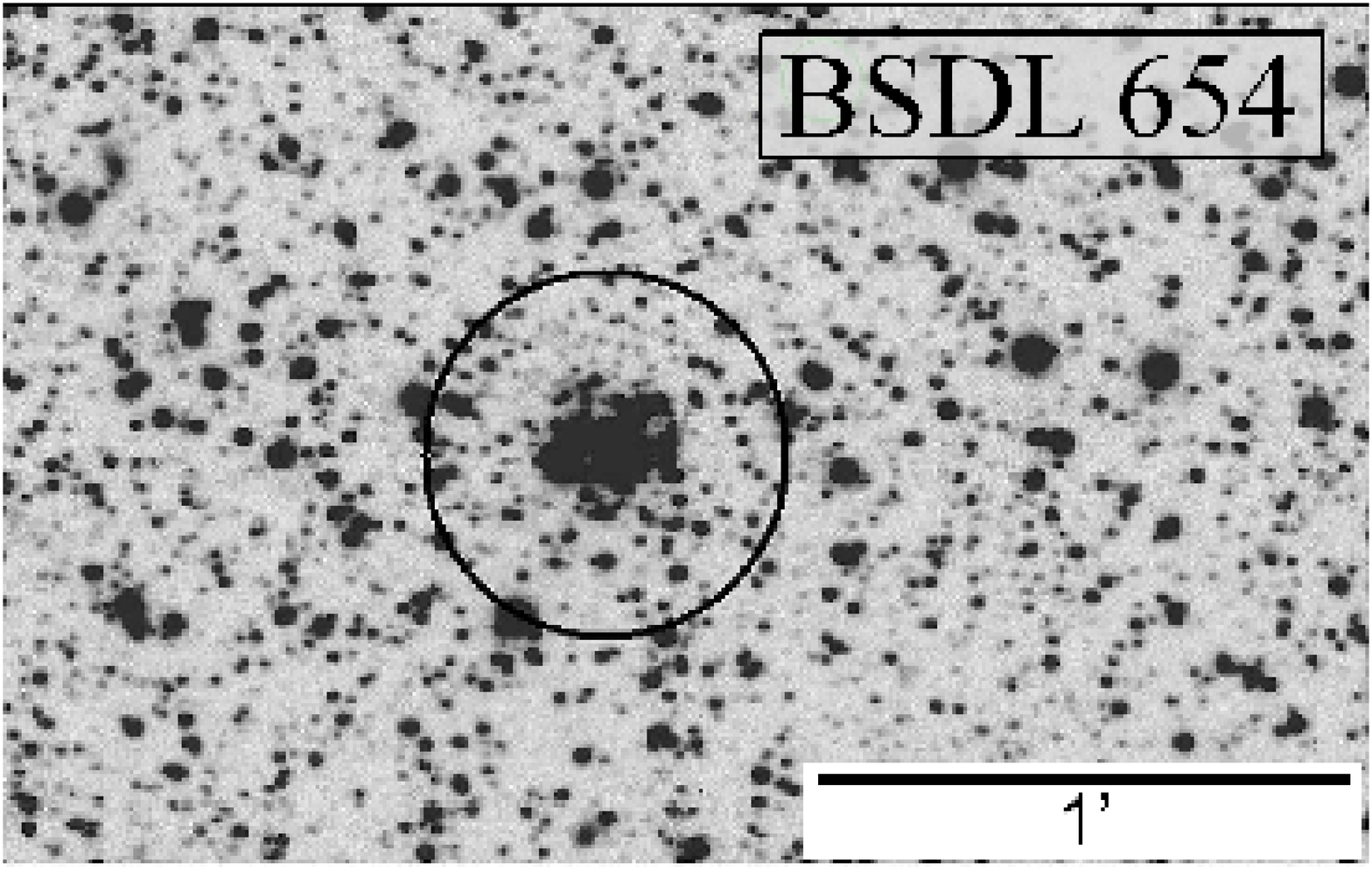}
 \includegraphics[width=30mm]{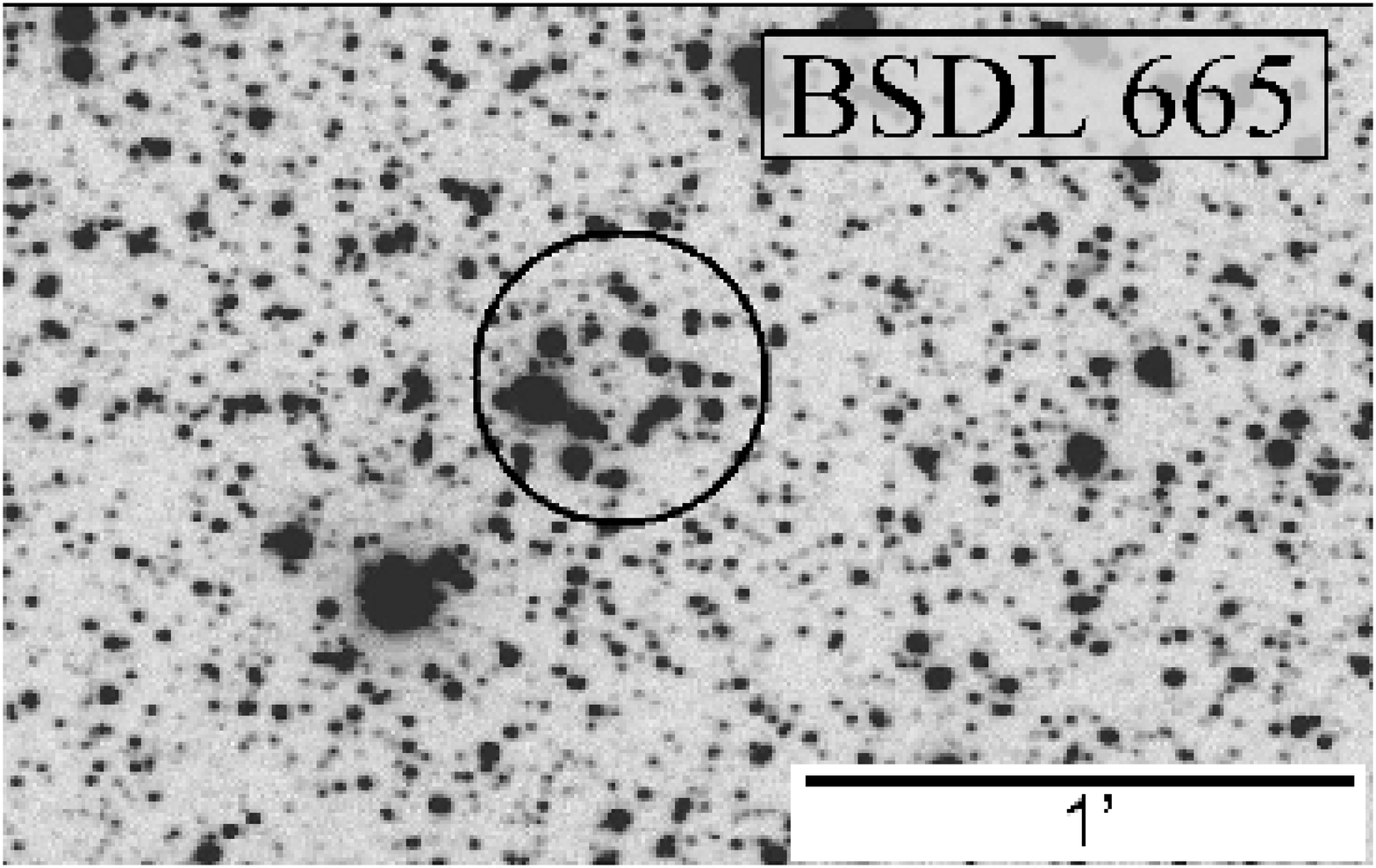}
 \includegraphics[width=30mm]{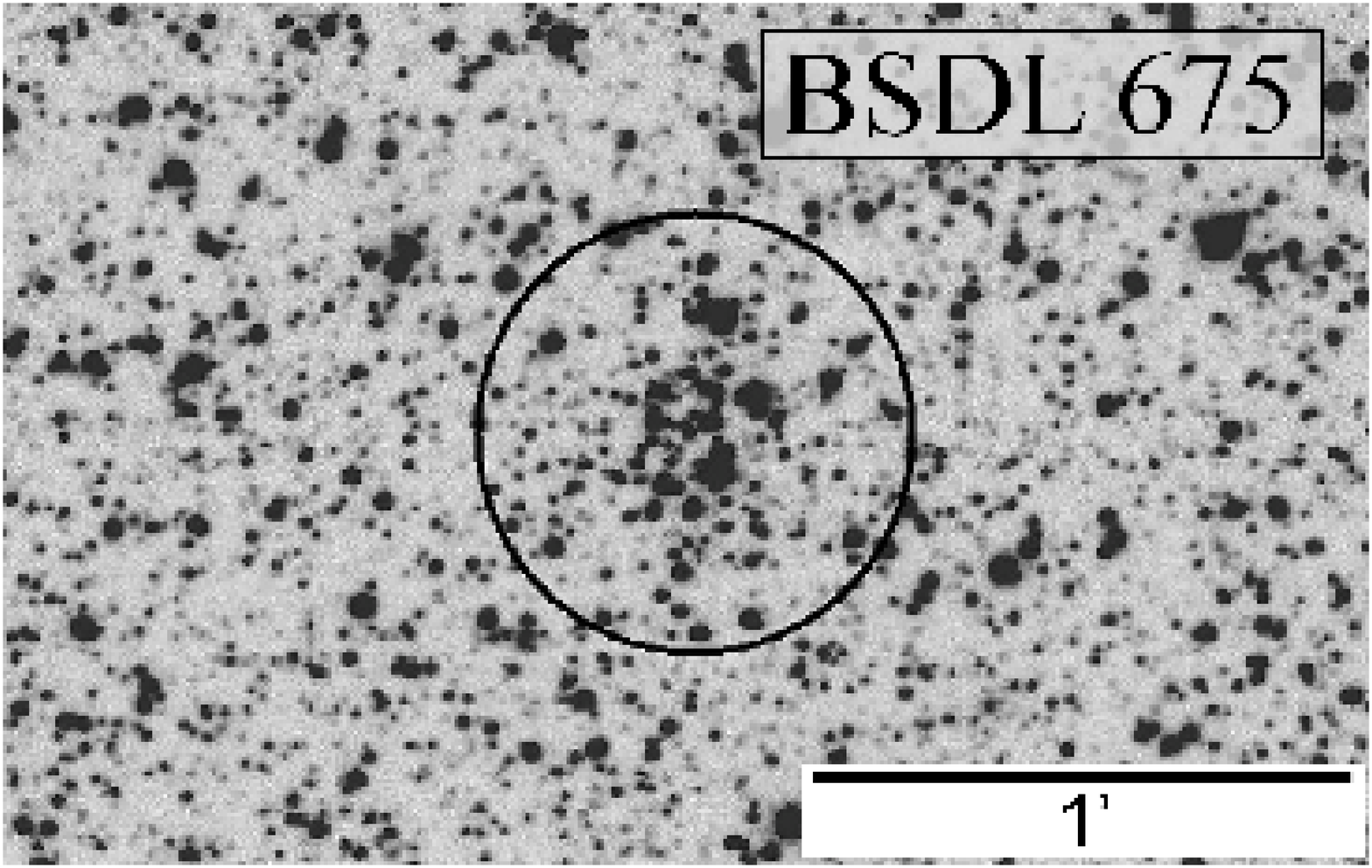}
 \includegraphics[width=30mm]{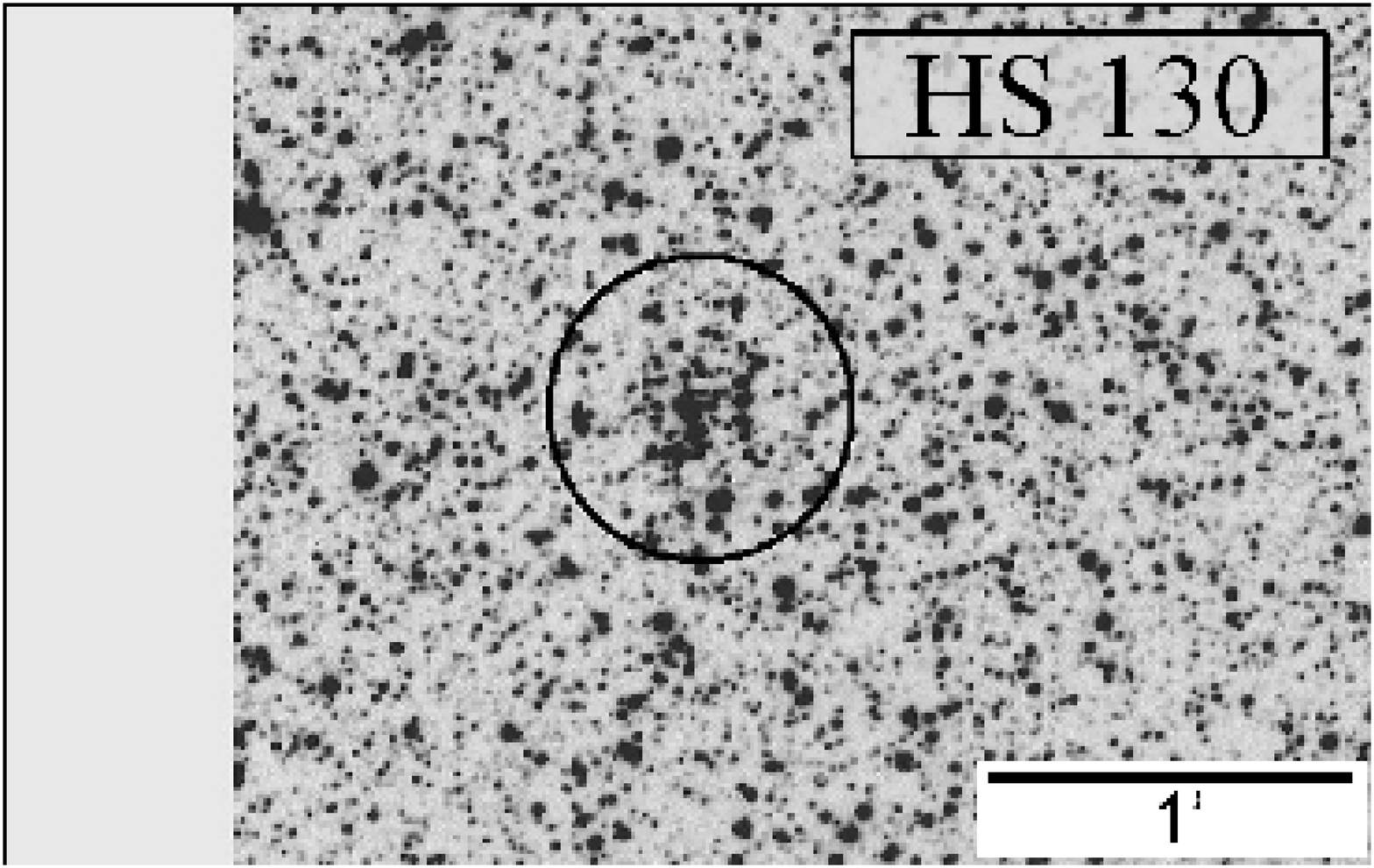}
 \includegraphics[width=30mm]{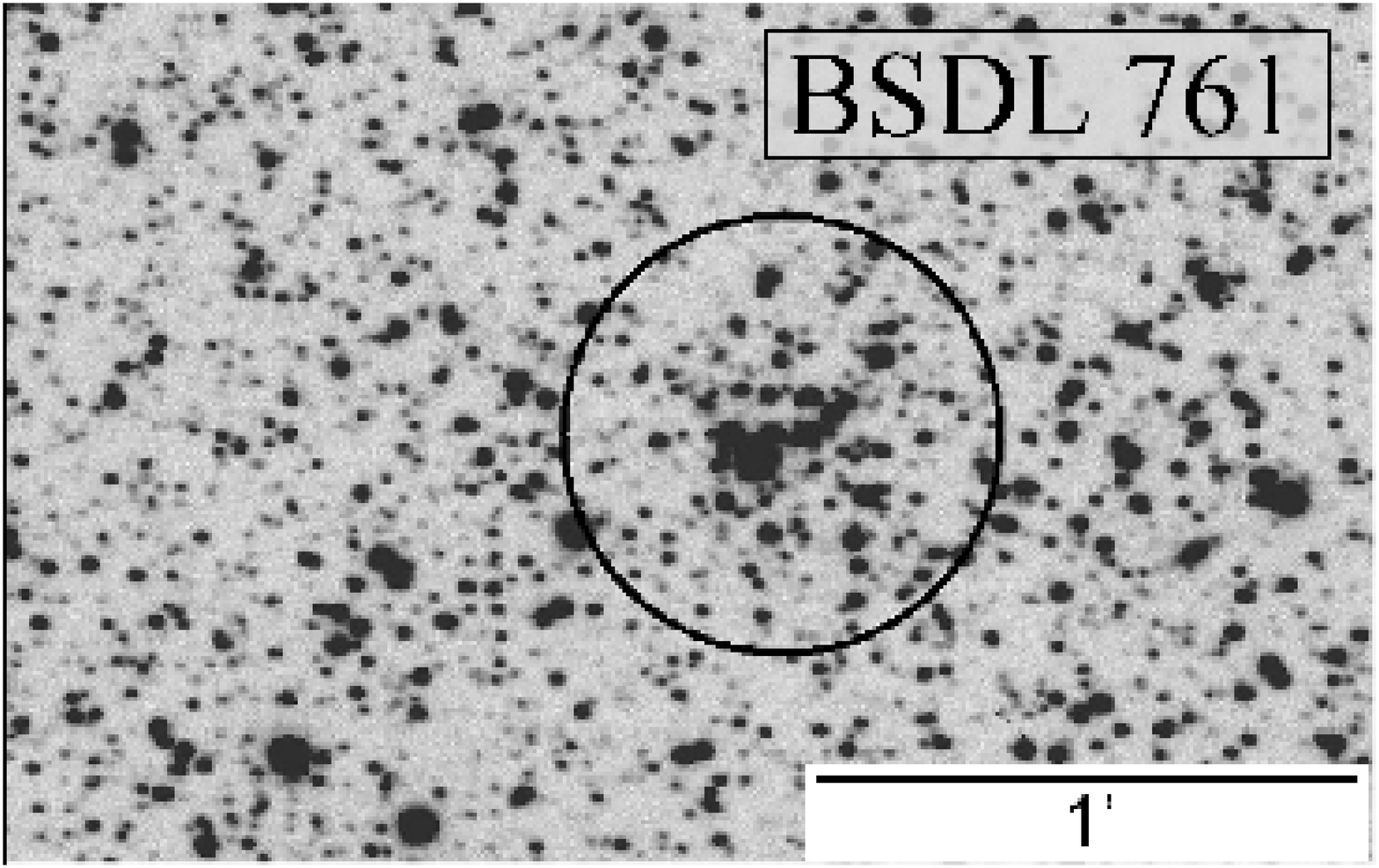}
 \includegraphics[width=30mm]{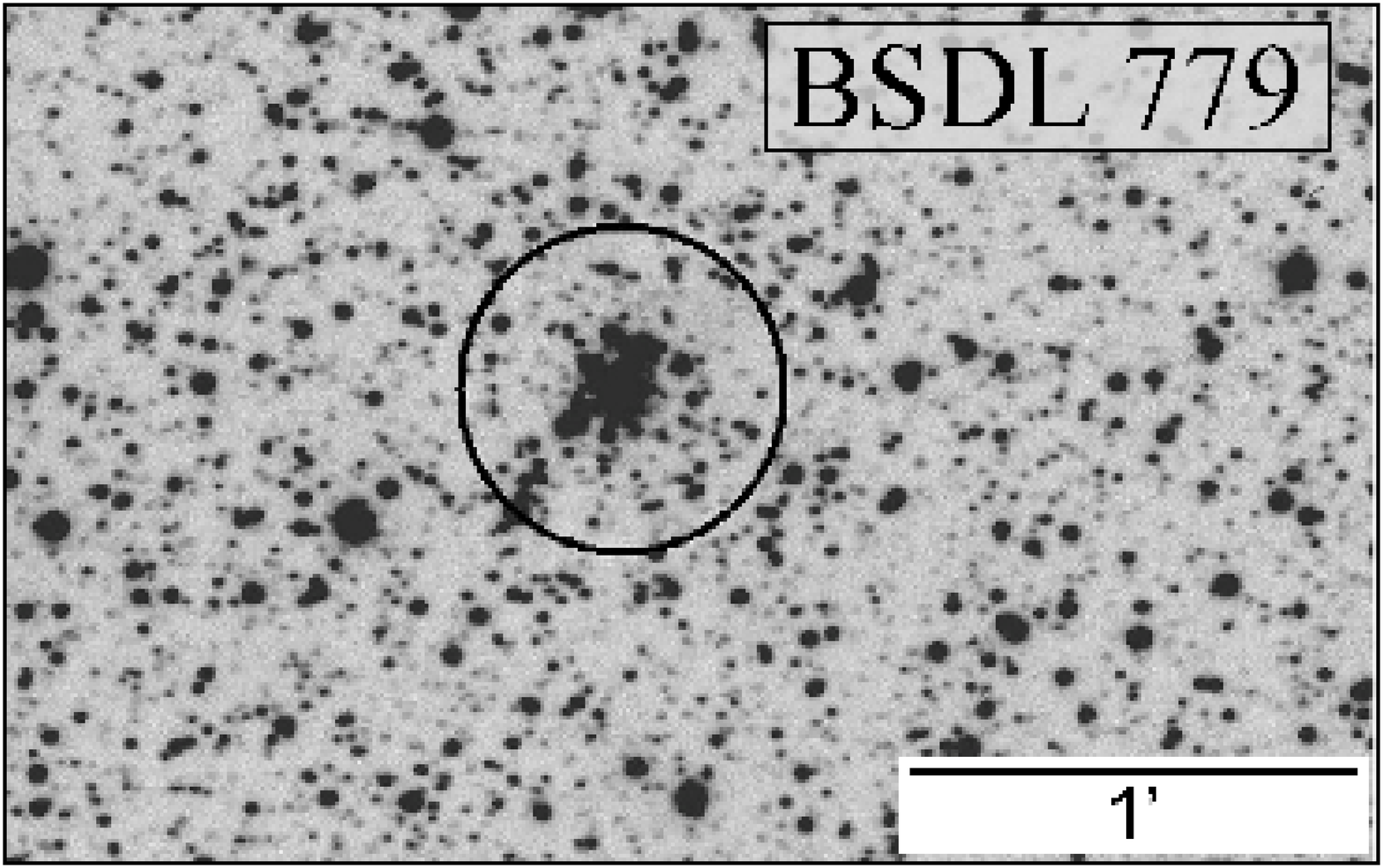}
 \includegraphics[width=30mm]{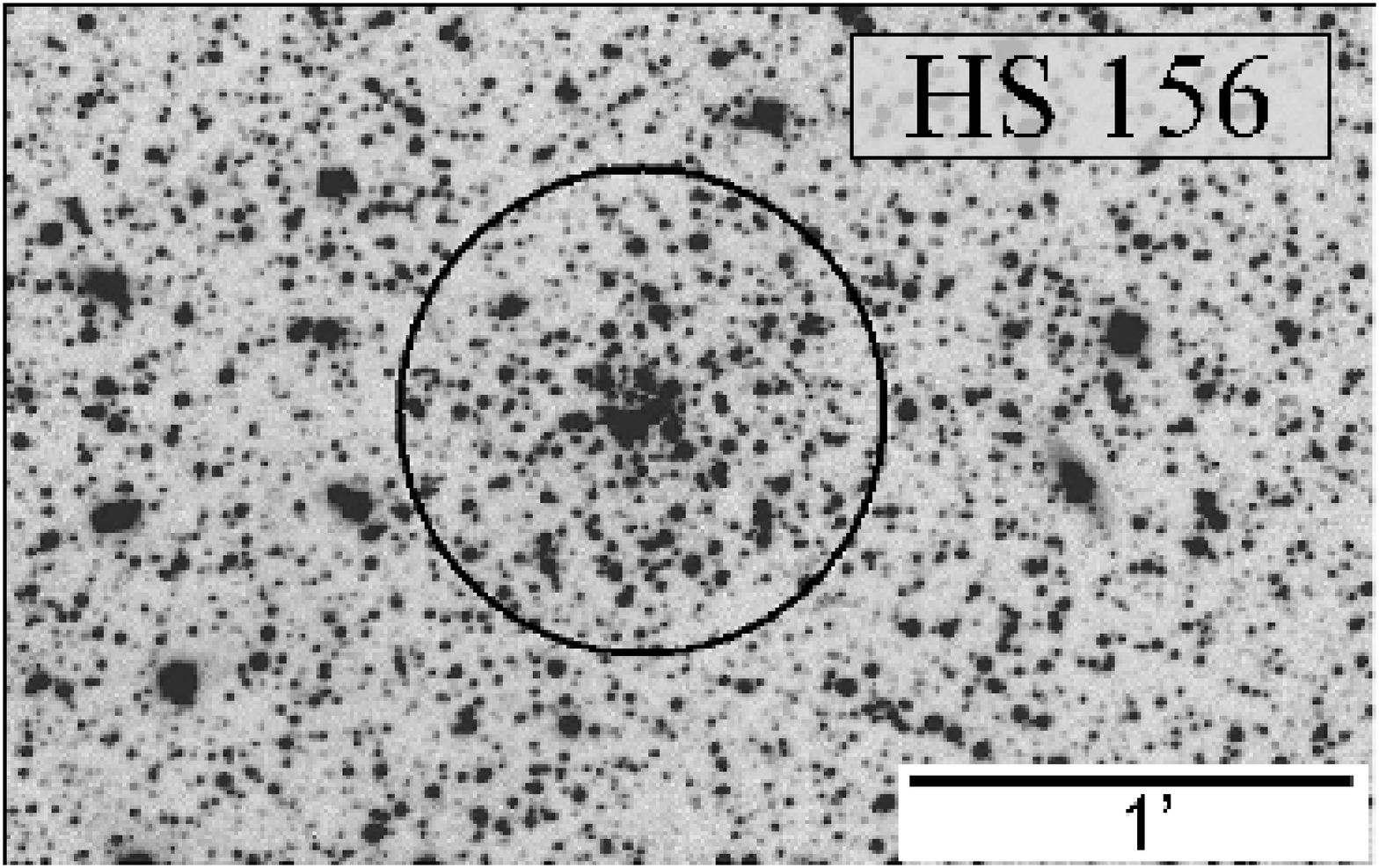}
 \includegraphics[width=30mm]{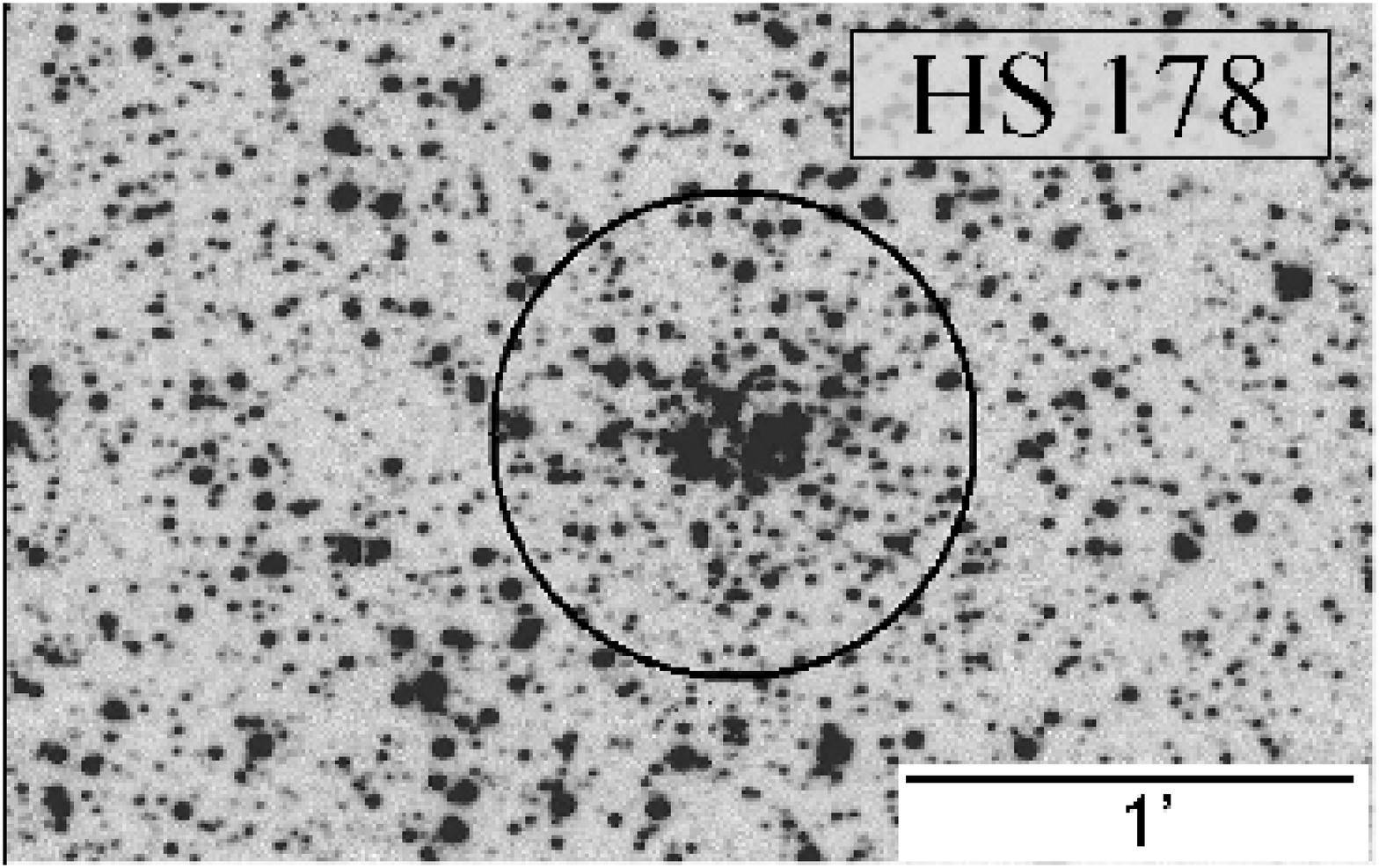}
 \includegraphics[width=30mm]{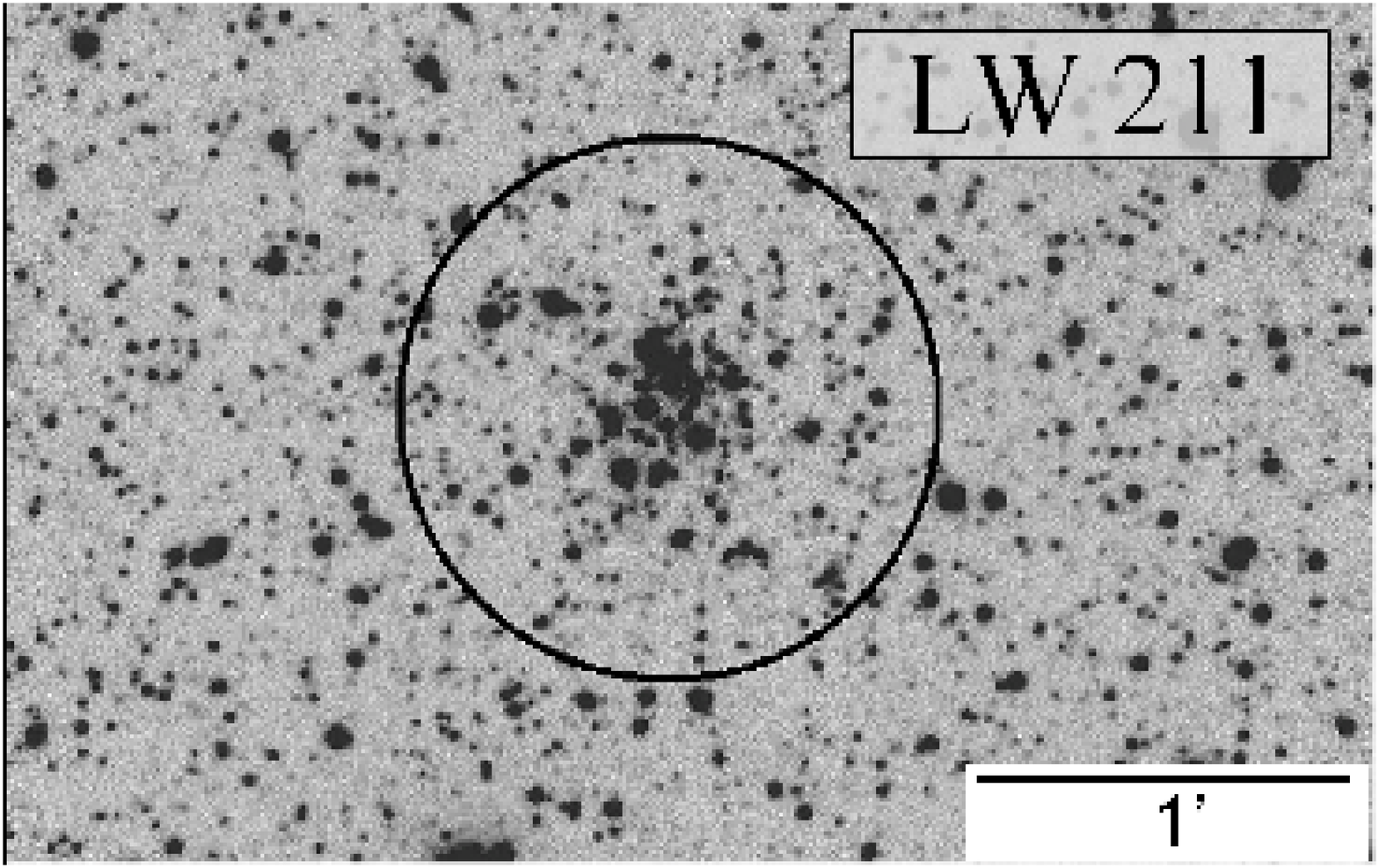}
 \includegraphics[width=30mm]{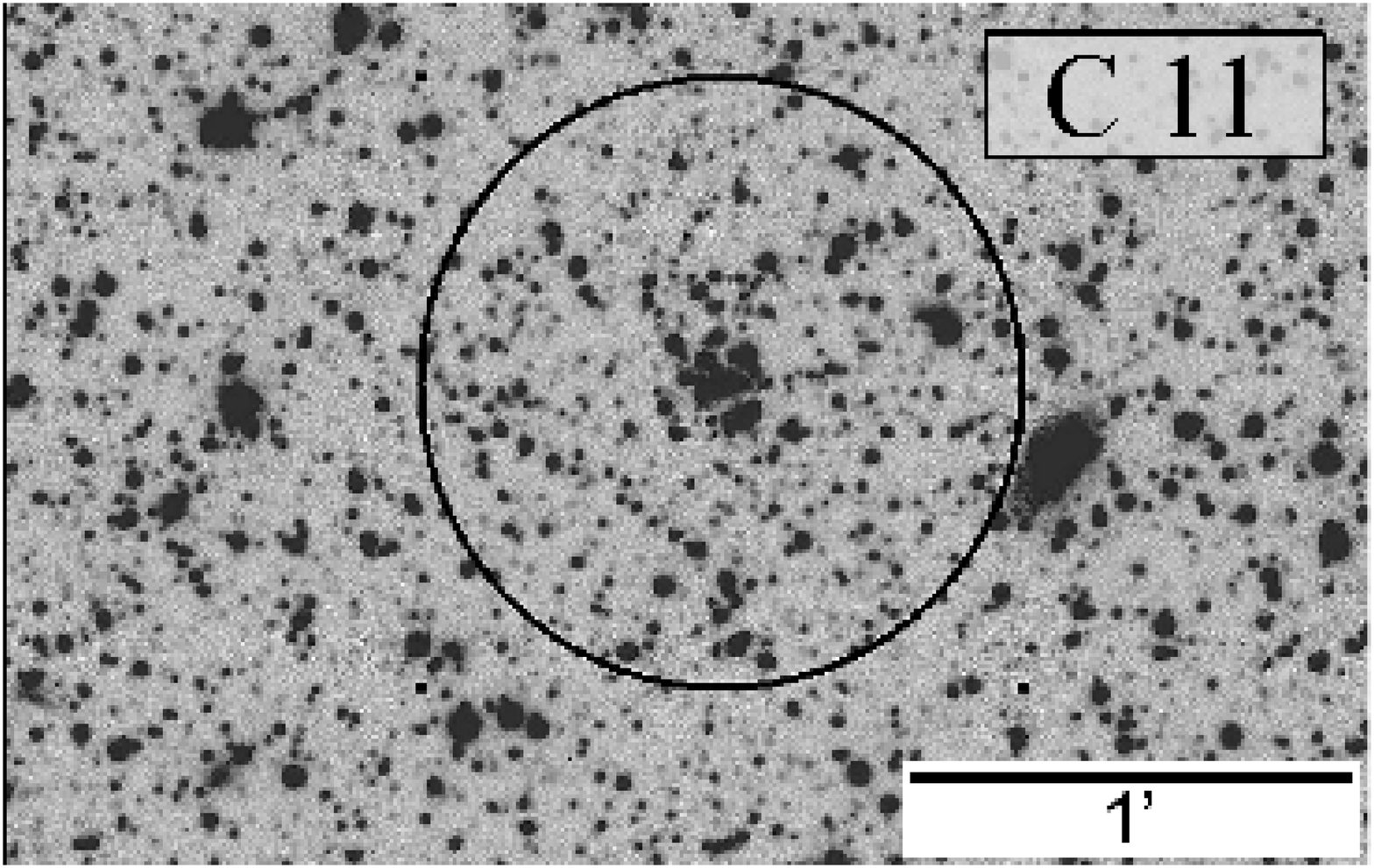}
 \includegraphics[width=30mm]{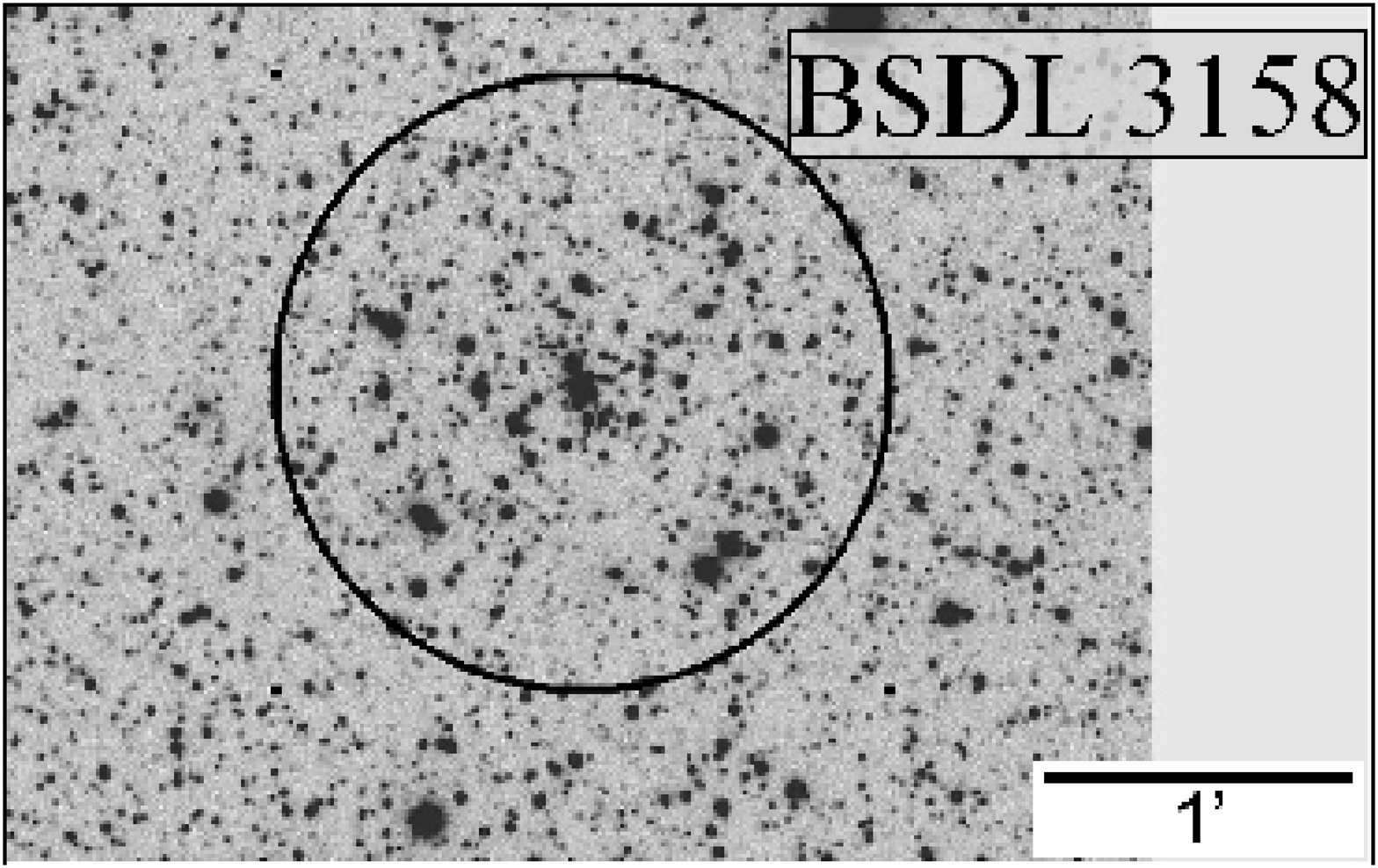}
 \includegraphics[width=30mm]{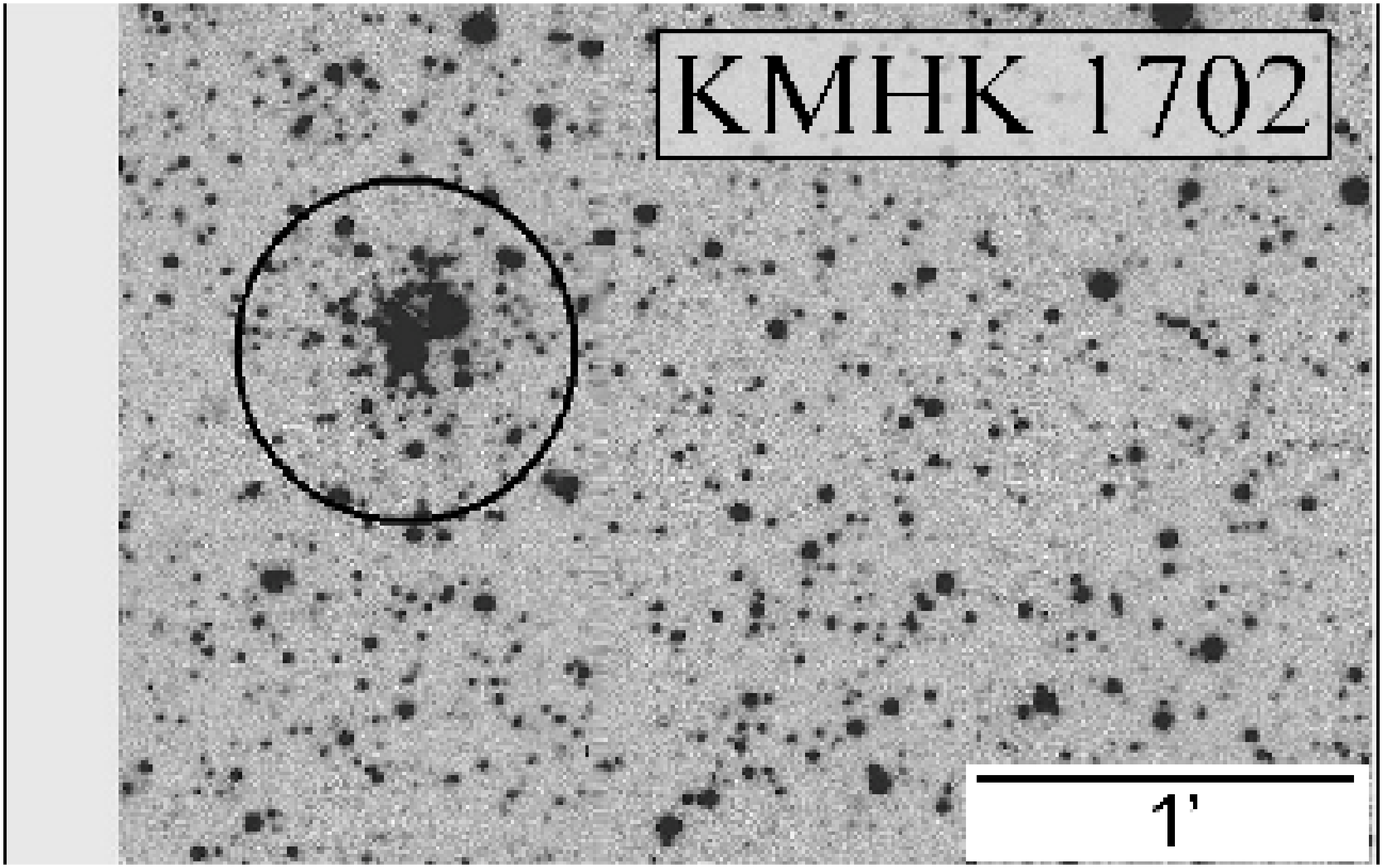} 
 \includegraphics[width=30mm]{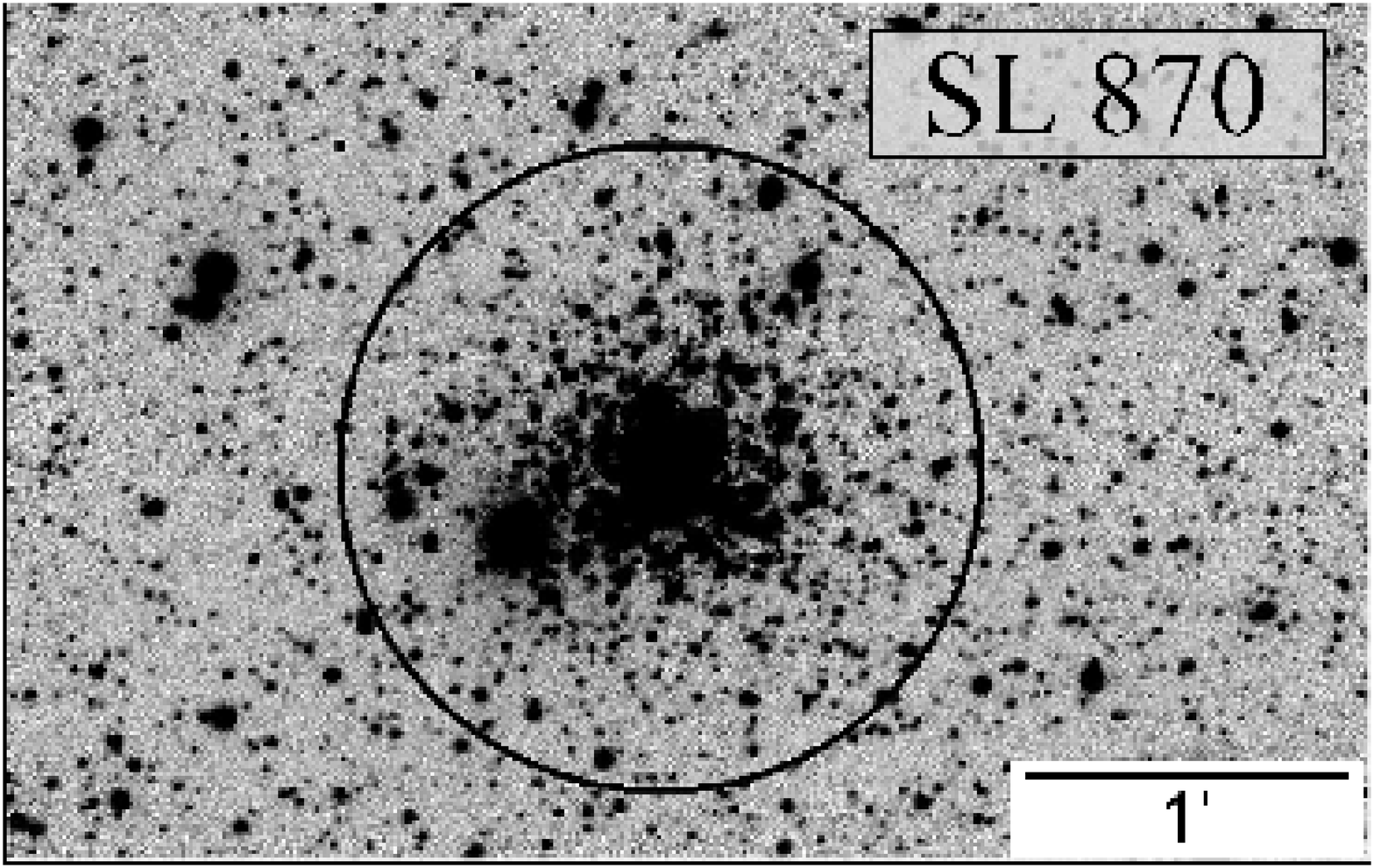}
  \caption{Schematic images of the stars observed in the fields of our target clusters. North is up and east is to the left. The circle in each figure has been drawn with the same radius (in pixels) as the radius obtained for each cluster.} 
  \label{f:chart}
 \end{figure*}

\subsection{Cluster properties from CMDs}

\begin{figure}
\includegraphics[width=70mm]{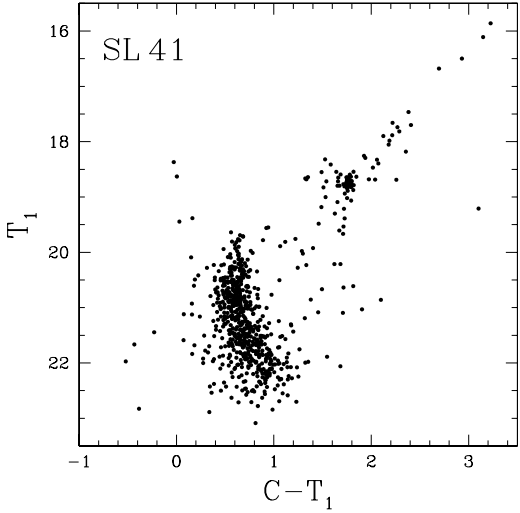}
\caption{Washington $(T_1,C-T_1)$ CMD for all the stars measured in the field of SL\,41, one of the most populated clusters of our sample. } 
\label{f:fig5}
\end{figure}

We show in Fig. \ref{f:chart} schematic images of the stars observed in the cluster fields, while in Fig. \ref{f:fig5} we show the $(T_1,C-T_1)$ CMD of all stars measured in the field of SL\,41, which is one of the most populated fields observed. The 
remaining cluster fields exhibit CMDs whose features vary from one cluster to another, mainly depending on age.  \\

\begin{figure}
\includegraphics[width=70mm]{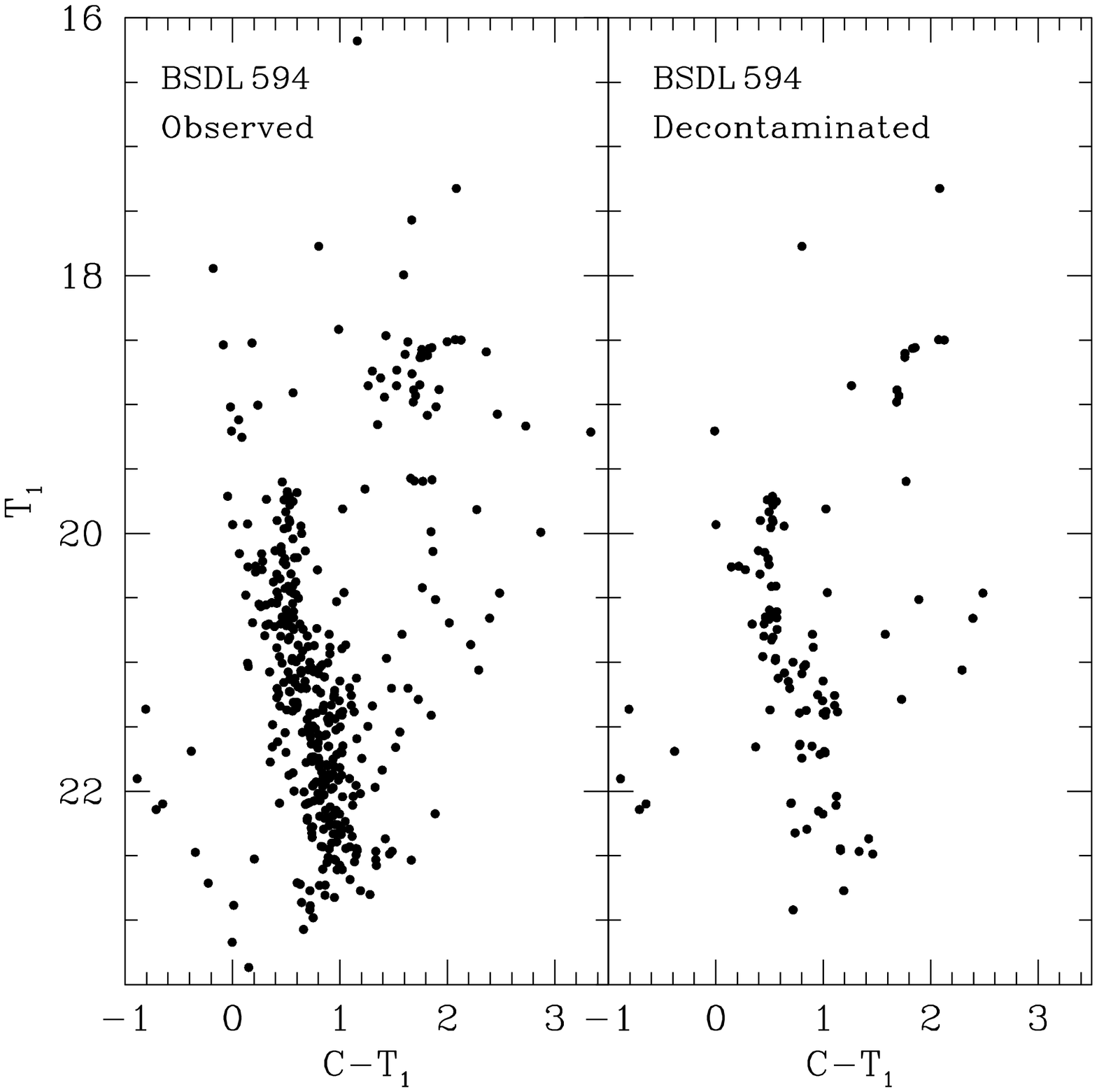}
\includegraphics[width=70mm]{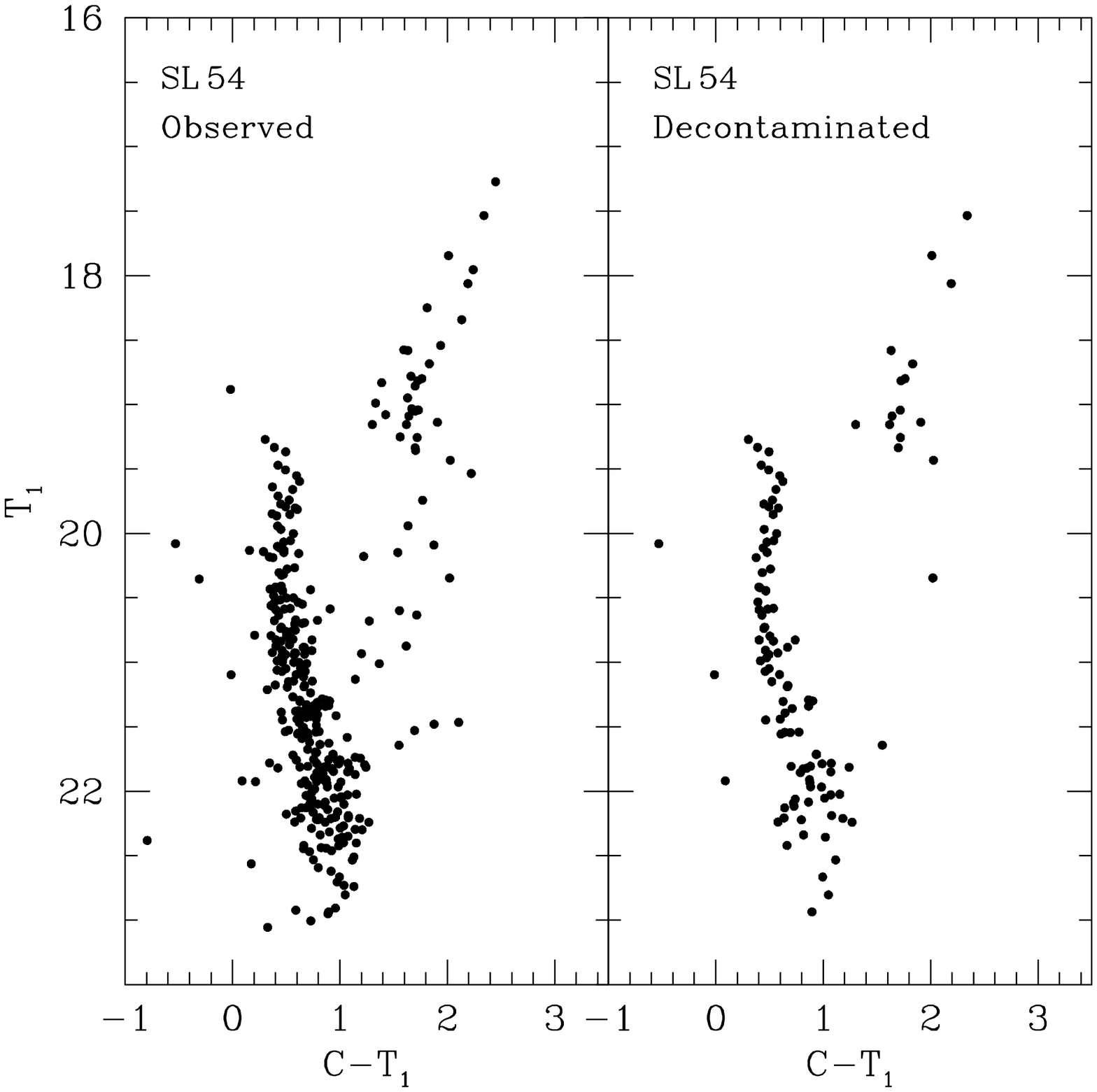}
\caption{Observed and cleaned CMDs of BSDL\,594 and SL\,54. } 
\label{f:ext}
\end{figure}

Since the fundamental cluster parameters estimation requires us to minimize the field-star contamination, we applied a statistical method recently developed in \citetads[]{pi12} that is described in detail in \citetads[]{pb12}. This procedure is briefly summarized as follows: It consists of selecting four field regions at a distance of between two and four times the cluster radius for each cluster. Then, their respective $(T_1,C-T_1)$ CMDs are obtained. The sizes of the areas of each field regions must be equal to the cluster area (generally taken as twice as large as the obtained cluster radius). Next, we count the stars lying within different intervals of magnitude-colour [$\delta T_1,\delta(C-T_1)$] in the CMD of each selected region. This new method includes variable intervals, depending on how populated the studied region is. The intervals happen to be bigger in more ``deserted'' regions in the CMD diagrams, such as in clump regions. Conversely, they appear to be smaller in more populated regions, such as in the main sequence. Finally, the number of stars counted for each interval [$\delta T_1,\delta(C-T_1)$] in the CMD of the surrounding field region is subtracted from the number of stars of the cluster region. For more details see \citetads[]{pb12}.  To illustrate the statistical cleaning procedure, we show in Fig. \ref{f:ext} the observed and cleaned CMDs for two faint star clusters of our sample BSDL\,594 and SL\,54.\\

Cluster-reddening values were estimated by interpolating the extinction maps of \citetads[]{bh82}. These maps were obtained from HI (21 cm) emission data for the southern sky and provide us with foreground E(B-V) colour excesses, which depend on the Galactic coordinates. 
As shown in column (3) of Table \ref{t:results}, the resulting E(B-V) values range between 0.03 and 0.12,  values typical for the LMC. 
For the distance, we adopted the value of the LMC true distance modulus (m-M)$_0$ = 18.50 $\pm$ 0.10 reported by \citetads[]{s10}. On the other hand, \citetads[]{ss09} found that the average depth for the LMC disc is 3.44 $\pm$ 0.16 kpc. If we consider that any cluster of 
the present sample could be placed in front of or behind the main body of the LMC, we conclude that the difference in apparent distance modulus could be as large as $\Delta (V-Mv)\sim$ 0.3 mag. Because we estimate an uncertainty of 0.2 - 0.3 mag when adjusting 
the isochrones to the cluster CMDs, our assumption of adopting one single value for the distance modulus of all the clusters should not dominate the error budget in our final results. In fact, when we overplotted the zero-age main sequence on the observed cluster CMDs, 
previously shifted by the $E(B-V)$ of Table \ref{t:results} and by $(m-M)_0$ = 18.50, we generally found an excellent match. \\

\begin{figure*}
\centering
 \includegraphics[width=40mm]{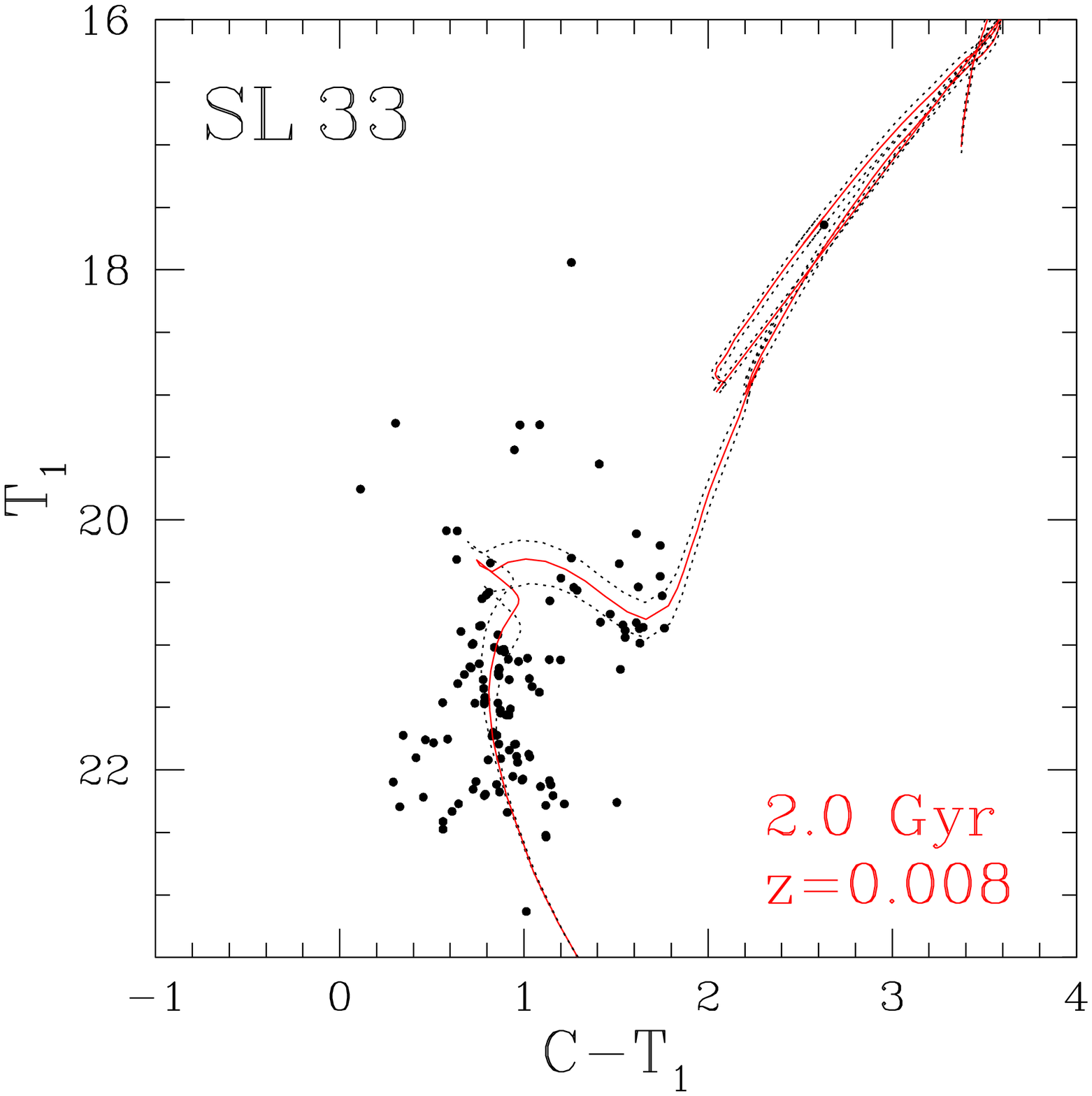} 
 \includegraphics[width=40mm]{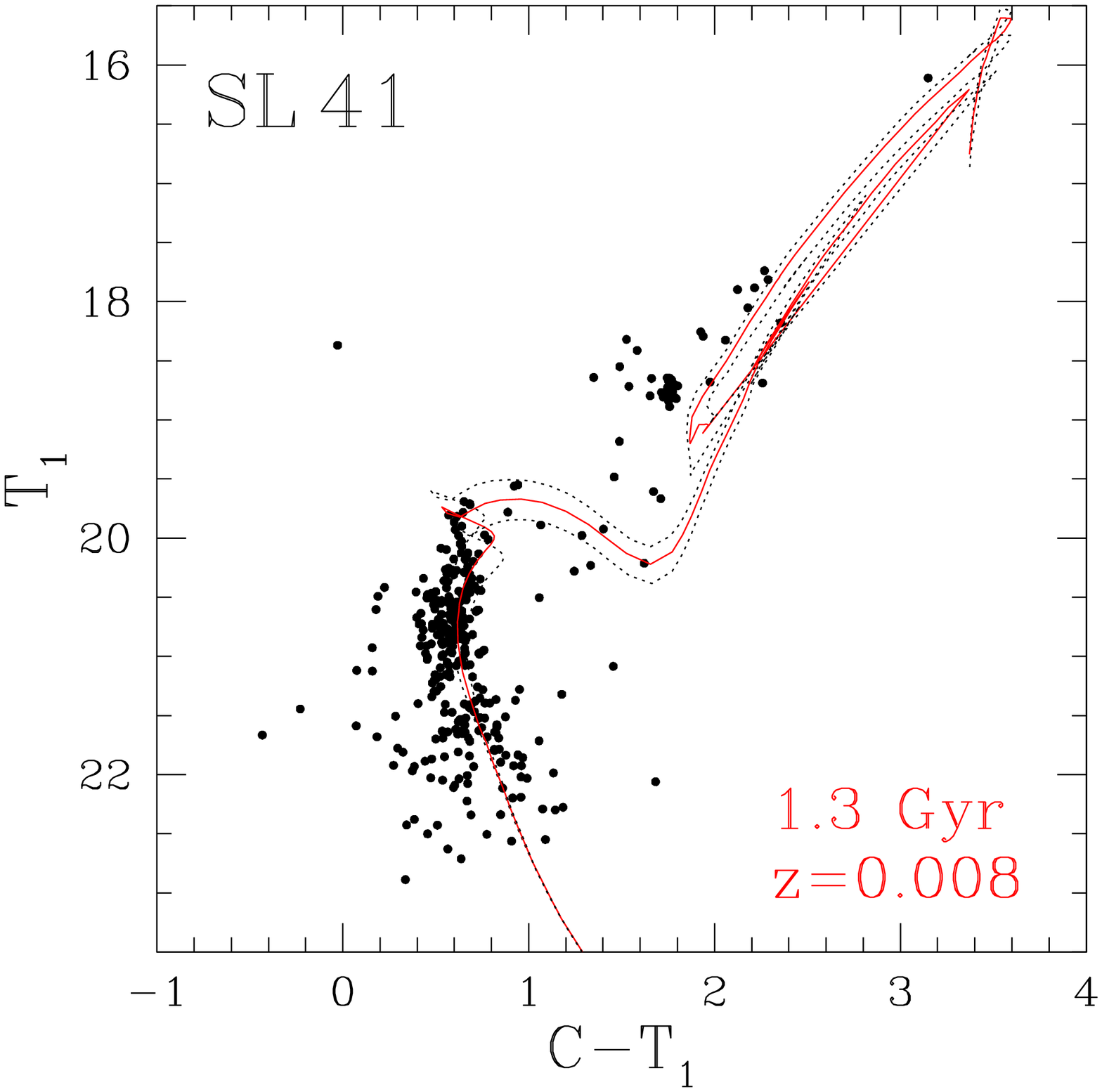}
 \includegraphics[width=40mm]{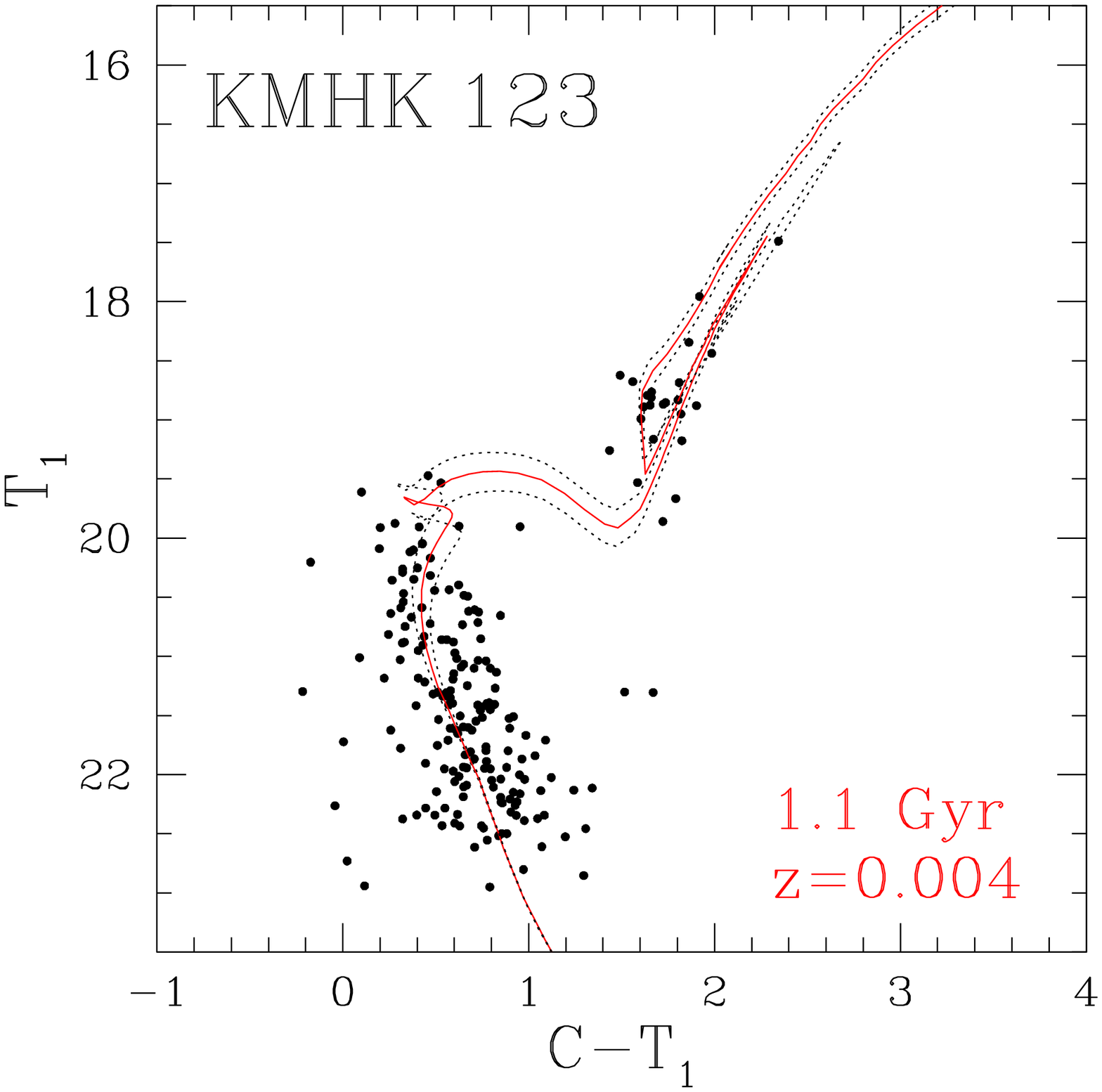}
 \includegraphics[width=40mm]{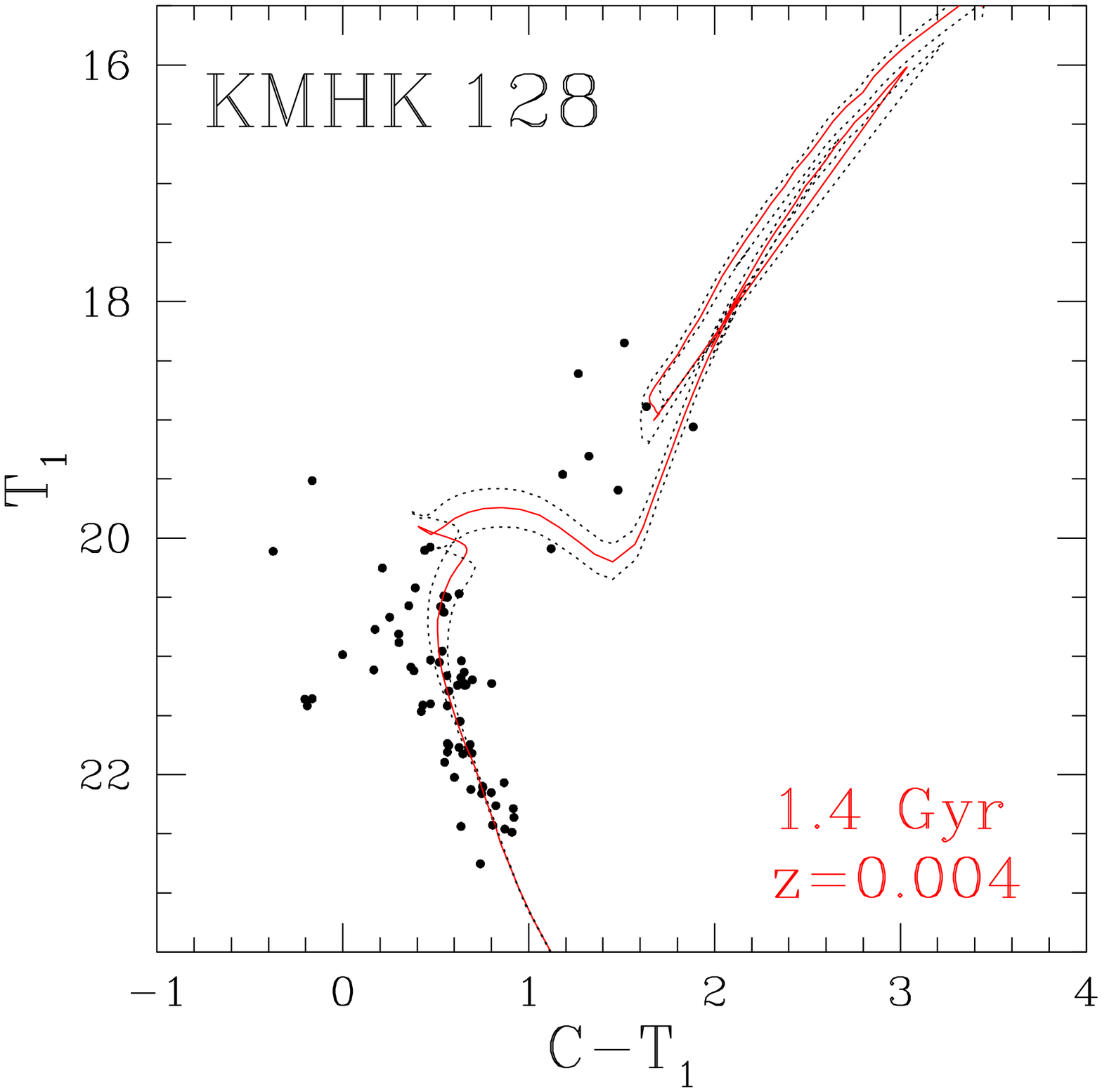}
 \includegraphics[width=40mm]{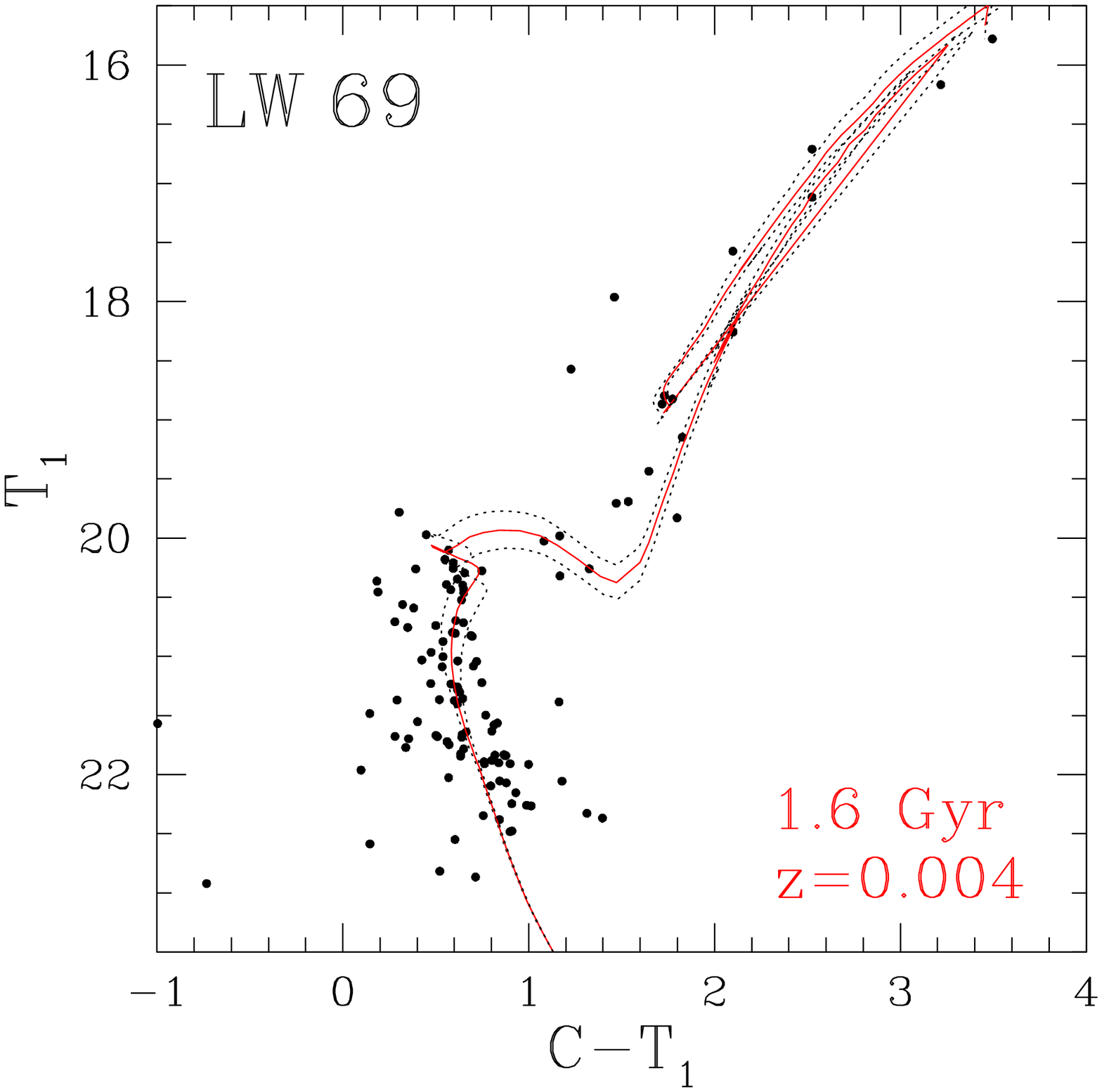}
 \includegraphics[width=40mm]{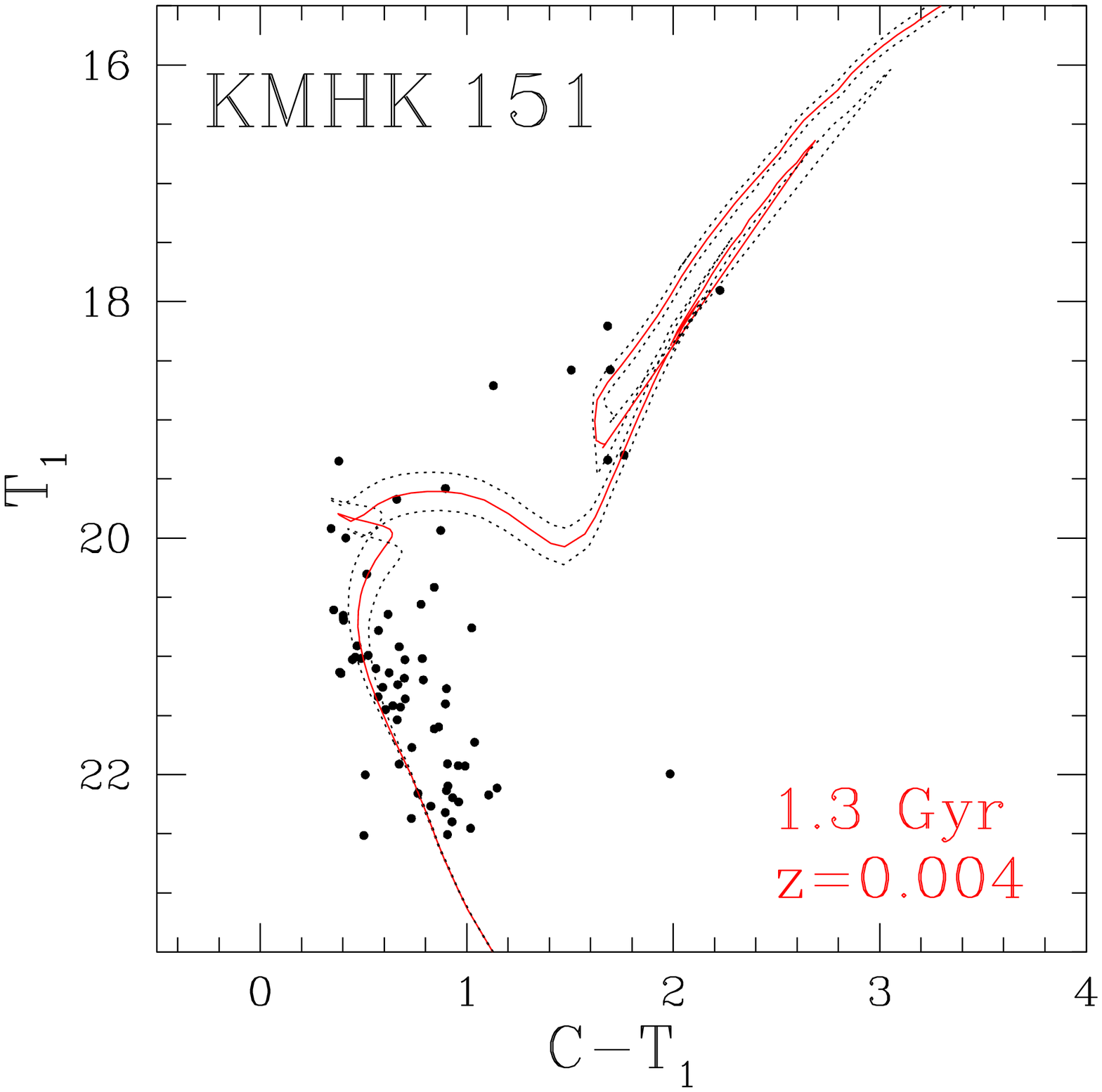}
 \includegraphics[width=40mm]{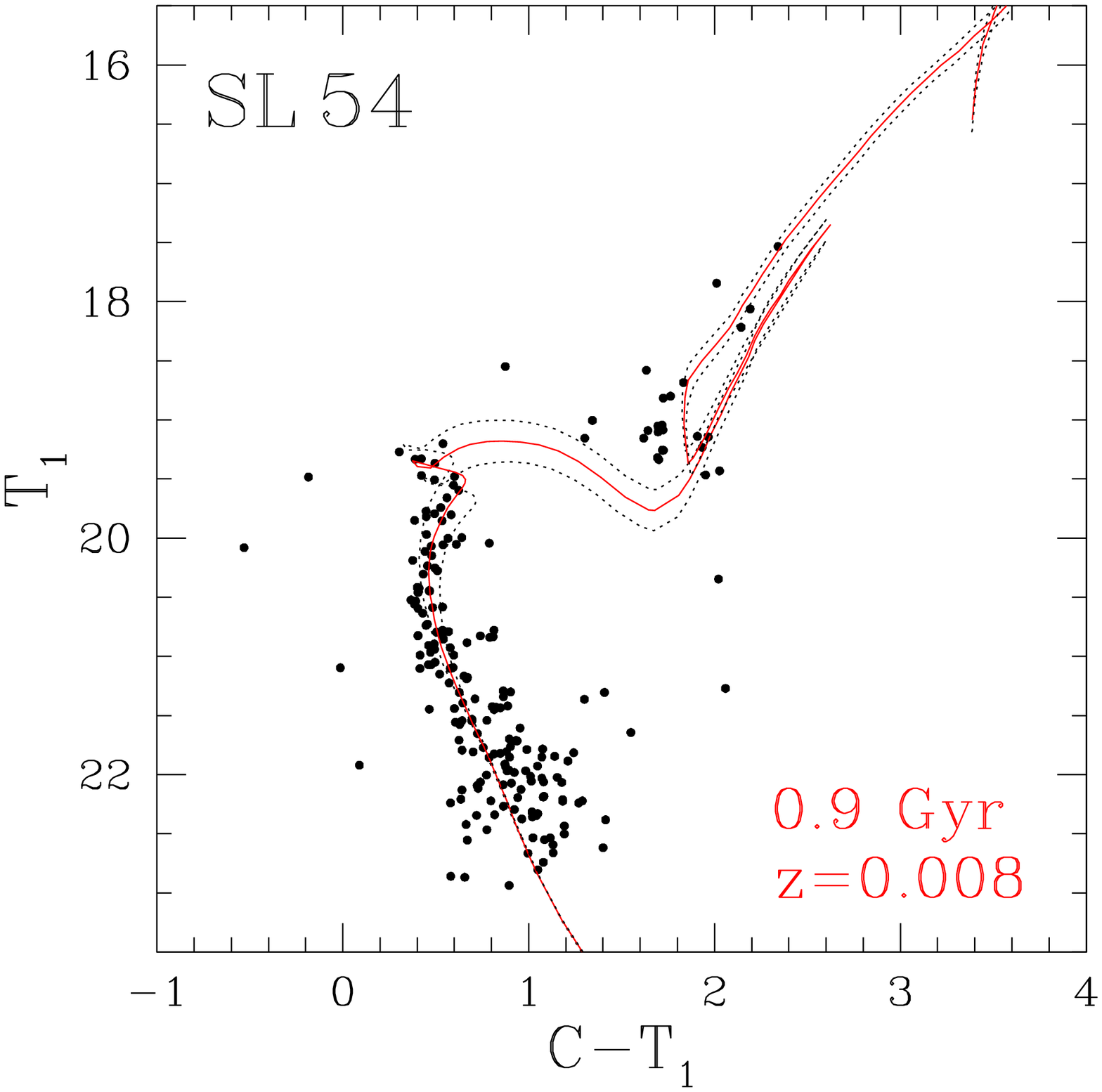}
 \includegraphics[width=40mm]{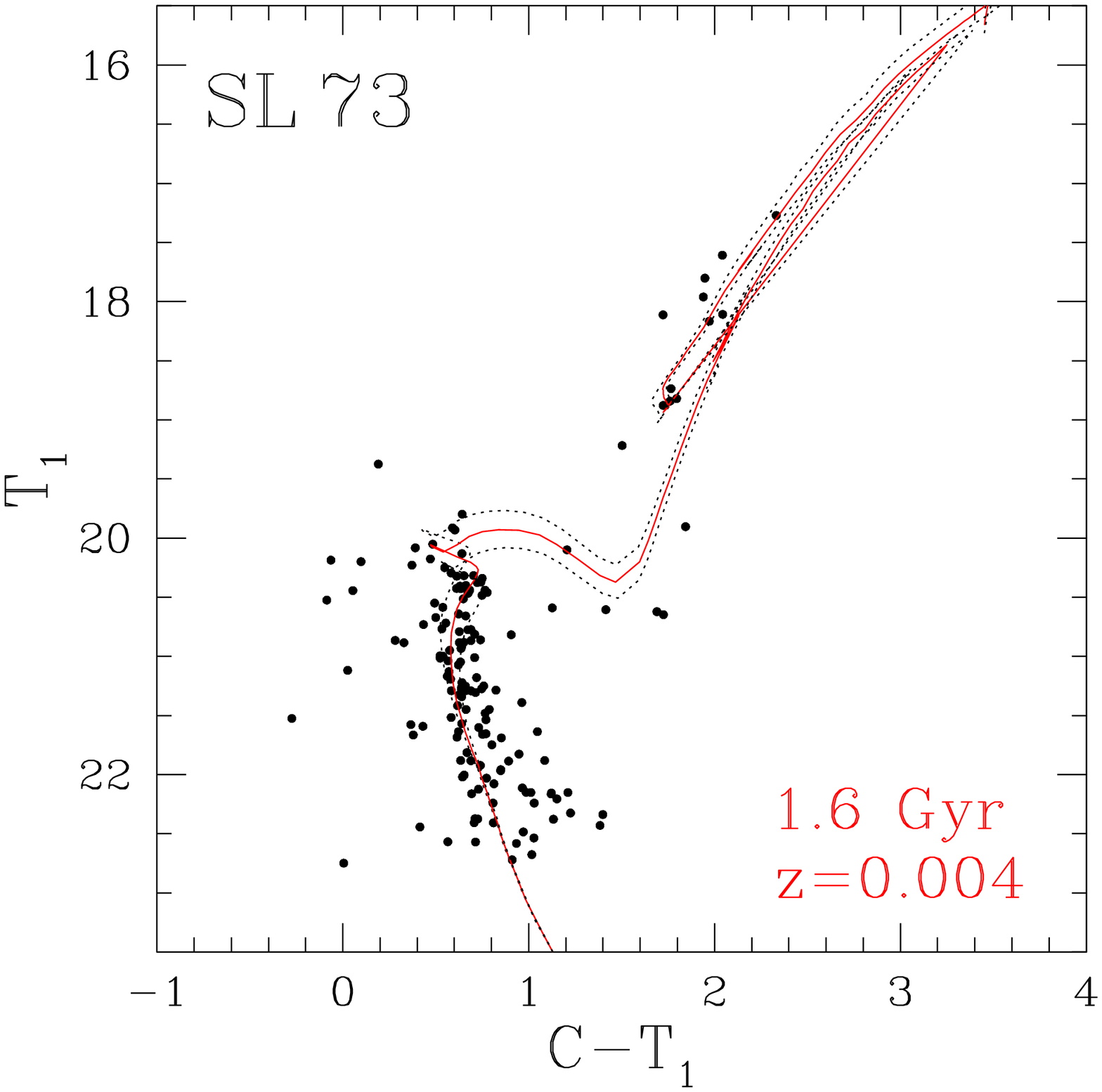}
 \includegraphics[width=40mm]{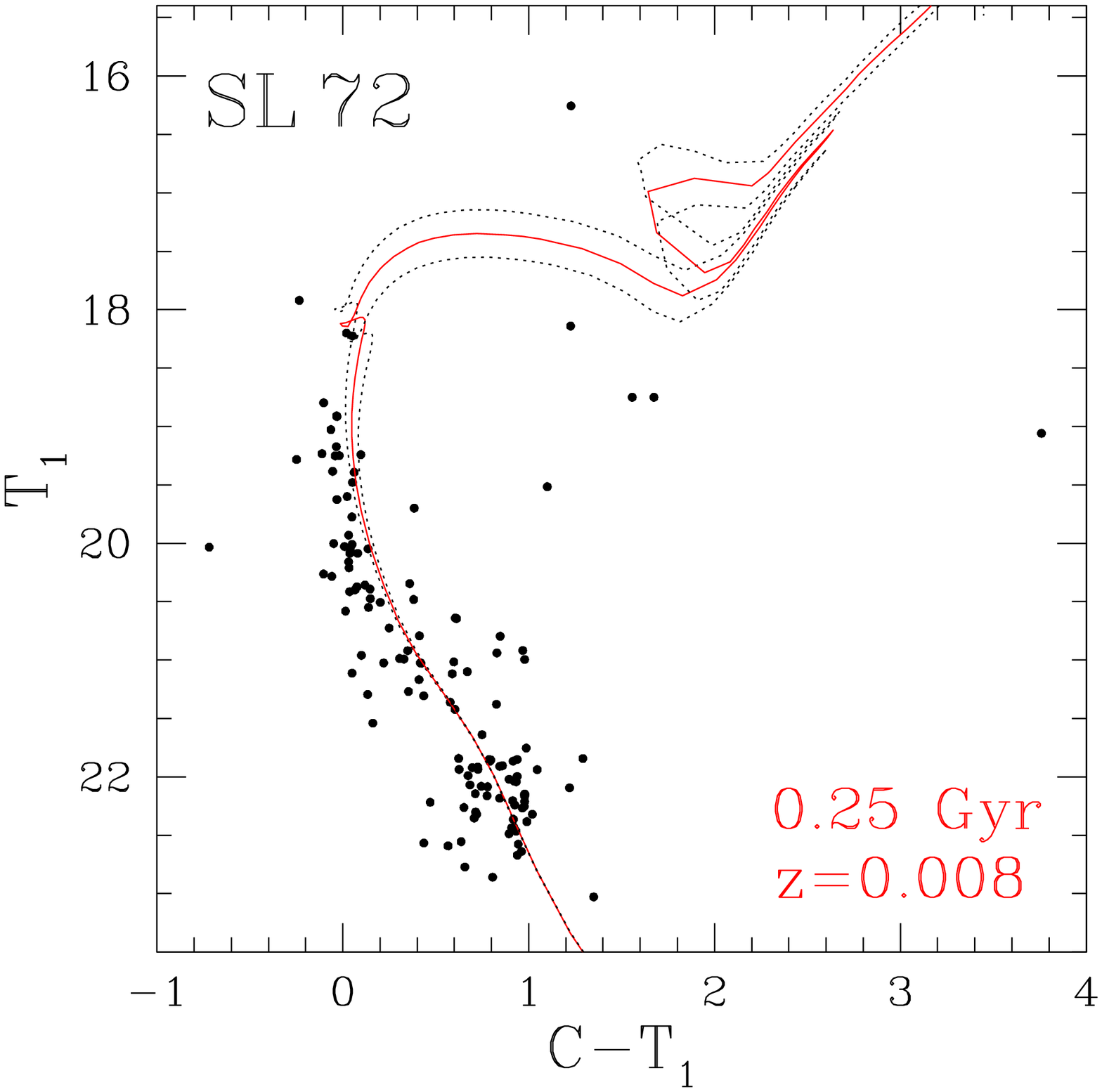}
 \includegraphics[width=40mm]{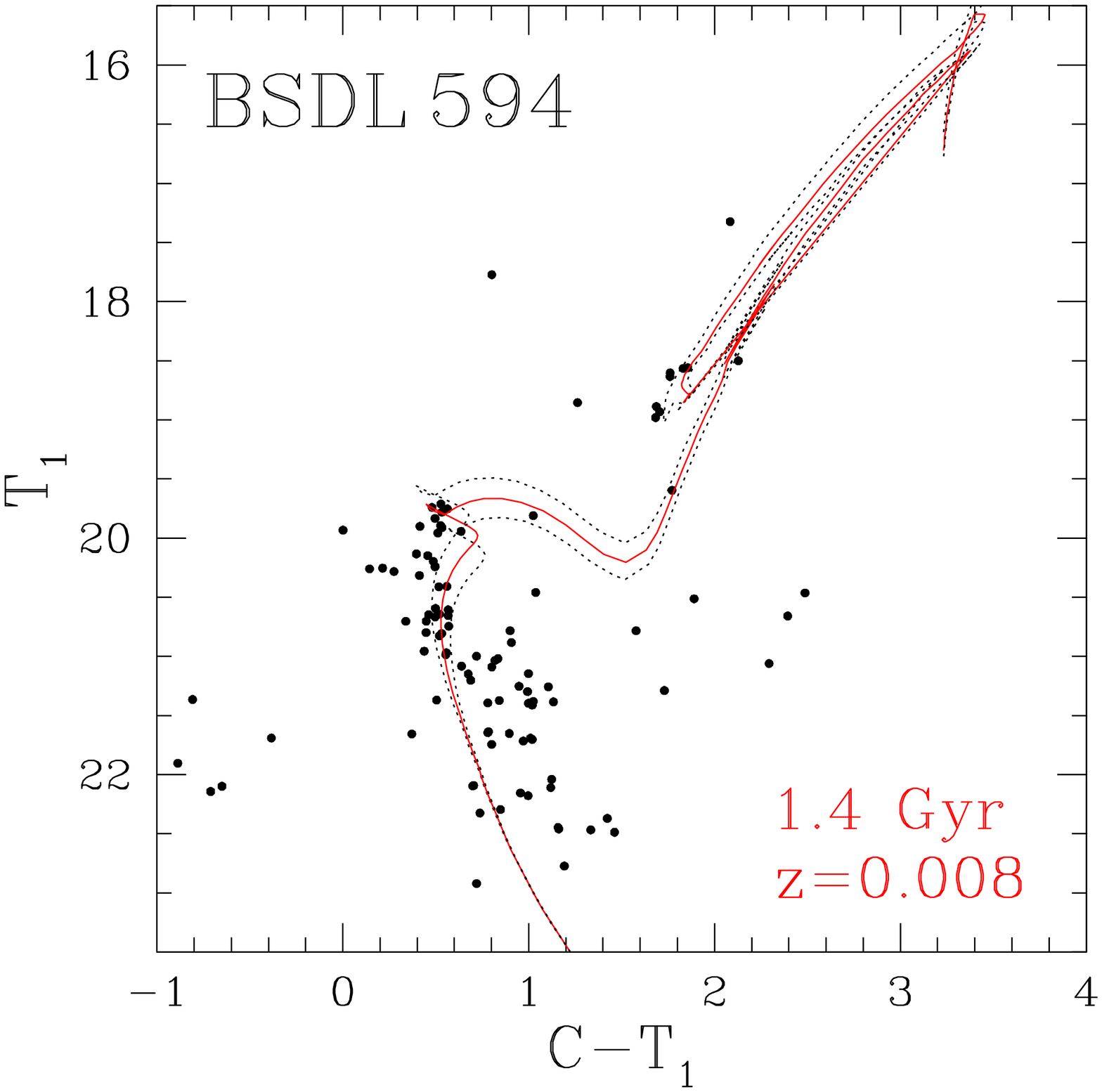}
 \includegraphics[width=40mm]{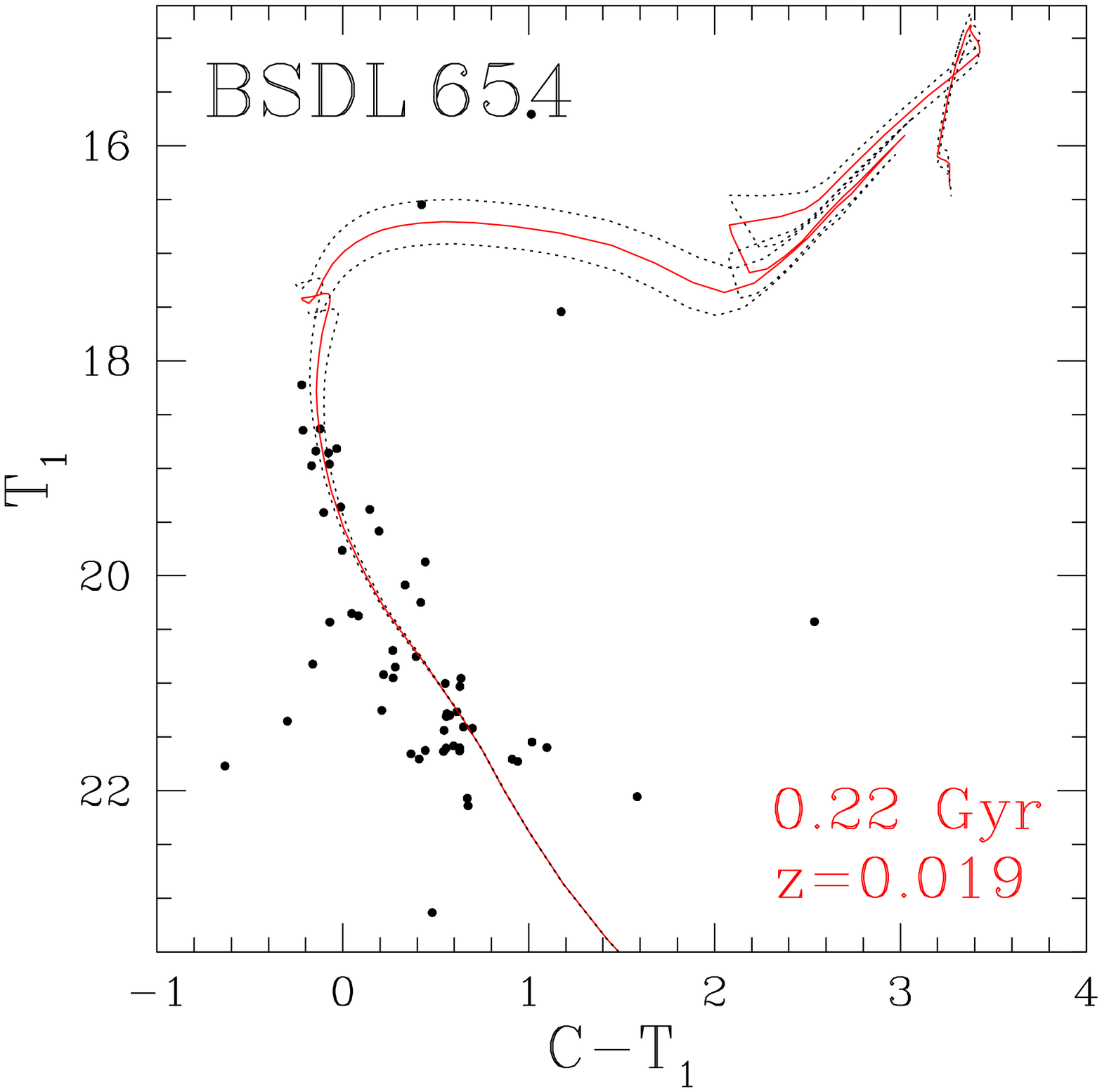}
 \includegraphics[width=40mm]{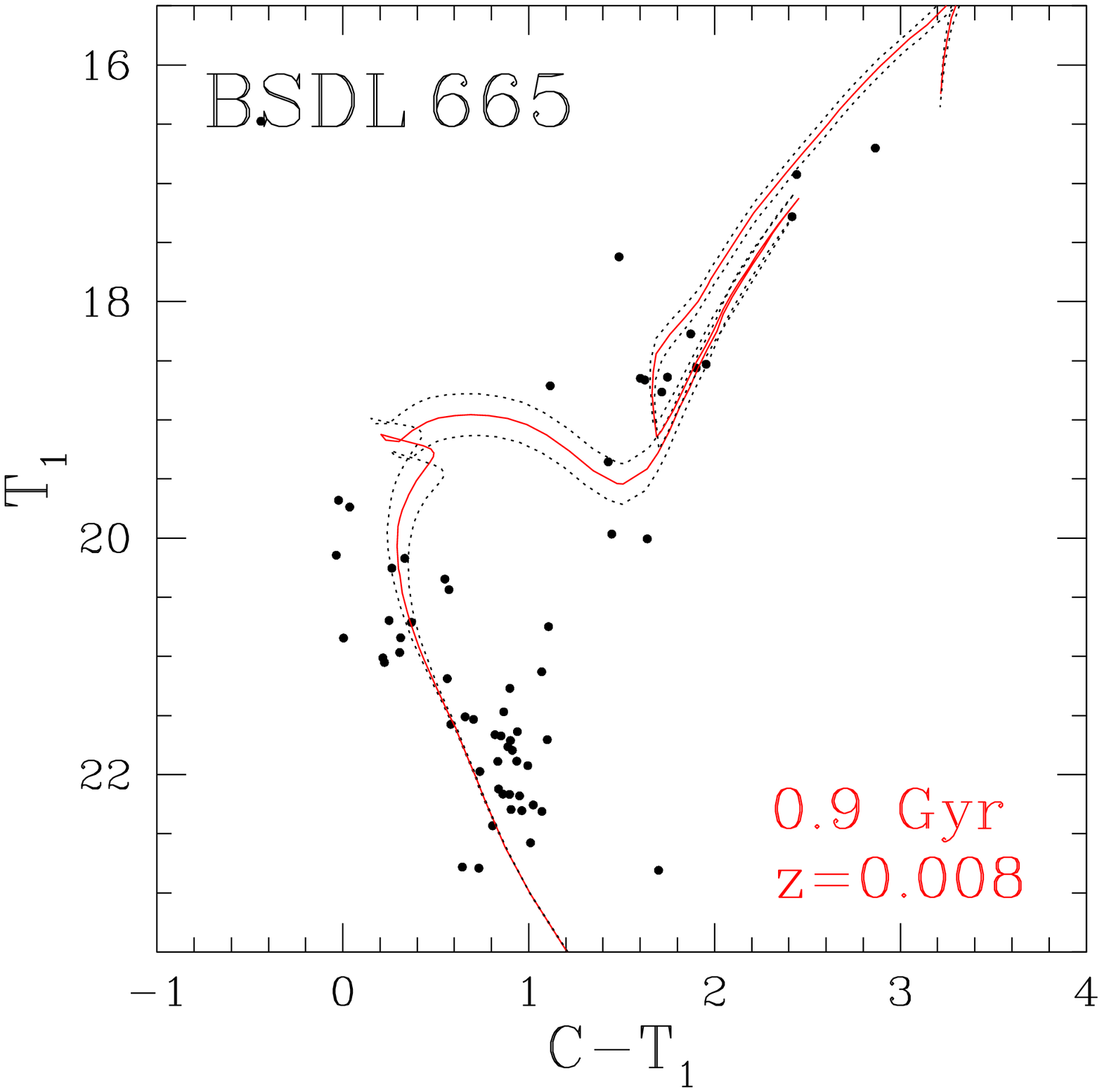}
 \includegraphics[width=40mm]{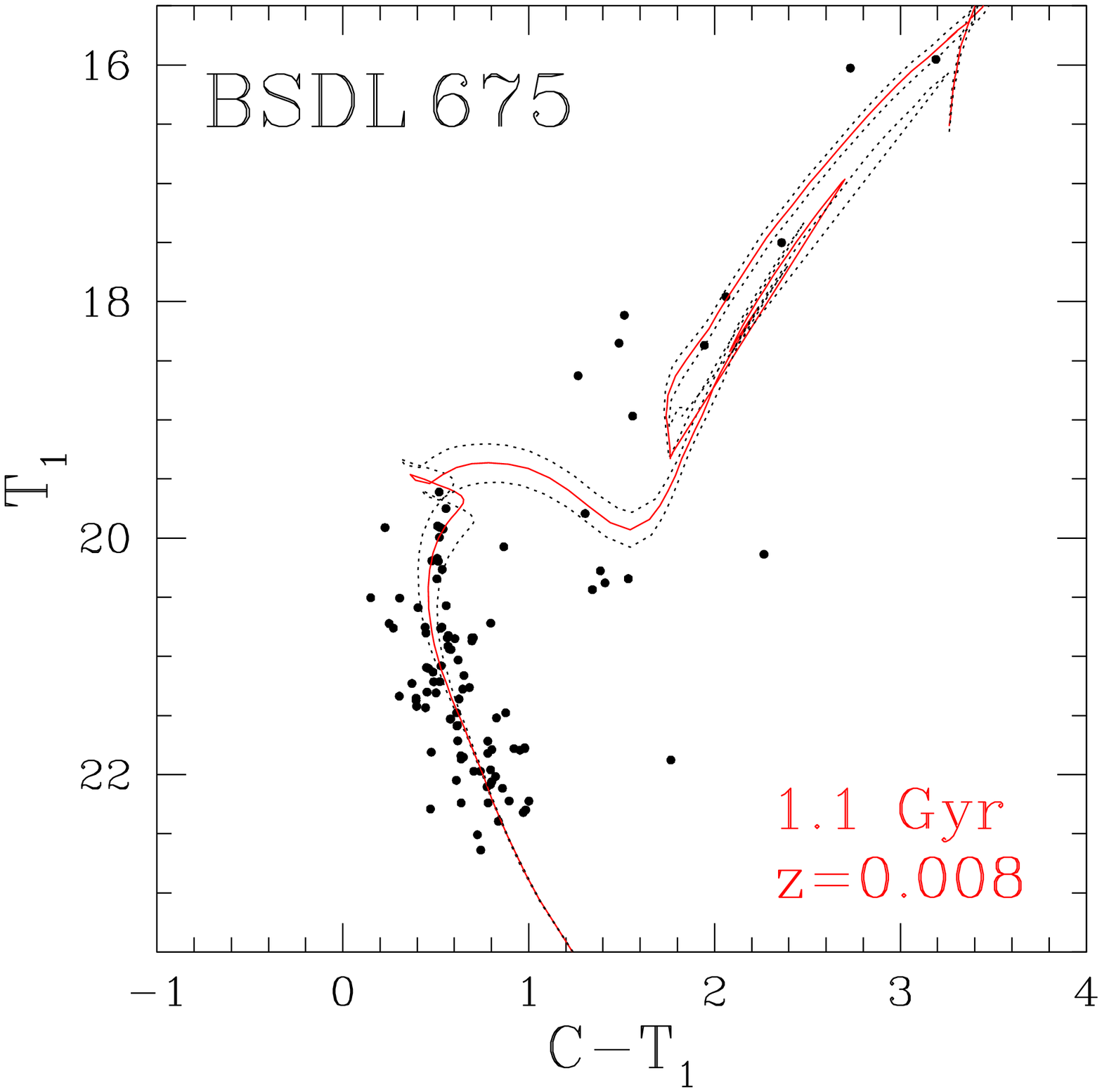}
 \includegraphics[width=40mm]{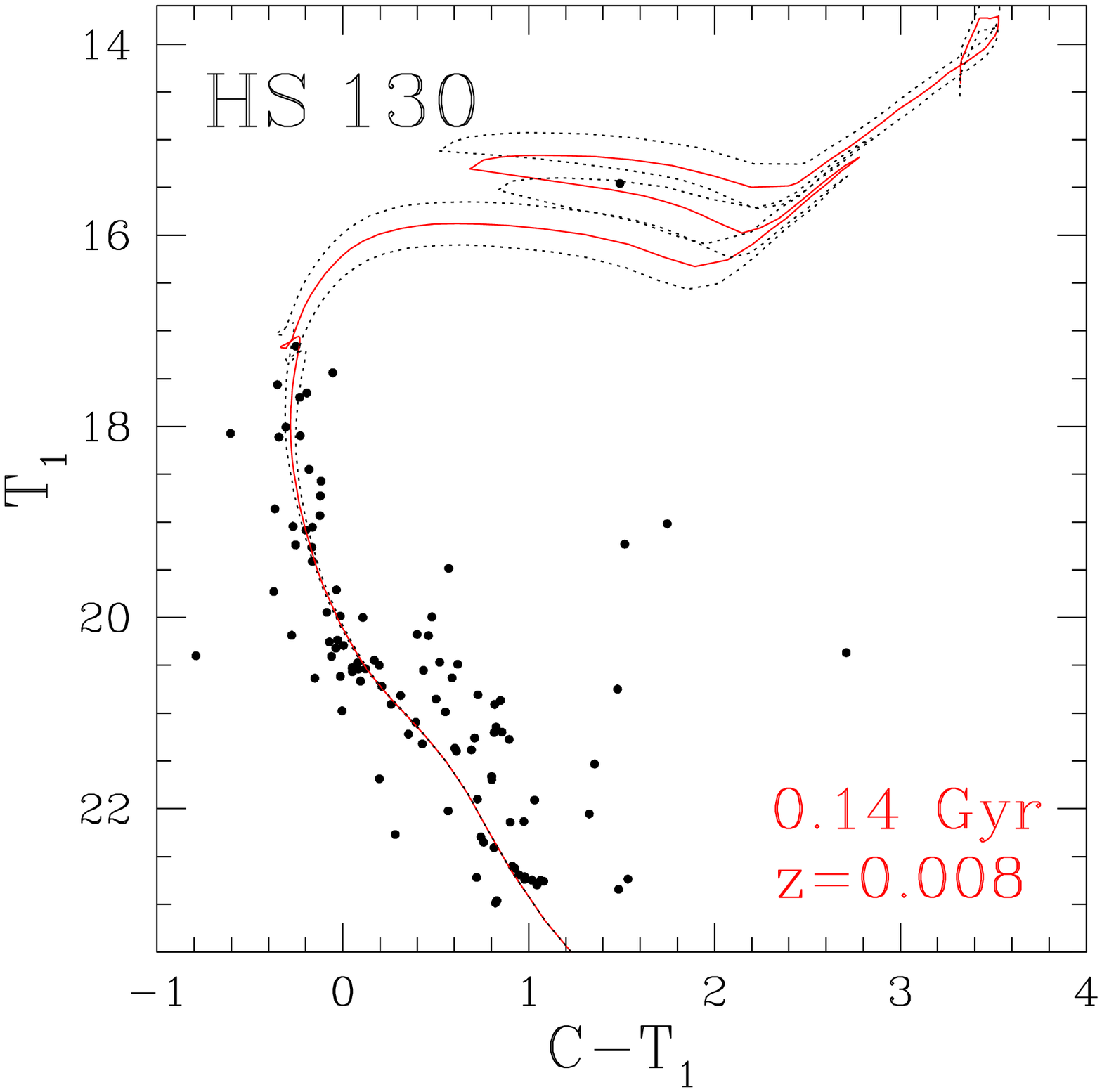}
 \includegraphics[width=40mm]{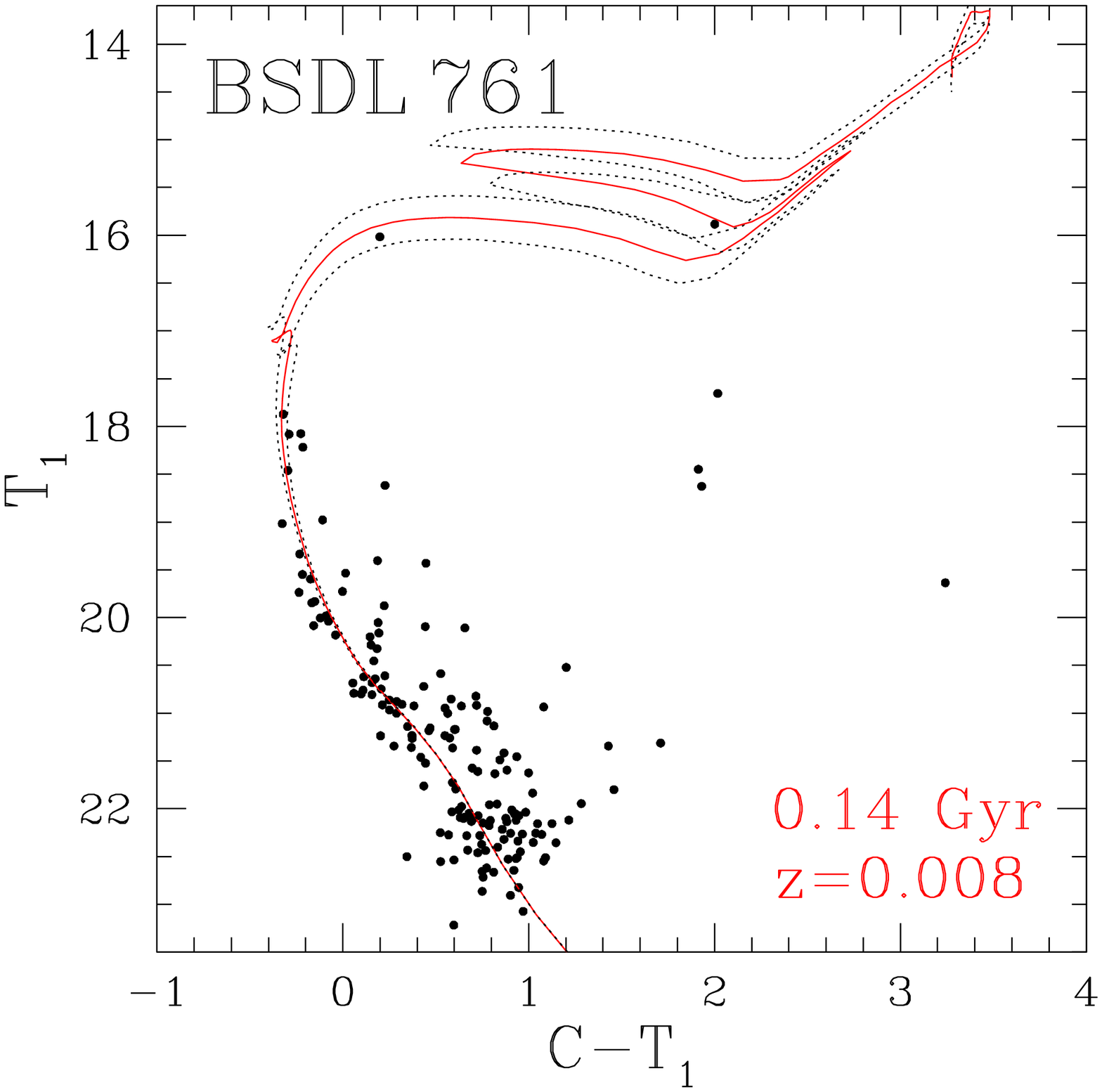}
 \includegraphics[width=40mm]{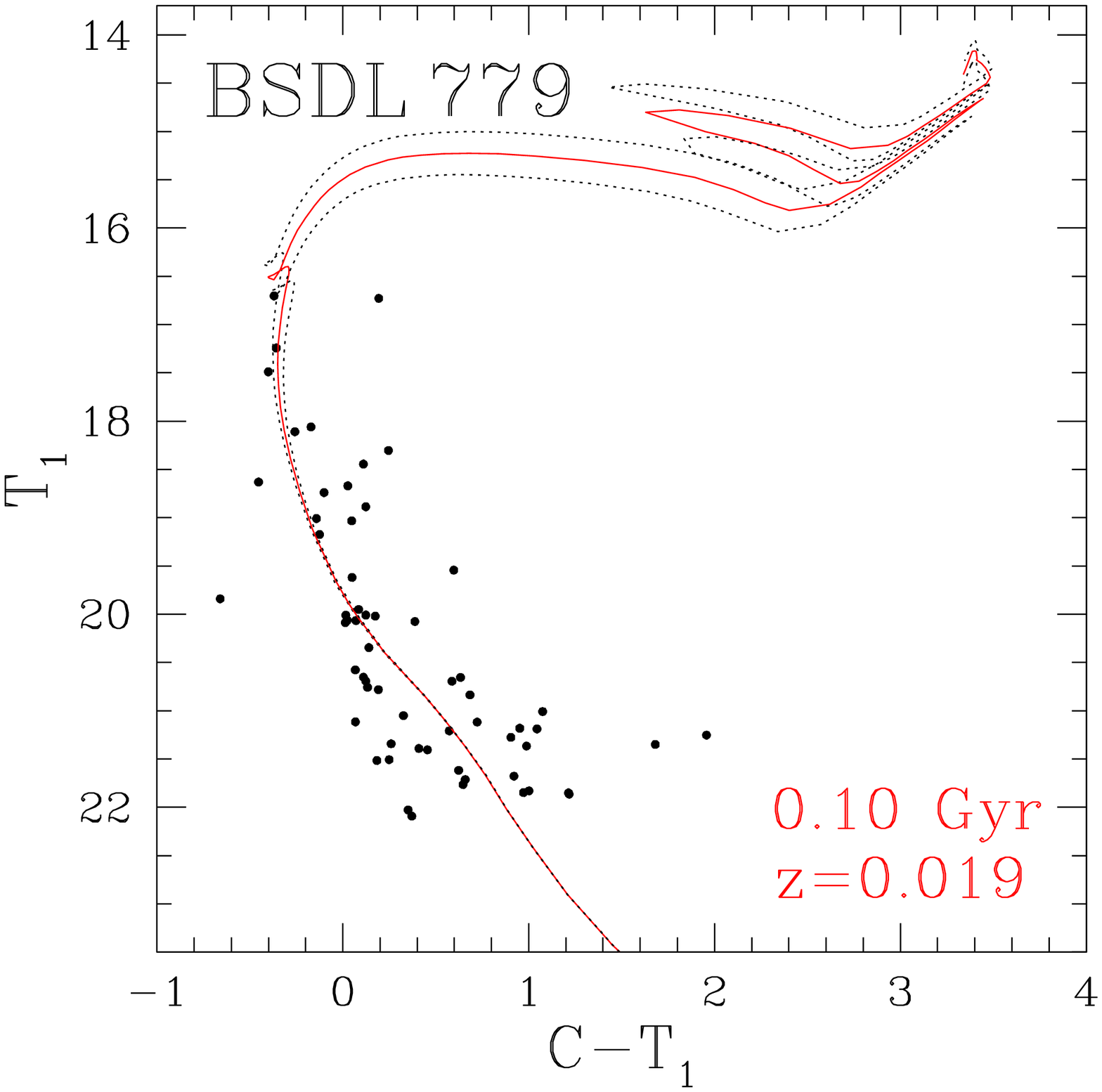}
 \includegraphics[width=40mm]{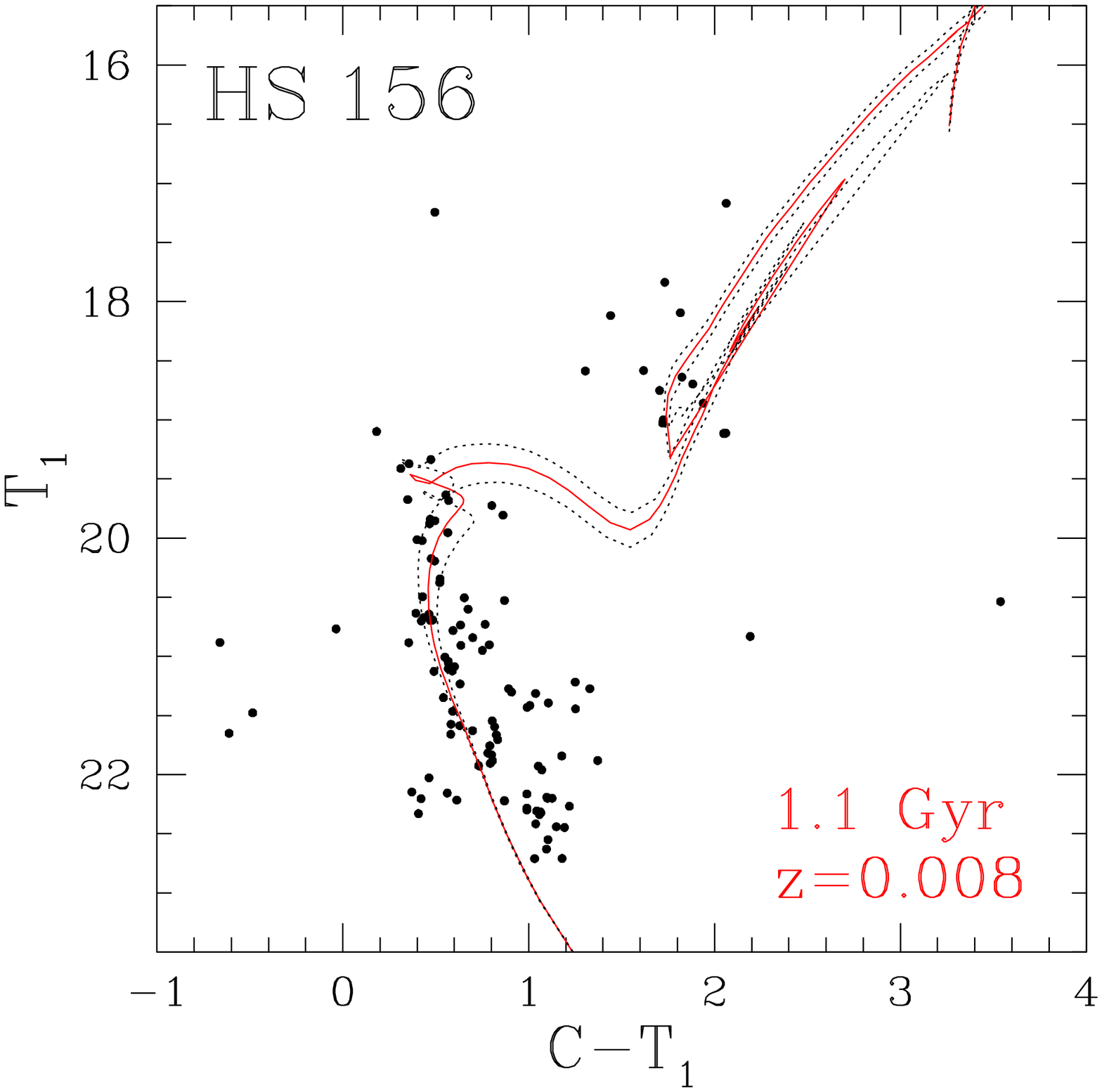}
 \includegraphics[width=40mm]{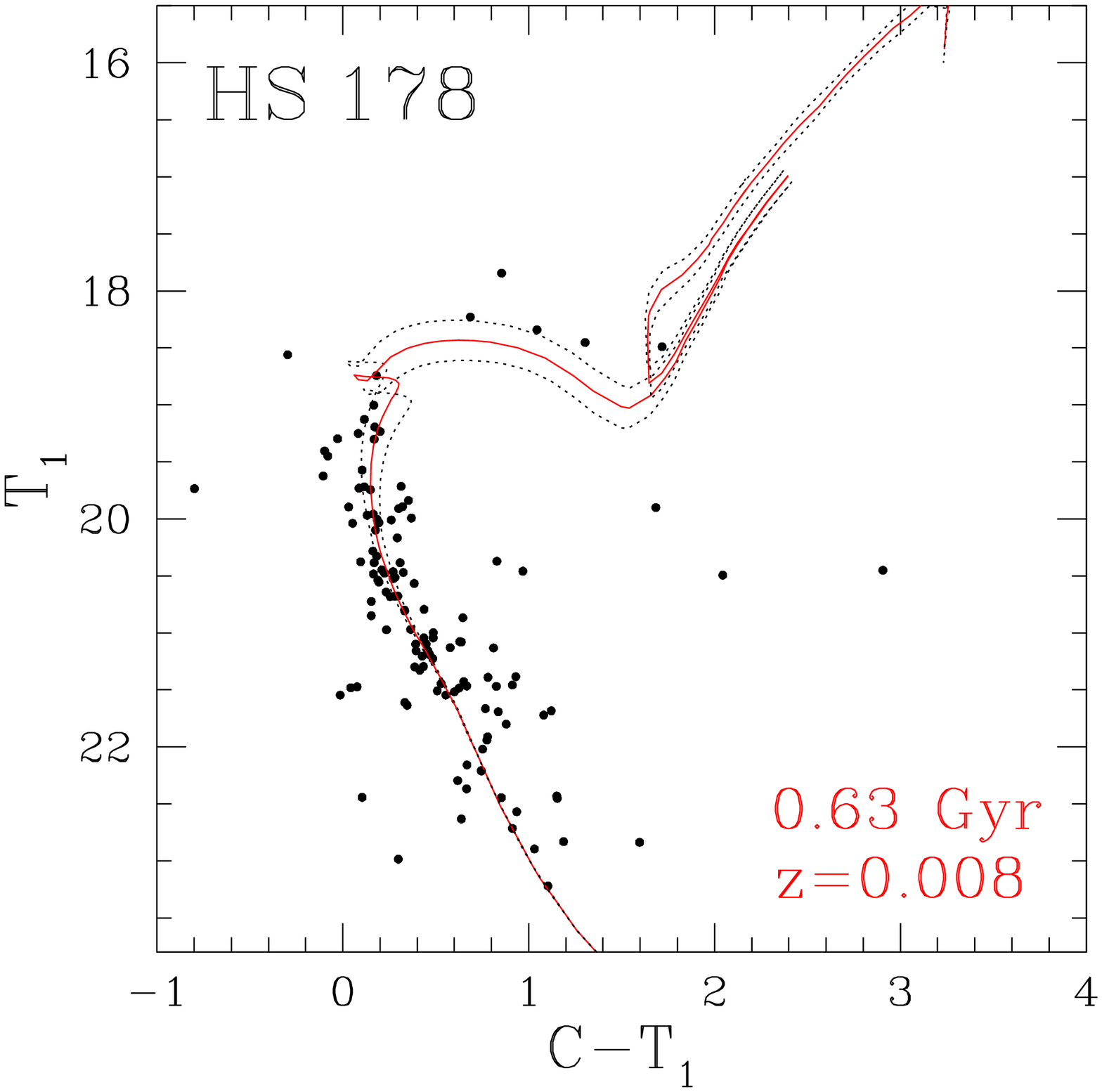}
 \includegraphics[width=40mm]{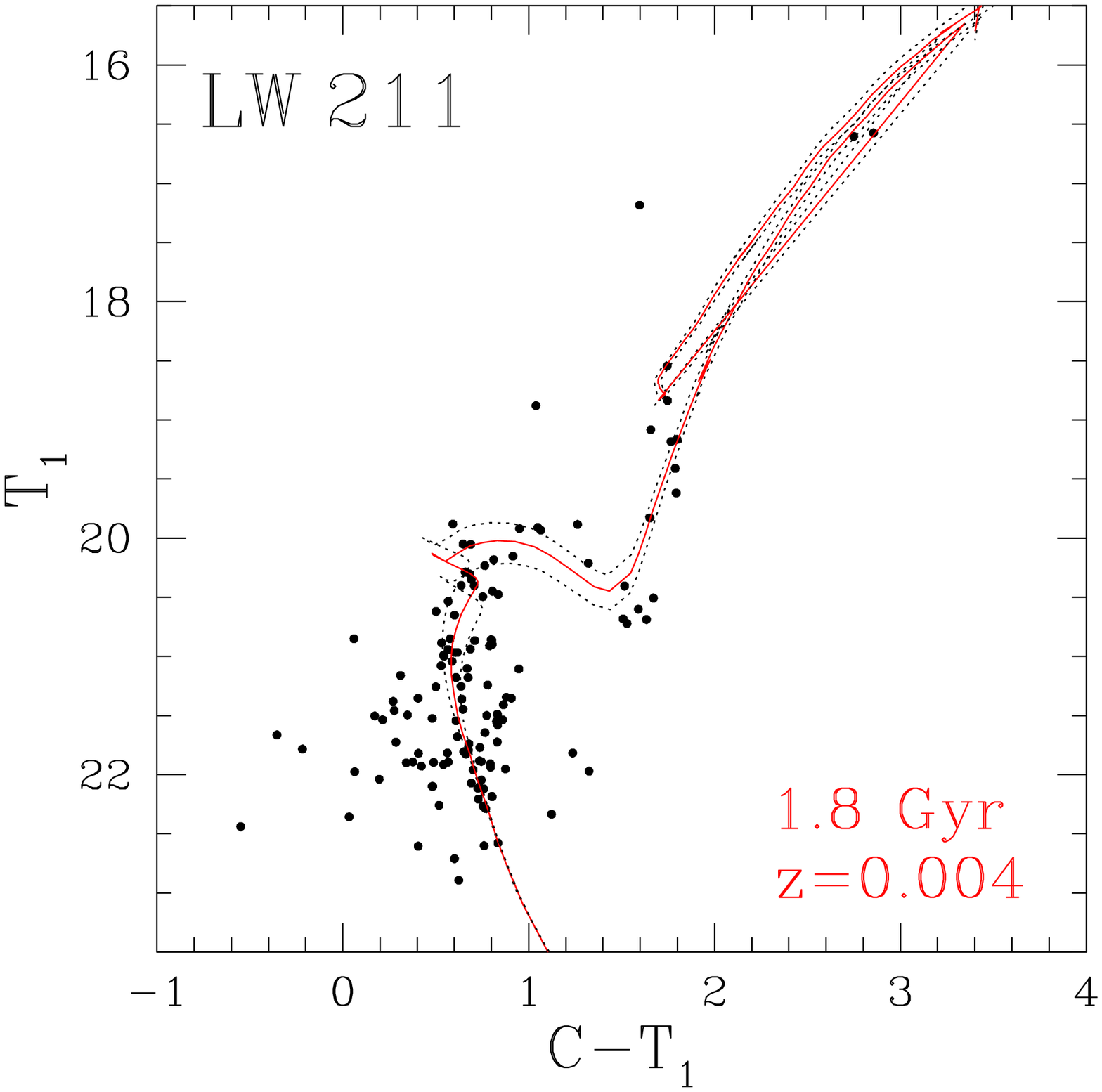}
 \includegraphics[width=40mm]{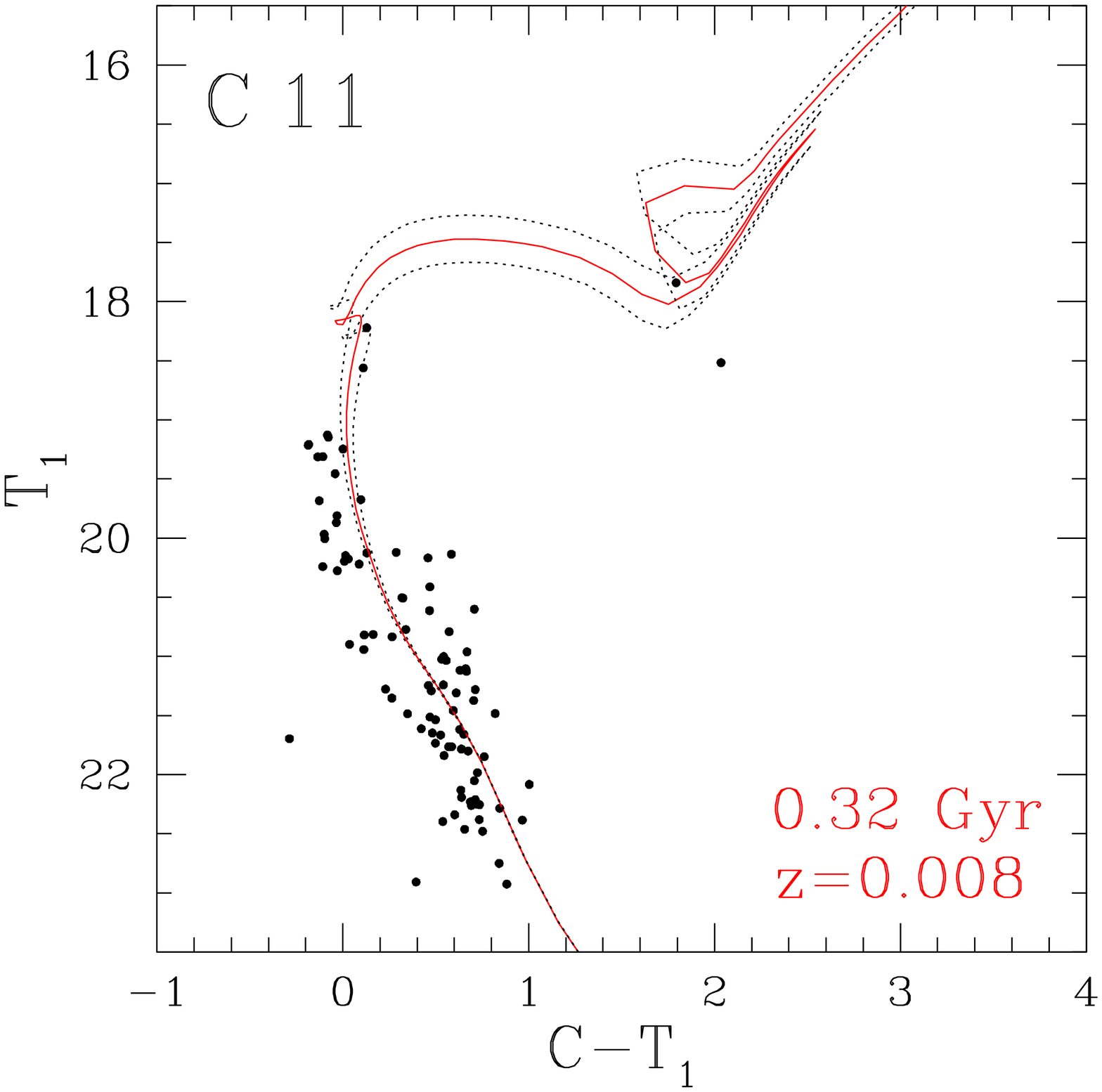}
 \includegraphics[width=40mm]{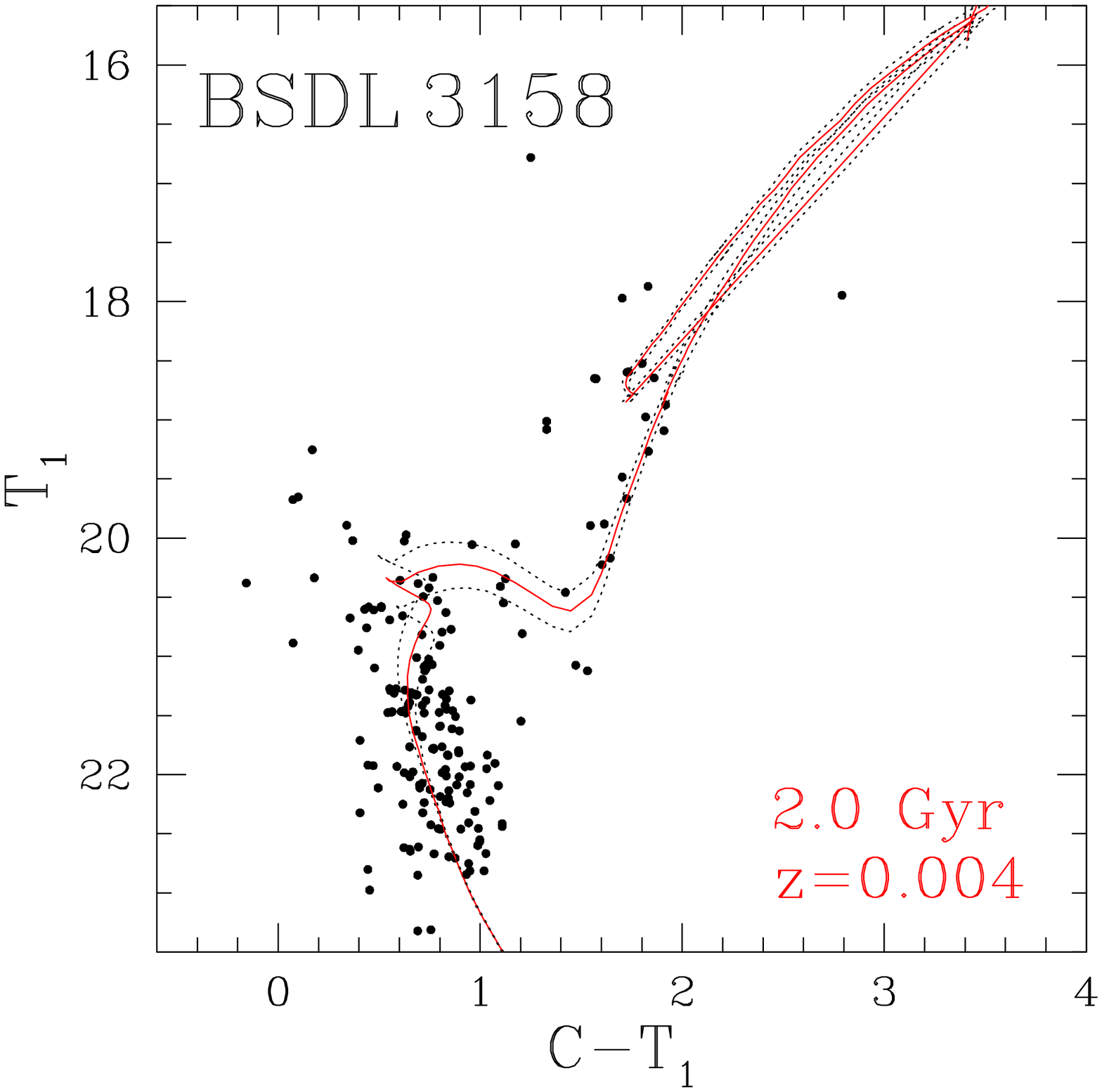}
 \includegraphics[width=40mm]{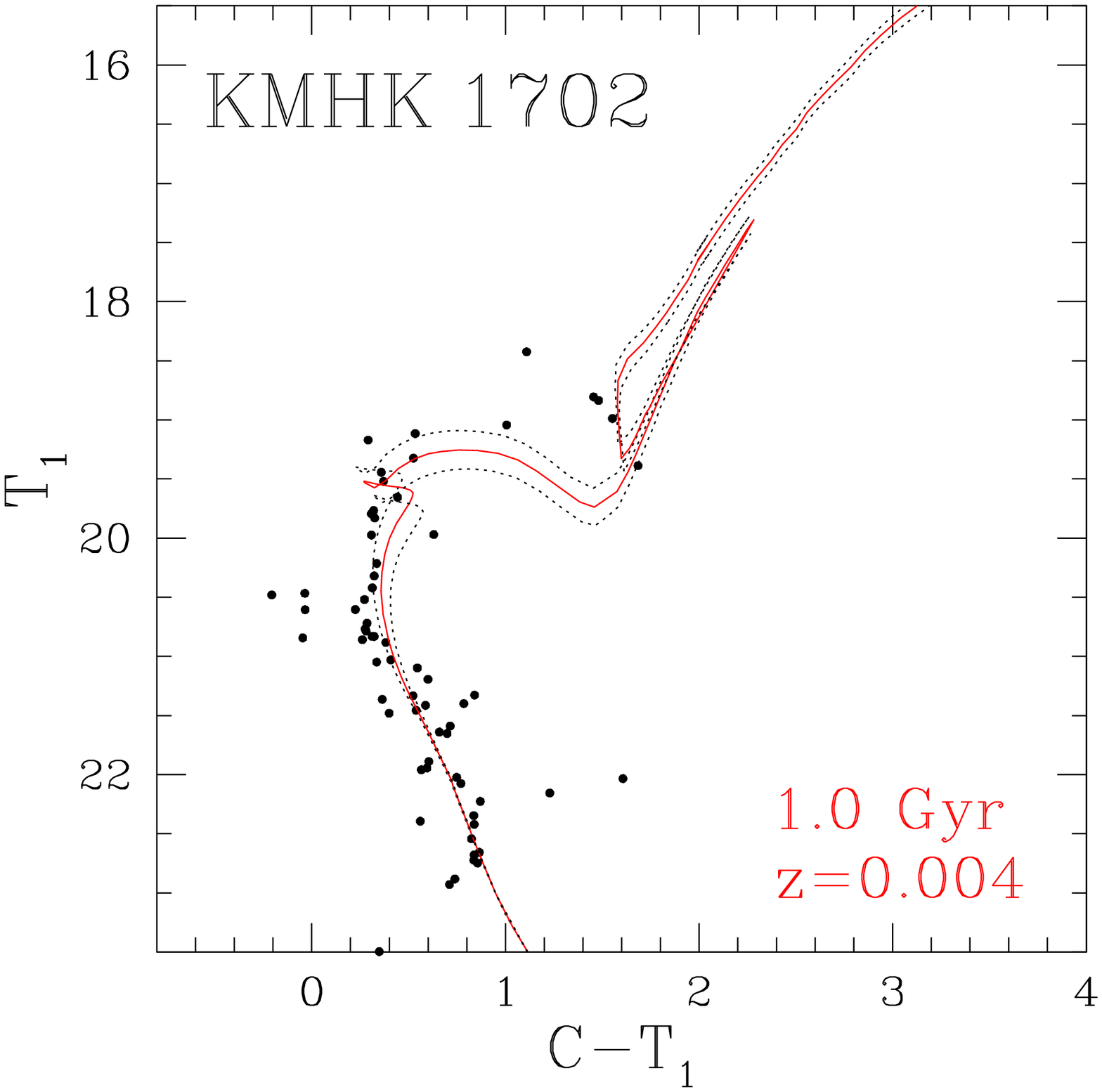}
 \includegraphics[width=40mm]{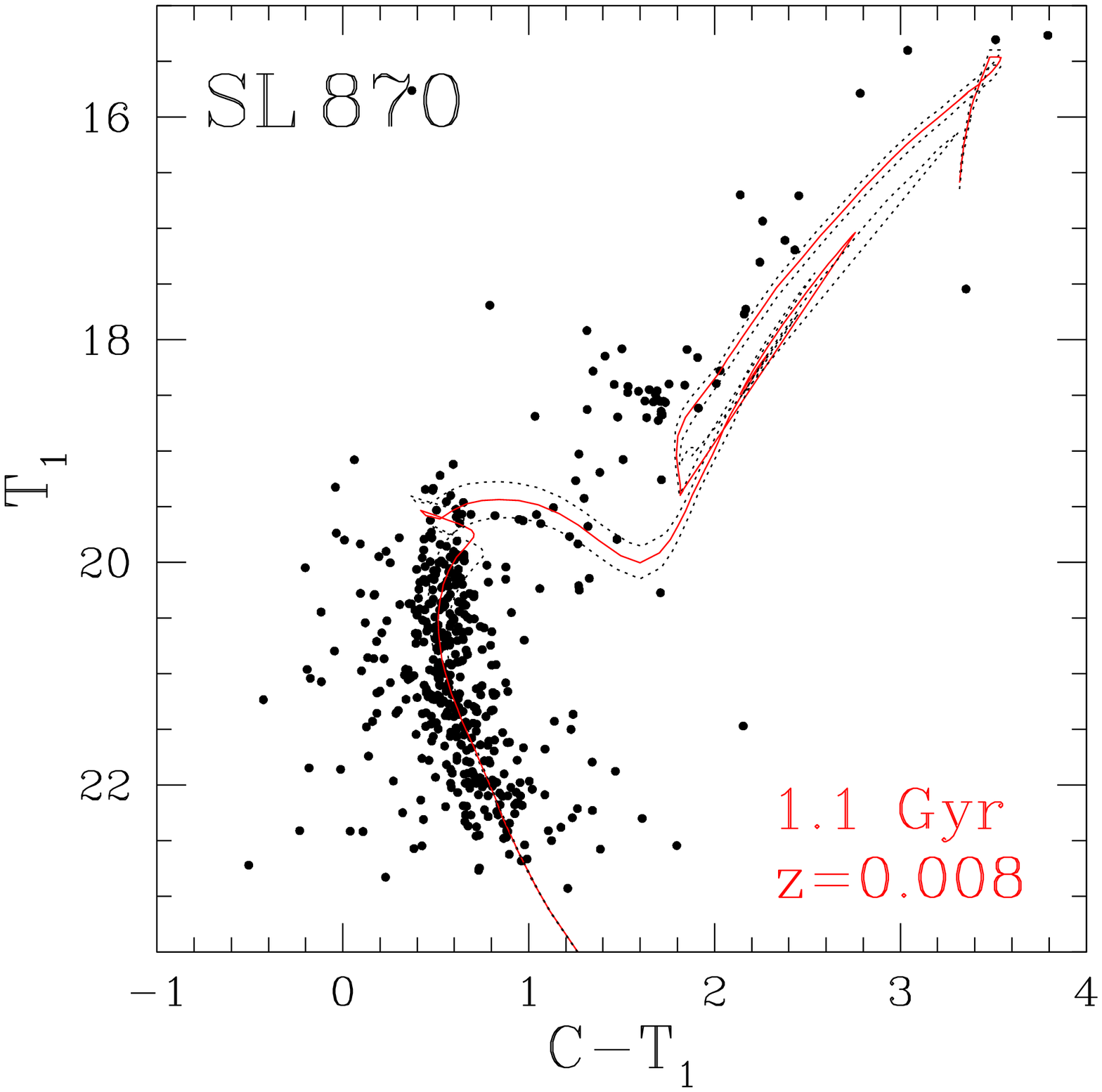}
  \caption{Cleaned Washington ($T_1,C-T_1)$ CMDs for the studied clusters. Isochrones from \citetads[]{gir02}, computed taking into account overshooting, are overplotted. The red solid and black dashed lines correspond to the derived cluster age and to the ages of the 
  nearest younger and older isochrones.} 
  \label{f:cmd}
 \end{figure*}

To estimate cluster ages, we used the theoretical isochrones computed by the Padova group \citepads[]{gir02} for the Washington photometric system. These isochrones include core-overshooting effects. Although we initially used the isochrones derived by the Geneva group \citepads[]{lsch01} which lead to nearly the same results, we finally decided to adopt Padova isochrones because they fit the fainter magnitudes of the main sequence (MS) better. We used chemical compositions of Z = 0.019, 0.008, and 0.004, equivalent to [Fe/H] = 0.0, -0.4, and -0.7, respectively, for the isochrone sets in steps of $\Delta$log t = 0.05 dex. Then, we selected a set of isochrones, 
along with the equations $E(C-T_1) = 1.97E(B-V)$ and $M_{T_1} = T_1 + 0.58E(B-V) - (V-M_V)$ given in \citetads[]{gs99}, and superimposed them on the cluster $(T_1,C-T_1)$ CMDs, once they were properly shifted by the corresponding $E(B-V)$ colour excesses and by the LMC 
distance modulus. In the matching procedure, we employed different isochrones, ranging from slightly younger to slightly older than the derived cluster age. We finally adopted as the cluster age the one corresponding to the isochrone that best matched the shape 
and position of cluster MSs, particularly at the turn-off (TO) level. We also took into account the $T_1$ magnitude of the red giant clump (RGC). The age error was estimated considering the isochrones that encompassed those features best. Columns 6 and 7 of Table 
\ref{t:results} and Fig. \ref{f:cmd} show the results of the isochrone fittings. For each cluster CMD, we plotted the isochrone of the adopted cluster age (solid line) and two additional isochrones bracketing the derived age (dotted lines). Note that the theoretically 
computed bluest stage during the core He-burning phase appears to be redder than the observed RGC in the CMDs of SL\,41 and SL\,870, a behaviour that has also been detected in previous studies \citep[see, e.g.,][]{g03,p04,cla07}. \\

A second method to derive cluster ages is based on the $\delta(T_1)$ parameter, defined as the difference in $T_1$ magnitude between the RGC and the MSTO in the Washington $(T_1,C-T_1)$ CMD. The age is obtained from the following equation given in \citetads[]{g97}:
\begin{equation}
Age(Gyr) = 0.23+2.31\times \delta T_1-1.80\times {\delta T_1}^2+0.645\times{\delta T_1}^3 ,  
\end{equation}
with a typical error of $\pm$0.3 Gyr. Age determination via $\delta T_1$, however, is applicable only to intermediate-age (IACs) and/or old clusters, i.e. generally older than 1 Gyr. Although some clusters seemed to be IACs (1-3 Gyr), it was not possible to determine 
their ages from $\delta T_1$ because the RGC in their CMDs was not clearly detected. This is because sometimes the central regions of the clusters appear to be saturated, perhaps there are just very few RC stars in some faint clusters, or else they are not photometrically well resolved in our images. In these cases, the red giant stars are missed and no clump is visible in the CMDs. This is why we derived ages of only ten clusters older than 1 Gyr based on the $\delta T_1$ parameter. The mean $\delta T_1$ values were estimated from the average of independent measurements made by two authors. These measurements agreed in general very well. The resulting $\delta T_1$ values and the corresponding cluster ages are listed in columns (4) and (5) of Table \ref{t:results}. 
As can be seen, $\delta T_1$ ages agree well with those estimated from the isochrone fitting, which confirms that both procedures allow us to estimate ages on a similar scale.\\

\citetads[]{pandey} studied the integrated magnitudes and colours for some LMC clusters from synthetic models, among which are BSDL\,654, BSDL\,675, and BSDL\,779. They report ages of 0.16 and 1 Gyr for BSDL\,654 and BSDL\,675, respectively. These age estimates agree well with ours within the errors (see Table \ref{t:results}). For BSDL\,779, however, these authors reported an age of 0.03 Gyr, i.e., substantially younger than the one obtained here. On the other hand, two clusters from our sample, SL\,870 and SL\,41, were classified in \citetads[]{b96} as belonging to SWB V type \citepads[]{swb}, based on their integrated (B-V) and (U-B) colours. This SWB V type is compatible with clusters belonging to the 0.8-2.0  Gyr range, which agrees well with the age values derived here for these two clusters.\\

\begin{figure}
 \centering
 \includegraphics[width=44mm]{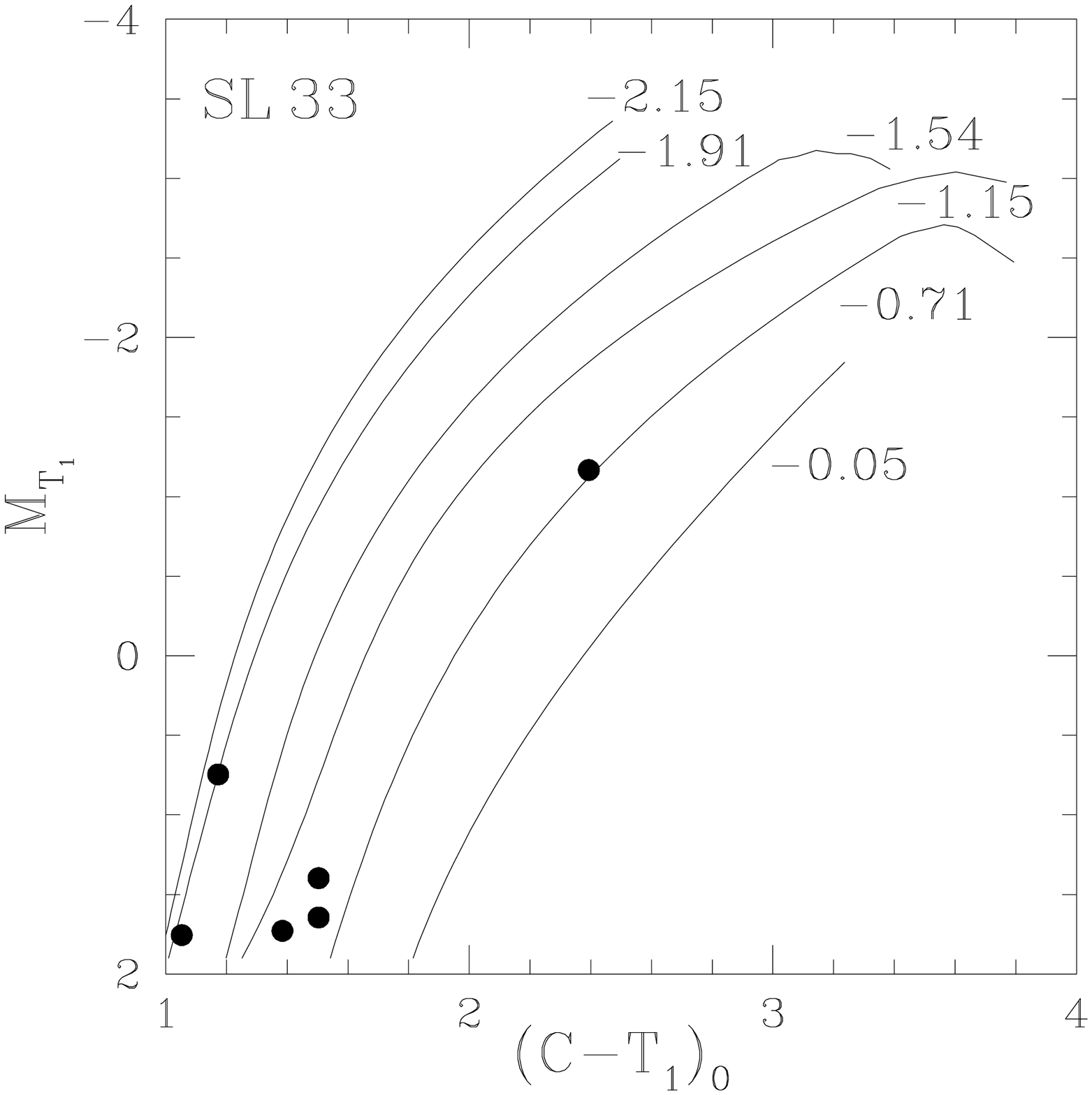}
 \includegraphics[width=44mm]{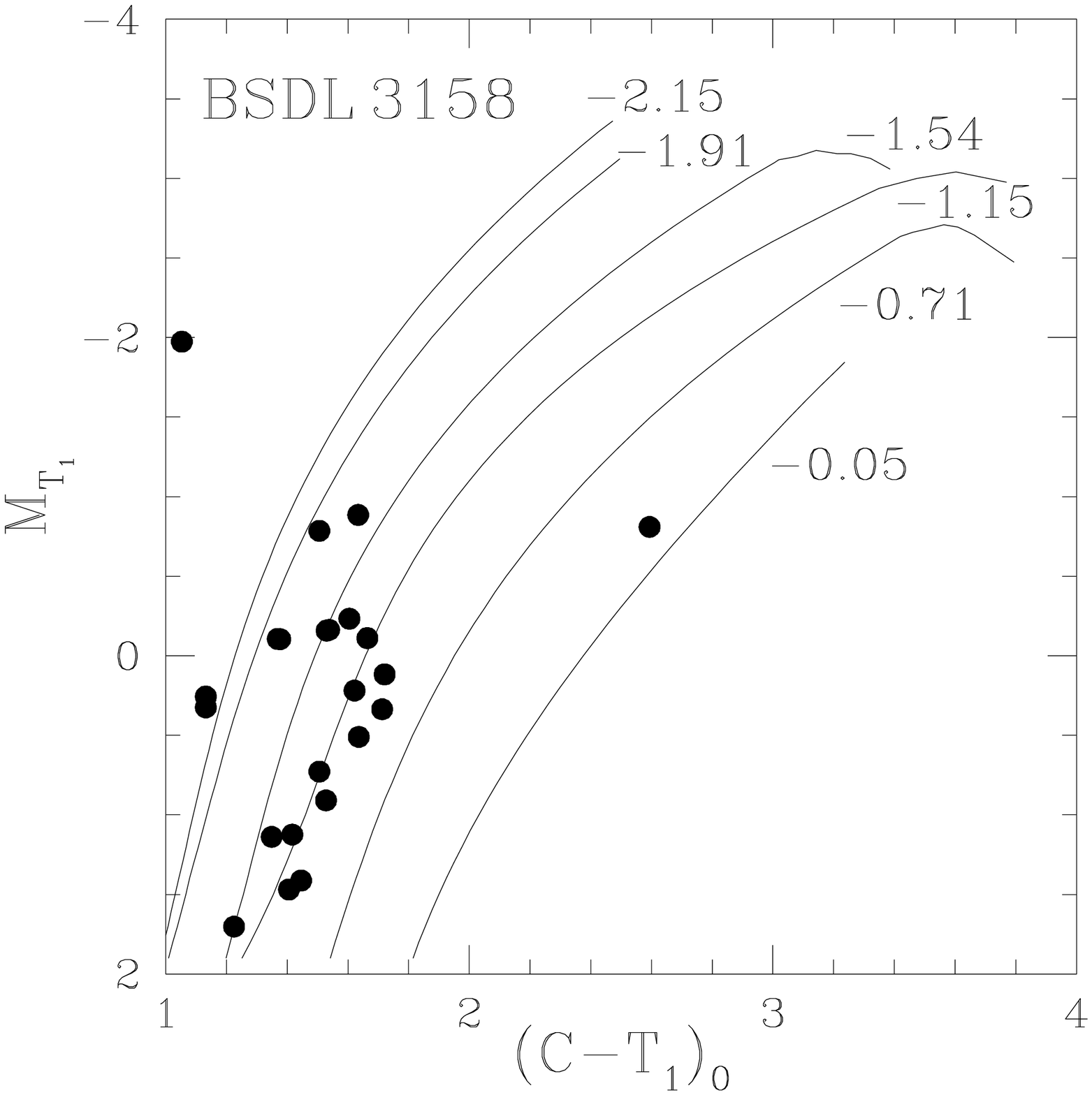}
  \caption{Washington $M_{T_1}$ vs. $(C-T_1)_0$ diagram of upper RGB stars in two LMC star clusters, with SGBs taken from \citetads[]{gs99} superimposed. An age-dependent correction to the indicated metallicities was applied for these clusters. } 
  \label{f:gs}
 \end{figure}

Metallicities for only the two oldest clusters of our sample (SL\,33 and BSDL\,3158) were also obtained using the [$M_{T_1},(C-T_1)_0$] plane with the standard giant branches (SGBs) of \citetads[]{gs99}. 
As these authors claim, this method should be applied only to star clusters aged 2 Gyr or older. Geisler and Sarajedini demonstrated that the metallicity sensitivity of the SGBs (each giant branch corresponds to an isoabundance curve) is three times higher than that of the V, I technique \citepads[]{da90} and that, consequently, it is possible to determine metallicities three times more precisely for a given photometric error. We followed the SGB procedure of inserting absolute $M_{T_1}$ magnitudes and intrinsic $(C-T_1)_0$ colours for the clusters into Fig. 4 of \citetads[]{gs99} to roughly derive their metal abundances ([Fe/H]) by interpolation (Fig. \ref{f:gs}). The derived metallicities were corrected for age effects following the prescriptions given in \citetads[]{g03}. The final age-corrected metallicities for SL\,33 and BSDL\,3158 are listed in column (8) of Table \ref{t:results}. They agree very well with the Z values associated with the isochrones that best resemble the cluster features. The \citetads[]{gir02} models are computed for [Fe/H] = -1.3 and -0.7 dex, but not for intermediate-metallicity values. We finally adopted for our cluster sample averaged values of the ages and metallicities derived from the two different procedures.\\

\begin{table*}
  \caption{Fundamental parameters of LMC clusters}
  \centering
  \begin{tabular}{lccccccc}
  \hline \hline
   Name  &  Deprojected & $E(B-V)$ & $\delta T_1$ & $\delta T_1$ Age & Isochrone Age & [Fe/H]$_{isochrone}$ & [Fe/H]$_{SGB}$  \\
         &   Distance ($^{\circ}$) &  &  & (Gyr)  & (Gyr) &  &  \\
 \hline
SL\,33     & 5.1 & 0.12 & -- & -- & 2.0 $\pm$ 0.2 & -0.4 & -0.6 \\  
SL\,41     & 5.1 & 0.12 & 1.0 & 1.4 $\pm$ 0.4 & 1.3 $\pm$ 0.1 & -0.4 & -- \\
KMHK\,123  & 5.0 & 0.12 & -- & -- & 1.1 $\pm$ 0.1 & -0.7 & -- \\
KMHK\,128  & 5.4 & 0.11 & -- & -- & 1.4 $\pm$ 0.2 & -0.7 & -- \\     
LW\,69     & 4.6 & 0.12 & 1.4 & 1.7 $\pm$ 0.4 & 1.6 $\pm$ 0.2 & -0.7 & -- \\
KMHK\,151  & 5.1 & 0.12 & 1.1 & 1.4 $\pm$ 0.4 & 1.3 $\pm$ 0.1 & -0.7 & -- \\
SL\,54     & 4.9 & 0.12 & -- & -- & 0.9 $\pm$ 0.1 & -0.4 & -- \\
SL\,73     & 4.7 & 0.12 & 1.3 & 1.6 $\pm$ 0.4 & 1.6 $\pm$ 0.2 & -0.7 & -- \\
SL\,72     & 4.4 & 0.13 & -- & -- & 0.25 $\pm$ 0.03 & -0.4 & -- \\     
BSDL\,594  & 3.4 & 0.05 & 0.8 & 1.3 $\pm$ 0.4 & 1.4 $\pm$ 0.2 & -0.4 & --   \\
BSDL\,654  & 3.6 & 0.03 & -- & -- & 0.22 $\pm$ 0.03 & 0.0 & -- \\
BSDL\,665  & 3.6 & 0.03 & -- & -- & 0.9 $\pm$ 0.1 & -0.4 & -- \\
BSDL\,675  & 3.0 & 0.06 & 1.0 & 1.4 $\pm$ 0.4 & 1.1 $\pm$ 0.1 & -0.4 & -- \\ 
HS\,130    & 2.5 & 0.06 & -- & -- & 0.14 $\pm$ 0.02 & -0.4 & -- \\
BSDL\,761  & 3.6 & 0.04 & -- & -- & 0.14 $\pm$ 0.02 & -0.4 & -- \\
BSDL\,779  & 3.3 & 0.04 & -- & -- & 0.10 $\pm$ 0.01 & 0.0 & -- \\
HS\,156    & 2.5 & 0.06 & 0.9 & 1.3 $\pm$ 0.4 & 1.1 $\pm$ 0.1 & -0.4 & --\\
HS\,178    & 3.5 & 0.04 & -- & -- & 0.63 $\pm$ 0.07 & -0.4 & -- \\
LW\,211    & 4.7 & 0.10 & -- & -- & 1.8 $\pm$ 0.2 & -0.7 & -- \\ 
C11        & 3.4 & 0.10 & -- & -- & 0.32 $\pm$ 0.05 & -0.4 & -- \\
BSDL\,3158 & 3.5 & 0.10 & 1.7 & 2.1 $\pm$ 0.4 & 2.0 $\pm$ 0.2 & -0.7 & -0.8 \\
KMHK\,1702 & 5.3 & 0.11 & 0.7 & 1.2 $\pm$ 0.4 & 1.0 $\pm$ 0.1 & -0.7 & --\\   
SL\,870    & 5.4 & 0.09 & 0.7 & 1.2 $\pm$ 0.4 & 1.1 $\pm$ 0.1 & -0.4 & --\\ 
\hline 
\label{t:results}
\end{tabular}
\end{table*}

\section{Discussion}

The present sample includes 23 mostly unstudied LMC clusters within the narrow age range of 0.1~-~2.1 Gyr, with metallicities between 0.0 and -0.8 dex. To search for possible variations of ages and metallicities as a function of cluster position in the LMC, we calculated deprojected angular distances for our sample using equation (1) of \citetads[]{cla05}, 
assuming that all clusters are part of the LMC inclined disc. We adopted i $\approx$ 35.8$^{\circ}$ and p’ = 145$^{\circ}$ for the tilt of the LMC plane and the position angle of the line of nodes, respectively \citepads[]{os02}. The resulting deprojected angular distances are listed in column (2) of Table \ref{t:results}. \\

\begin{figure}
 \includegraphics[width=85mm]{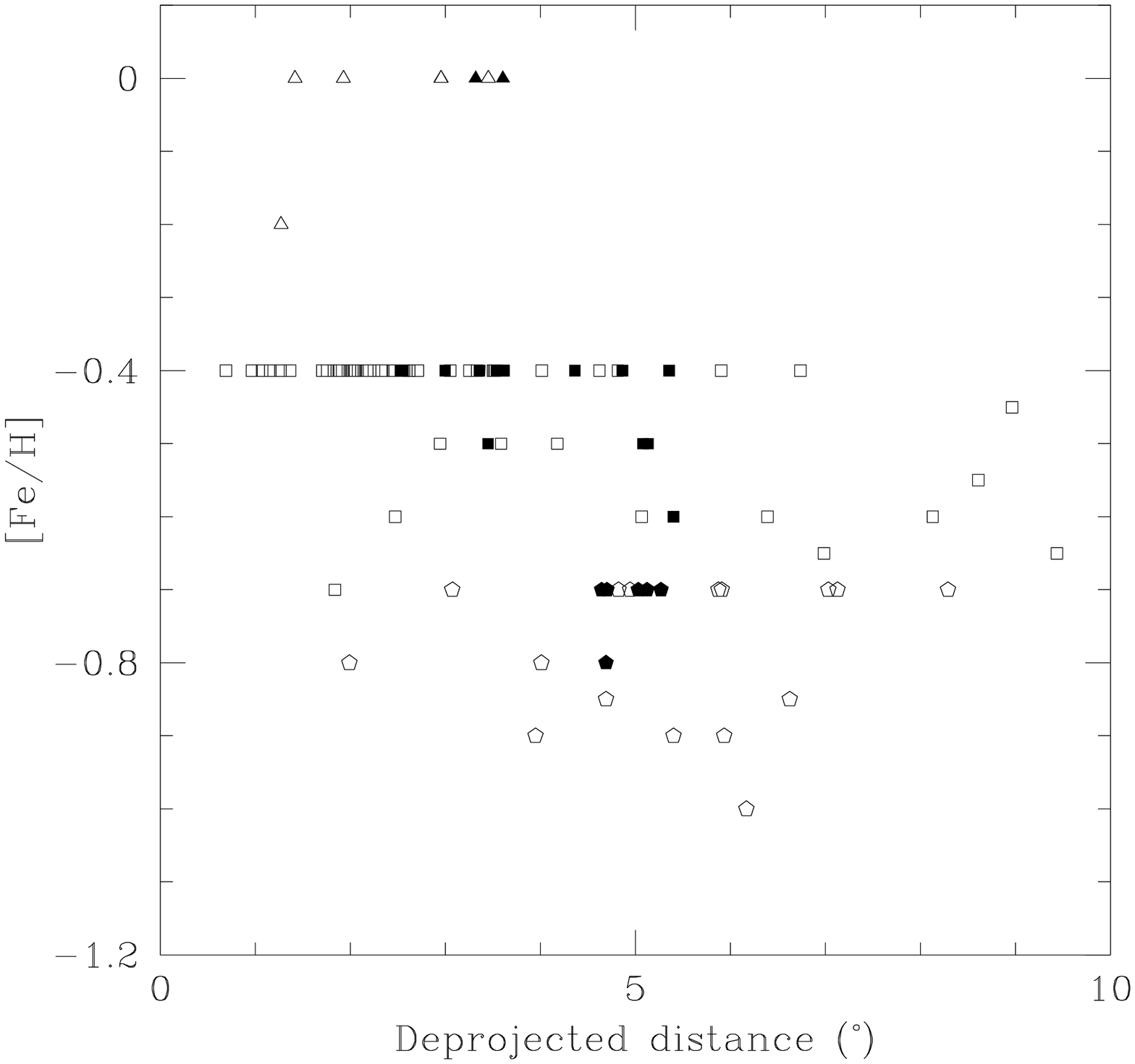}
 \caption{Metallicities as a function of deprojected angular distances from the LMC centre for the 23 studied clusters (filled symbols). Clusters included in Table \ref{t:wcls} are represented by open symbols. Triangles, squares, and pentagons correspond to the following [Fe/H] intervals: [Fe/H]~$>$~-0.4; -0.4~$\ge$~[Fe/H]~$>$~-0.7, and [Fe/H]~$\le$~-0.7, respectively. The old 
 cluster ESO\,121-03 has not been plotted.} 
 \label{f:dist1}
\end{figure}

\begin{figure}
 \includegraphics[width=85mm]{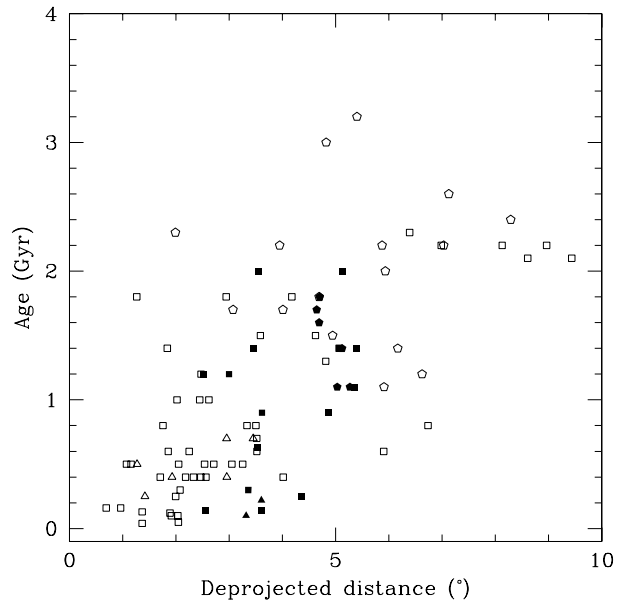}
 \caption{Ages as a function of deprojected angular distances from the LMC centre for the 23 studied clusters (filled symbols). Clusters included in Table 5 are represented by open symbols. Triangles, squares, and pentagons correspond to the following [Fe/H] intervals: [[Fe/H]~$>$~-0.4; -0.4~$\ge$~[Fe/H]~$>$~-0.7, and [Fe/H]~$\le$~-0.7, respectively. The old cluster ESO\,121-03 has not been plotted.} 
 \label{f:dist2}
\end{figure}

Fig. \ref{f:dist1} shows that in the comparatively small range of deprojected angular distances considered here (2.5$^{\circ}$~-~6$^{\circ}$), the most metal-rich clusters of our sample (filled symbols) tend to lie closer to NGC\,1928, whose position was adopted as the LMC centre ($\alpha_{2000}$ = 5$^h$20$^m$57$^s$, $\delta_{2000}$ = -69$^{\circ}$28'41''). Filled symbols of Fig.  \ref{f:dist2} show how our derived cluster ages vary as a function of the deprojected distances. As the clusters become older, their corresponding deprojected distances tend to be proportionally larger, thus supporting previous results in similar ranges of deprojected angular distances for clusters \citepads[e.g.,][]{p09} and field stars \citepads[]{pg12}. \\                                                                         

To examine what the above mentioned position-age-metallicity relationships are like when the range of deprojected angular distances is enlarged, we searched in the literature for previous LMC cluster studies in which ages and metallicities were determined on a similar scale as that of the present cluster sample. We found a total of 76 LMC star clusters with ages and metallicities 
derived from the Washington $C,T_1$ technique using CTIO telescopes and applying the same methods as in the current study (Table \ref{t:wcls}). Our 23 clusters increase the current total amount of LMC clusters using Washington photometry by $\sim$ 30\%. We calculated deprojected distances for these additional 76 clusters and included them in column (2) of Table \ref{t:wcls}. We plot in Figs. \ref{f:dist1} and \ref{f:dist2} their ages and metallicities as a function of their deprojected distances for three metallicity intervals, [Fe/H]~$>$~-0.4 (triangles), -0.4~$\ge$~[Fe/H]~$>$~-0.7 (squares), and [Fe/H]~ $\leq$~-0.7. When the new 76 clusters are added, the range of deprojected distances increases by a factor of $\sim$ 2.5. Fig. \ref{f:dist1} now reveals that although the most metal-rich clusters tend to be located closer to the galaxy centre, clusters with [Fe/H]~$<$~-0.4 appear to be distributed over the entire LMC disc, with a high dispersion. This fact reinforces the idea of the nonexistence of a radial metallicity gradient in the LMC, as suggested in several previous studies \citepads[e.g.,][]{gro06,ca08,p09}. Fig. \ref{f:dist2} includes clusters younger than 3.2 Gyr, since ESO\,121-03 ($\sim$ 8.5 Gyr) has not been plotted. As expected, the most metal-poor clusters turn out to be also the older ones. Moreover, cluster formation seems to be concentrated in the inner LMC disc, since younger clusters were formed closer to the LMC centre than the older ones.\\

\section{Summary and conclusions}

We continued the systematic study of star clusters of the LMC carried out using the Washington photometric system. We presented $(T_1,C-T_1)$ CMDs for 23 star clusters that lie within the inner disc and outer regions of the LMC. Only five of these clusters have previous age estimates in the literature. However, no metallicity estimate whatsoever has been found for any cluster of our sample. Ages and metallicities were determined by two different methods. We compared the CMDs with theoretical isochrones in the Washington system and also estimated ages using the magnitude difference between the RGC and the MSTO, and derived metallicities by comparing the giant branches with standard calibrating clusters. The two methods applied for determining ages and metallicities agree well with each other. Fourteen clusters are found to be IACs (1-3 Gyr), with [Fe/H] values ranging from -0.4 to -0.7. The remaining nine objects are younger than 1 Gyr, with metallicities between 0.0 and -0.4. By combining the current results with those of a sample of 76 additional clusters with ages and metallicities derived on a scale similar to that of the present work, we confirmed previous findings regarding the chemical evolution of the LMC. We found no evidence of a metallicity gradient and also found that the younger clusters were formed closer to the LMC centre than the older 
ones.\\

\begin{acknowledgements}
We thank the staff and personnel at CTIO for hospitality and assistance during the observations. We gratefully acknowledge financial support from the Argentinian institutions CONICET, FONCYT and SECYT (Universidad Nacional de C\'ordoba). D.G. gratefully acknowledges support from the Chilean BASAL Centro de Excelencia en Astrof\'isica y Tecnolog\'ias Afines (CATA) grant PFB-06/2007. This work is based on observations made at Cerro Tololo Inter-American Observatory , which is operated by AURA, Inc., under cooperative agreement with the National Science Foundation. T.P. wishes to thank L. Bassino, R. Leiton and L. Macri for their valuable guidance and discussion in the reduction process of MOSAIC data. We appreciate the valuable comments and suggestions of the anonimous referee, which helped us to improve the manuscript. This research has made use of the SIMBAD database, operated at CDS, Strasbourg, France; also the SAO/NASA Astrophysics data (ADS).
\end{acknowledgements}

\bibliographystyle{aa}    
\bibliography{palma}        

\newpage
\begin{longtable}{lcccc}
  \caption{\label{t:wcls} Ages and metallicities of LMC star clusters derived from Washington photometry .}\\
  \hline \hline
   Cluster name  & Deprojected &Age (Gyr) & [Fe/H] & Reference \\
         &  distance ($^\circ$) &  &   &   \\
  \hline
  \endfirsthead
  \caption{continued.} \\
  \hline \hline
   Cluster name  & Deprojected &Age (Gyr) & [Fe/H] & Reference \\
         &  distance ($^\circ$) &  &   &   \\
  \hline 
  \endhead
  \hline
  \endfoot
  SL\,8       & 4.2 & 1.6 / 1.8 & -0.5 &  1,2  \\
  NGC\,1697   & 3.5 & 0.7$\pm$0.1 & 0.0 & 3  \\
  HS\,38      & 4.0 & 0.4$\pm$0.1 & -0.4 & 4  \\
  KMHK\,229   & 2.6 & 1.0$\pm$0.2 & -0.4 & 4  \\
  H88-26      & 3.3 & 0.8$\pm$0.2 & -0.4 & 4  \\
  H88-40      & 3.5 & 0.7$\pm$0.2 & -0.4 & 4  \\
  SL\,126     & 8.9 & 1.9 / 2.2 & -0.45 & 1,2  \\
  SL\,133     & 5.9 & 2.2$\pm$0.3 & -0.7 & 3 \\
  H88-55      & 3.3 & 0.5$\pm$0.1 & -0.4 & 4  \\
  SL\,154     & 3.0 & 0.5$\pm$0.1 & -0.4 & 4  \\
  KMHK\,506   & 2.2 & 0.6$\pm$0.1 & -0.4 & 4  \\
  SL\,218     & 2.0 & 0.05$\pm$0.01 & -0.4 & 5  \\
  NGC\,1836   & 1.9 & 0.4$\pm$0.1 & 0.0 & 6  \\
  BRHT\,4b    & 1.9 & 0.10$\pm$0.02 & -0.4 & 5  \\
  NGC\,1839   & 1.9 & 0.12$\pm$0.02 & -0.4 & 5  \\
  NGC\,1838   & 2.0 & 0.10$\pm$0.02 & -0.4 & 5  \\
  SL\,229     & 2.1 & 0.3$\pm$0.1 & -0.4 & 4 \\
  SL\,244     & 1.8 & 1.6 / 1.4$\pm$0.3 & -0.7 & 1,8  \\
  BSDL\,716   & 2.2 & 0.4$\pm$0.1 & -0.4 &  4 \\
  SL\,262     & 8.6 & 2.1 & -0.55 & 1,2 \\
  HS\,151     & 1.8 & 0.8$\pm$0.2 & -0.4 & 4 \\  
  NGC\,1860   & 1.4 & 0.25$\pm$0.5 & 0.0 & 6  \\
  H88-188     & 2.7 & 0.5$\pm$0.1 & -0.4 & 4  \\
  HS\,154     & 2.5 & 0.5$\pm$0.1 & -0.4 & 4  \\
  SL\,293     & 2.5 & 0.4$\pm$0.1 & -0.4 & 4  \\
  NGC\,1863   & 1.4 & 0.04$\pm$0.01 & -0.4 & 5  \\
  SL\,300     & 2.6 & 0.4$\pm$0.1 & -0.4 & 4  \\
  NGC\,1865   & 1.3 & 0.9 / 0.5$\pm$0.1 & -0.2 & 1,6  \\
  BSDL\,1024  & 1.0 & 0.16$\pm$0.03 & -0.4 & 4  \\
  BSDL\,1035  & 1.2 & 0.5$\pm$0.1 & -0.4 & 4  \\
  H88-245     & 0.7 & 0.16$\pm$0.04 & -0.4 & 4  \\
  SL\,351     & 1.1 & 0.5$\pm$0.1 & -0.4 & 4  \\
  SL\,359     & 1.3 & 1.8$\pm$0.3 & -0.4 & 1,8  \\
  SL\,388     & 7.0 & 2.6 / 2.2 & -0.65 & 1,2  \\
  IC\,2134    & 6.9 & 1.0 & ---   &  2  \\
  SL\,451     & 7.0 & 2.2 & -0.7  & 1,2   \\
  SL\,446A    & 2.0 & 2.3$\pm$0.3 & -0.8 & 1,8 \\
  SL\,444     & 2.0 & 0.5$\pm$0.1 & -0.4 & 6  \\
  SL\,490     & 4.8 & 1.3$\pm$0.3 & -0.4 & 7  \\
  SL\,505     & 2.5 & 1.6 / 1.2$\pm$0.3 & -0.6 & 1,8 \\
  SL\,510     & 1.4 & 0.13$\pm$0.03 & -0.4 & 4  \\
  SL\,509     & 6.6 & 1.4 / 1.2 & -0.85 & 1,2  \\
  LW\,224     & 3.0 & 0.7$\pm$0.1 & 0.0 & 6  \\
  LW\,231     & 6.7 & 0.8$\pm$0.3 & -0.4 & 7  \\
  NGC\,1997   & 7.1 & 2.6$\pm$0.5 & -0.7 & 3 \\
  SL\,548     & 3.0 & 0.4$\pm$0.1 & 0.0 & 6 \\
  SL\,555     & 3.1 & 1.6 / 1.7$\pm$0.3 & -0.7 & 1,8  \\
  SL\,549     & 5.9 & 1.3 / 2.0$\pm$0.3 & -0.9 & 1,8  \\
  KMHK\,1045  & 1.9 & 0.6$\pm$0.1 & -0.4 & 4  \\
  KMHK\,1055  & 2.0 & 1.0$\pm$0.2 & -0.4 & 4  \\
  SL\,588     & 1.7 & 0.4$\pm$0.1 & -0.4 & 4  \\
  NGC\,2093   & 2.0 & 0.25$\pm$0.05 & -0.4 & 4  \\
  SL\,663     & 4.8 & 3.0$\pm$0.8 & -0.7 & 3  \\
  SL\,674     & 3.9 & 2.1 / 2.2$\pm$0.3 & -0.9 & 1,8  \\
  SL\,678     & 4.0 & 1.8$\pm$0.3 & -0.8 & 8  \\
  H88-333     & 2.3 & 0.4$\pm$0.1 & -0.4 & 4  \\
  BSDL\,2995  & 2.4 & 1.0$\pm$0.2 & -0.4 & 4  \\
  H7          & 3.2 & 1.4 & --- & 1 \\
  SL\,769     & 2.9 & 1.8 & -0.5  & 2  \\
  KMHK\,1507  & 3.5 & 0.8$\pm$0.3 & -0.4 & 7  \\
  SL\,775     & 3.5 & 0.6$\pm$0.3 & -0.4 & 7  \\
  OHSC\,28    & 8.3 & 2.4$\pm$0.5 & -0.7 & 3  \\
  NGC\,2161   & 5.9 & 1.1$\pm$0.3 & -0.7 & 9  \\
  NGC\,2153   & 4.6 & 1.3 & ---  & 1  \\
  NGC\,2155   & 5.4 & 3.2$\pm$0.6 & -0.9 & 10  \\
  SL\,817     & 3.6 & 2.5 / 1.5 & -0.5 & 1,2  \\
  ESO\,121-03 & 10.4 & 8.5 & -1.05 & 1,2  \\
  SL\,842     & 8.1 & 1.9 / 2.2 & -0.6 & 1,2  \\
  NGC\,2213   & 4.6 & 1.5 & -0.4 & 11  \\
  SL\,862     & 4.7 & 1.8 & -0.85 & 1,2  \\
  OHSC\,33    & 6.2 & 1.2 / 1.4 & -1.0 & 1,2  \\
  SL\,874     & 4.9 & 1.5$\pm$0.3 & -0.7 & 9  \\
  KMHK\,1719  & 5.1 & 1.4$\pm$0.3 & -0.6 & 9  \\
  LW\,469     & 5.9 & 0.6$\pm$0.1 & -0.4 & 4  \\
  SL\,896     & 6.4 & 2.3$\pm$0.3 & -0.6 & 10  \\
  OHSC\,37    & 9.4 & 2.1 & -0.65 & 1,2  \\  
\end{longtable}
\tablebib{
(1) \citetads[]{g97}; (2) \citetads[]{b98}; (3) \citetads[]{p09}; (4) \citetads[]{p12}; (5) \citetads[]{p03a}; (6) \citetads[]{p03b}; (7) 
\citetads[]{pal11}; (8) \citetads[]{g03}; (9) \citetads[]{petal11}; (10) \citetads[]{p02}; (11) \citetads[]{g87}. 
}

\end{document}